\documentclass[aps, nofootinbib,superscriptaddress, preprintnumbers, longbibliography]{article}

\usepackage{jcappub}
\usepackage[utf8]{inputenc}
\usepackage{amsmath,amsfonts,amssymb}
\usepackage{graphicx}
\usepackage[toc,page]{appendix}
\usepackage{cleveref}
\usepackage[loose]{units}
\usepackage{siunitx} 
\usepackage{booktabs}
\usepackage{orcidlink}

\newcommand{\beqn}{\begin{equation}}
\newcommand{\eeqn}{\end{equation}}
\newcommand{\dd}{\mathrm{d}}
\newcommand{\gmn}{g_{\mu\nu}}
\newcommand{\fmn}{f_{\mu\nu}}

\newcommand{\mFP}{m_\mathrm{FP}}
\newcommand{\ba}{\bar\alpha}

\newcommand{\OFP}{\Omega_\mathrm{FP}}
\newcommand{\OL}{\Omega_\Lambda}

\begin{document}

\title{
Combining cosmological and local bounds on bimetric theory}

\author[a,b]{Angelo Caravano,}
\author[a]{Marvin L\"uben \orcidlink{0000-0002-1259-7356},}
\author[b,c]{Jochen Weller \orcidlink{0000-0002-8282-2010}}

\affiliation[a]{Max-Planck-Institut f\"ur Physik (Werner-Heisenberg-Institut),\\
 F\"ohringer Ring 6, 80805 Munich, Germany}
\affiliation[b]{
	Universit\"ats-Sternwarte M\"unchen, Fakult\"at f\"ur Physik, Ludwig-Maximilians Universit\"at,\\ Scheinerstr. 1, 81679 M\"unchen, Germany}
\affiliation[c]{Max Planck Institute for Extraterrestrial Physics,\\
Giessenbachstr. 1, 85748 Garching, Germany}

\emailAdd{caravano@usm.lmu.de}
\emailAdd{mlueben@mpp.mpg.de}
\emailAdd{jochen.weller@usm.lmu.de}

\begin{flushright} 
\texttt{MPP-2021-6}
\end{flushright}

\abstract{
Ghost-free bimetric theory describes two nonlinearly interacting spin-2 fields, one massive and one massless, thus extending general relativity.
We confront bimetric theory with observations of Supernovae type 1a, Baryon Acoustic Oscillations and the Cosmic Microwave Background in a statistical analysis, utilising the recently proposed physical parametrisation.
This directly constrains the physical parameters of the theory, such as the mass of the spin-2 field and its coupling to matter.
We find that all models under consideration are in agreement with the data.
Next, we compare these results to bounds from local tests of gravity.
Our analysis reveals that all two- and three parameter models are observationally consistent with both cosmological and local tests of gravity.
The minimal bimetric model (only $\beta_1$) is ruled out by our combined analysis.
}

\maketitle

\section{Introduction}

In our current era of precision cosmology, cosmological tests of gravity with remarkable accuracy become available.
On the one hand, the direct detection of gravitational waves by the LIGO/Virgo collaboration~\cite{Abbott:2016blz} has confirmed some of the most fundamental predictions of general relativity -- gravitational waves~\cite{Einstein:1918btx}.
Moreover, the joint detection of gravitational waves with their electromagnetic counterpart~\cite{TheLIGOScientific:2017qsa} show that gravitational waves indeed travel at the speed of light to large precision~\cite{Monitor:2017mdv}.
On the other hand, measurements of the cosmic microwave background~\cite{Aghanim:2018eyx,Aiola:2020azj} and local observables~\cite{Wong:2019kwg,Riess:2019cxk,Freedman:2020dne,Pesce:2020xfe} challenge the standard model of cosmology.
In fact, the discrepancy between the inferred value for the Hubble parameter today ($H_0$) from early and late-time measurements represents a substantial tension~\cite{Bernal:2016gxb}.
From a theoretical perspective, the cosmological constant problems remain unsolved~\cite{Weinberg:1988cp,Martin:2012bt}.
In addition, concerns about the quantum consistency of a positive cosmological constant have been raised, in the context of quantum breaking~\cite{Dvali:2013eja,Dvali:2014gua,Dvali:2017eba,Dvali:2018fqu} and the swampland\footnote{For a critical perspective on the de Sitter conjectures and the cosmological implications, see e.g.~\cite{Akrami:2018ylq}.}~\cite{Obied:2018sgi,Agrawal:2018own,Garg:2018reu,Ooguri:2018wrx}.

Therefore, it is worth investigating theories beyond the standard model of cosmology, which modify the gravitational theory or replace the cosmological constant by some dark energy~\cite{Joyce:2016vqv}.
To modify gravity, one has to break one of the axioms of Lovelock's theorem~\cite{Lovelock:1971yv}.
A particularly simple possibility is to add new degrees of freedom to the gravitational sector.
The resulting scalar-tensor~\cite{Horndeski:1974wa,Deffayet:2009wt} and vector-tensor~\cite{Tasinato:2014eka,Heisenberg:2014rta} theories are already tightly constrained by gravitational wave observations and forced into their simplest forms~\cite{Creminelli:2017sry,Sakstein:2017xjx,Ezquiaga:2017ekz,Baker:2017hug}.
Also massive gravity~\cite{deRham:2010kj,Hassan:2011hr} is tightly constrained observationally~\cite{deRham:2016nuf}.
Contrarily, bimetric theory~\cite{Hassan:2011zd} remains mostly unconstrained by the determination of the propagation speed of gravitational waves~\cite{Max:2017flc,Max:2017kdc}.
In the present paper, we will further investigate the observational viability of the latter theory.

Historically, the question of whether the graviton can have a finite mass goes back to Fierz and Pauli, who presented a linear theory of a massive graviton in~\cite{Fierz:1939ix}.
However, this theory failed even basic solar-system tests of gravity due to the van Dam-Veltman-Zakharov (vDVZ) discontinuity~\cite{Zakharov:1970cc,vanDam:1970vg}.
Shortly after, Vainshtein argued that nonlinear terms might cure the discontinuity and render a nonlinear theory of a massive graviton observationally viable~\cite{Vainshtein:1972sx}.
However, Boulware and Deser argued that a fully nonlinear theory of a massive graviton will inevitably introduce an additional degree of freedom (d.o.f.) with negative norm\footnote{This additional degree of freedom is commonly referred to as Boulware-Deser ghost, which represents a Laplacian instability~\cite{Ostrogradsky:1850fid,Pais:1950za,Boulware:1973my}.
Such an instability spoils unitarity and renders the theory ill-defined.}.
Rather recently, de Rham, Gabadadze and Tolley identified a loop-hole in the argument, which lead to the construction of a fully nonlinear and ghost-free theory of massive gravity~\cite{deRham:2010ik,deRham:2010kj,Hassan:2011hr}\footnote{Since then, the ghost-freedom of massive gravity and its extensions was confirmed in a series of papers utilising various methods, see e.g.~\cite{deRham:2011rn,Hassan:2011tf,Hassan:2011ea,Hassan:2012qv,Hassan:2018mbl}.
For a pedagogical paper on how to consistently count the number of propagating degrees of freedom in first-order field theories, see~\cite{ErrastiDiez:2020dux}.}.
The construction of a nonlinear theory for a massive graviton requires the introduction of a reference metric, which is kept fixed.
A natural extension is to promote this fixed reference metric to be dynamical, which leads to ghost-free bimetric theory~\cite{Hassan:2011zd}.

While massive gravity describes a single massive graviton, bimetric theory contains a massless and a massive spin-2 field~\cite{Hassan:2011zd}.
This yields distinct phenomenological features.
The mass $\mFP$ of a single massive graviton is tightly constrained.
While gravitational wave observations provide the upper bound of $\mFP\lesssim 10^{-22}\,{\rm eV}$~\cite{Baker:2017hug}, the most stringent constraints come from the solar system with $\mFP\lesssim 10^{-33}\,{\rm eV}$~\cite{deRham:2016nuf}.
Moreover, massive gravity does not give rise to viable homogenous and isotropic cosmological solutions~\cite{DAmico:2011eto,Gumrukcuoglu:2011ew,Gumrukcuoglu:2011zh,DeFelice:2012mx}.
Reviews on theoretical and phenomenological aspects of massive gravity can be found in~\cite{Hinterbichler:2011tt,deRham:2014zqa,deRham:2016nuf}.

Contrarily, bimetric theory has a viable and rich phenomenology.
In the context of cosmology, the theory allows for homogeneous and isotropic cosmological solutions with late-time de Sitter attractor.
The (self-)interactions of the massive and massless spin-2 fields give rise to dynamical dark energy, even in the absence of vacuum energy, i.e. so-called self-accelerating solutions~\cite{Volkov:2011an,vonStrauss:2011mq,Comelli:2011zm,Volkov:2012cf,Volkov:2012zb,Akrami:2012vf,DeFelice:2014nja,Volkov:2013roa,Konnig:2013gxa}.
Bimetric theory can alleviate the $H_0$-tension if the spin-2 mass is of the order of the Hubble constant~\cite{Mortsell:2018mfj,Luben:2019yyx}
Cosmological perturbations and cosmic structure formation remain challenging due to the gradient instability at early times~\cite{Comelli:2012db,Khosravi:2012rk,Berg:2012kn,Sakakihara:2012iq,Konnig:2014dna,Comelli:2014bqa,DeFelice:2014nja,Solomon:2014dua,Konnig:2014xva,Lagos:2014lca,Konnig:2015lfa,Akrami:2015qga}, which might be cured by nonlinear terms in analogy to the Vainshtein mechanism~\cite{Aoki:2015xqa,Mortsell:2015exa,Luben:2019yyx}.
The theory contains the massive spin-2 field as dark-matter candidate~\cite{Aoki:2016zgp,Babichev:2016hir,Babichev:2016bxi,Chu:2017msm}.

Spherically symmetric solutions~\cite{Comelli:2011wq} have been confronted with observations on galaxy cluster~\cite{Platscher:2018voh}, galactic~\cite{Enander:2013kza,Enander:2015kda,Platscher:2018voh}, and solar system scales~\cite{Enander:2015kda,Hohmann:2017uxe,Luben:2018ekw}.
The nonlinear Vainshtein mechanism suppresses modifications in the gravitational sector on small scales~\cite{Babichev:2013pfa,Enander:2015kda}.
Going further, gravitational waves were studied in~\cite{DeFelice:2013nba,Lagos:2014lca,Cusin:2014psa,Amendola:2015tua,Johnson:2015tfa,Sakakihara:2015naa,Fasiello:2015csa,Cusin:2015pya,Sakakihara:2016ubu,Biagetti:2017viz,Dimastrogiovanni:2018uqy,Jimenez:2019lrk}, with an application to binary mergers~\cite{Max:2017flc,Max:2017kdc}.
Summarising, bimetric theory is in agreement with the current corresponding observational bounds and even leads to possibly detectable effects, but it is not clear, whether the theory is consistent with all theoretical and observational constraints simultaneously.

To consistently combine the various observational and theoretical constraints on the parameter space, a new parametrisation of bimetric solutions was developed in~\cite{Luben:2020xll}.
The idea is to formulate solutions in terms of the following physical parameters: mass of the spin-2 field, its coupling constant to ordinary matter and the effective asymptotic cosmological constant.
The procedure automatically yields theoretical constraints that ensure a viable cosmic expansion history, i.e. real-valued, non-singular and devoid of the Higuchi ghost.

In this paper, we extend the analysis of~\cite{Luben:2020xll} to test bimetric cosmology against data from Supernovae type 1a (SN1a), Baryon Acoustic Oscillations (BAOs) and the Cosmic Microwave Background (CMB).
While bimetric theory has been tested against these data sets beforehand~\cite{vonStrauss:2011mq,Akrami:2012vf,Konnig:2013gxa,Dhawan:2017leu,Mortsell:2018mfj,Lindner:2020eez}, here we directly constrain the physical parameters.
We restrict our analysis to all theoretically viable bimetric models with up to three free interaction parameters\footnote{To be more precise, we consider all models with up to three free parameters $\beta_n$.
Strictly speaking, the parameters $\beta_{1,2,3}$ parametrise interactions, while the parameters $\beta_{0,4}$ parametrise vacuum energy.}.
In addition, we allow for non-zero spatial curvature in the statistical analysis.

Next, we compare the cosmological bounds to bounds from local tests of gravity.
We use tests of the Newtonian gravitational potential from laboratory to planetary scale as summarised in~\cite{Fischbach:1999bc,Adelberger:2003zx,Adelberger:2009zz,Murata:2014nra}.
The Yukawa parametrisation, where the gravitational potential is written as a superposition of the usual $1/r$-term and a Yukawa-term, is appropriate for our purposes.
We use the existing bounds to constrain the physical parameters of bimetric theory.
We add further bounds from tests of the scalar curvature, i.e. measurements of the deflection and time delay of light~\cite{Bertotti:2003rm,Will:2014kxa,Hohmann:2017uxe}.

The aforementioned frameworks are appropriate only in regimes without Vainshtein screening.
In screened spacetime regions these bounds are not trustworthy, because deviations from general relativity (GR) are suppressed.
However, only a subregion of the full bimetric parameter space supports Vainshtein screening.
In this paper, we relate the bounds of~\cite{Babichev:2013pfa,Enander:2015kda}, which ensure a working Vainshtein mechanism, to the physical parameters and identify the region of the physical parameter space that supports screening.
We will see that the Vainshtein mechanism does not affect the observational bounds from local tests of gravity, such that these are directly applicable.
Summarising, we consistently combine for the first time these observational and theoretical constraints from cosmology and local tests of gravity with each other.

This paper is organised as follows.
We start with a brief summary of bimetric theory, the cosmological solutions and the physical parameters in~\cref{sec:bimetric-theory}.
We proceed with the statistical analysis using cosmological data in~\cref{sec:statistical-analysis}.
In~\cref{sec:local-bounds}, we discuss the bounds from local tests of gravity and the Vainshtein screening, which we confront with the cosmological constraints.
We summarise our analysis and conclude in~\cref{sec:conclude}.

\section{Review of bimetric theory}
\label{sec:bimetric-theory}

In this section we briefly discuss bimetric theory, its mass spectrum, cosmological solutions and the physical parametrisation.
For a thorough introduction to the field we refer to~\cite{Schmidt-May:2015vnx}.

The construction of a nonlinear mass term involves two metric tensors, which we call $\gmn$ and $\fmn$.
The action of ghost-free bimetric theory for these metrics reads~\cite{deRham:2010ik,deRham:2010kj,Hassan:2011vm,Hassan:2011hr,Hassan:2011zd}
\begin{flalign}\label{eq:bimetric-action}
	S = m_{\rm g}^2\int \dd^4x \Bigg[ \sqrt{-g} R^{\rm g} + \alpha^2 \sqrt{-f}R^{\rm f} - 2\mathcal U\left(\sqrt{g^{-1}f}\right)\Bigg]
	+\int \dd^4x \sqrt{-g} \mathcal L_{\rm m}\left(g, \Phi\right)
\end{flalign}
where $R^{\rm g}$ and $R^{\rm f}$ are the Ricci scalars of $\gmn$ and $\fmn$, resp.
The parameter $m_{\rm g}$ is the Planck mass of $\gmn$ and $\alpha$ is the Planck mass of $\fmn$ normalised to $m_{\rm g}$.
The ghost-free bimetric potential $\mathcal U$ is defined in terms of the square-root matrix $S=\sqrt{g^{-1}f}$ and given by\footnote{Note
that bimetric theory is well-defined only if both metrics have a common time direction~\cite{Hassan:2017ugh}.}
\begin{flalign}
	\mathcal U \left( \sqrt{g^{-1}f} \right) = \sum_{n=0}^4\beta_n e_n\left(\sqrt{g^{-1}f}\right)\,.
\end{flalign}
Here, $\beta_n$ are real parameters with mass dimension two in our parametrisation.
The functions $e_n$ are the elementary symmetric polynomials~\cite{Hassan:2011vm}.
Matter fields are collectively denoted by $\Phi$, which minimally couple\footnote{This
setup is referred to singly-coupled bimetric theory.
More general ghost-free matter couplings exist in bimetric theory~\cite{deRham:2014naa,deRham:2014fha,Heisenberg:2014rka,Heisenberg:2015iqa,Hinterbichler:2015yaa,Luben:2018kll}.} to the metric $\gmn$ via the matter Lagrangian $\mathcal L_{\rm m}$.
Hence, $\gmn$ is the physical metric that defines the geometry in which the matter fields propagate.

Varying the action with respect to $g^{\mu\nu}$ and $f^{\mu\nu}$ yields the modified Einstein equations
\begin{flalign}\label{eq:mod-einstein}
	G^{\rm g}_{\mu\nu} + \mathcal U^{\rm g}_{\mu\nu} = \frac{1}{m_{\rm g}^2} T_{\mu\nu}\,,\qquad
	G^{\rm f}_{\mu\nu} + \alpha^{-2}\mathcal U^{\rm f}_{\mu\nu} = 0\,,
\end{flalign}
where $G^{\rm g,f}_{\mu\nu}$ are the Einstein tensors of $\gmn$ and $\fmn$, resp.
The terms coming from the variation of the bimetric potential are given by
\begin{flalign}
	\mathcal U^{\rm g}_{\mu\nu} = \sum_{n=0}^3 (-1)^n\beta_n g_{\mu\lambda} Y^\lambda_{(n)\nu}(S)\,,\qquad \mathcal U^{\rm f}_{\mu\nu} = \sum_{n=0}^3 (-1)^n\beta_{4-n} f_{\mu\lambda} Y^\lambda_{(n)\nu}(S^{-1})
\end{flalign}
where the matrices $Y^\lambda_{(n)\nu}$ are given in, e.g.~\cite{Hassan:2011vm}.
Since matter couples to the physical metric $\gmn$, the matter stress-energy tensor is given by
\begin{flalign}
	T_{\mu\nu} = \frac{-2}{\sqrt{-g}}\frac{\delta \sqrt{-g}\mathcal L_{\rm m}}{\delta g^{\mu\nu}}\,.
\end{flalign}
If the matter sector is invariant under diffeomorphisms, the stress-energy tensor of matter is conserved,
\begin{flalign}\label{eq:energy-conservation}
	\nabla^\mu\, T_{\mu\nu} = 0\,,
\end{flalign}
where $\nabla_\mu$ is the covariant derivative compatible with $\gmn$.
The Bianchi identity, $\nabla^\mu G^{\rm g}_{\mu\nu}=0$, then yields the so-called Bianchi constraint,
\begin{flalign}\label{eq:bianchi-constraint}
	\nabla^\mu \mathcal U^{\rm g}_{\mu\nu} =0\,.
\end{flalign}
The Bianchi identity in the $\fmn$-sector yields another Bianchi constraint, which however coincides with eq.~(\ref{eq:bianchi-constraint}) by diffeomorphism invariance.

\subsection{Vacuum solutions and mass eigenstates}

Before moving to cosmology, we briefly review an important class of solutions in bimetric theory~\cite{Hassan:2011zd,Hassan:2012wr}.
Let both metric tensors be proportional as $\fmn = c^2 \gmn$ with conformal factor $c$.
This ansatz is a solution to the modified Einstein equations only in vacuum, i.e. for $T_{\mu\nu}=0$.
The Bianchi constraint~(\ref{eq:bianchi-constraint}) implies that $c$ is a constant.
Upon this ansatz, the modified Einstein equations~(\ref{eq:mod-einstein}) simplify to
\begin{flalign}\label{e	q:mod-einstein-vac}
	G^{\rm g} + \Lambda_{\rm g}\, g_{\mu\nu} =0 \,, \qquad G^{\rm f}_{\mu\nu} + c^{-2}\Lambda_{\rm f}\, \fmn= 0\,,
\end{flalign}
where the effective cosmological constants are given in terms of the bimetric parameters as
\begin{flalign}
	\Lambda_{\rm g} = \beta_0 + 3\beta_1 c + 3\beta_2 c^2 + \beta_3 c^3\,,\qquad
	\Lambda_{\rm f} = \frac{1}{\alpha^2 c^2} \left( \beta_1 c + 3\beta_2 c^2 + \beta_3 c^3 +\beta_4 c^4 \right)\,.
\end{flalign}
Subtracting both equations in~(\ref{e	q:mod-einstein-vac}) and using that both metrics are conformally related yields $\Lambda_{\rm g} = \Lambda_{\rm f}\equiv \Lambda$, or in terms of the bimetric parameters
\begin{flalign}\label{eq:vacuum-solution}
	\alpha^2\beta_3 c^4 + (3\alpha^2\beta_2-\beta_4)c^3 + 3(\alpha^2\beta_1-\beta_3)c^2+(\alpha^2\beta_0-\beta_3)c-\beta_1 =0\,.
\end{flalign}
This is a polynomial in $c$ of degree 4.
Hence, there are up to four real-valued solutions for $c$ in terms of the bimetric parameters $\alpha$ and $\beta_n$, each corresponds to a vacuum solution.

Bimetric theory has a well-defined mass spectrum around proportional solutions.
Let us perturb the metrics around the background $\bar g_{\mu\nu}$ as
\begin{flalign}
	g_{\mu\nu} = \bar g_{\mu\nu} + \frac{1}{m_{\rm g}} \delta \gmn\,,\qquad \fmn = c^2\bar g_{\mu\nu}+\frac{c}{\alpha m_{\rm g}} \delta \fmn \,,
\end{flalign}
where $\delta\gmn \ll m_{\rm g}$ and $\delta \fmn \ll \alpha m_{\rm g}$ are small fluctuations around the proportional background $\bar f_{\mu\nu}=c^2 \bar g_{\mu\nu}$.
The mass eigenstates are linear superpositions of the metric fluctuations.
The spectrum contains a massless mode, $\delta G_{\mu\nu}$, and a massive mode, $\delta M_{\mu\nu}$, with Fierz-Pauli mass\footnote{The
mass term of $\delta M_{\mu\nu}$ has the Fierz-Pauli structure~\cite{Fierz:1939ix}.}
\begin{flalign}
	\mFP^2 = \left(1 + \frac{1}{\alpha^2 c^2}\right) c(\beta_1 + 2\beta_2 c+ \beta_3 c^2)\,.
\end{flalign}
We can express the metric fluctuations in terms of the mass eigenstates as
\begin{flalign}\label{eq:mixing-fluctuations}
	\delta \gmn = \frac{1}{\sqrt{1+\ba^2}}\left(\delta G_{\mu\nu} - \ba\, \delta M_{\mu\nu}\right)\,, \qquad \delta\fmn = \frac{1}{\sqrt{1+\ba^2}}\left(\delta M_{\mu\nu} + \ba\, \delta G_{\mu\nu}\right)\,.
\end{flalign}
Here, we defined $\ba = \alpha c$ that parametrises the mixing of the massless and the massive mode in the metric fluctuations.
From here we can already identify the GR and massive gravity limit of bimetric theory~\cite{Akrami:2015qga,Luben:2018ekw}.
The GR-limit is $\ba\rightarrow0$ because then the fluctuation of $\gmn$ is aligned with the massless mode while the fluctuation of $\fmn$ is aligned with the massive mode.
In the opposite limit $\bar\alpha\rightarrow\infty$ bimetric theory approaches massive gravity as then $\delta\gmn$ is aligned with the massive mode while $\delta\fmn$ coincides with the massless mode.

\subsection{FLRW solutions}

Assuming homogeneity and isotropy according to the cosmological principle, 
both metric can be cast into the Friedmann-Lema\^{i}tre-Robertson-Walker (FLRW)
form~\cite{Volkov:2011an,vonStrauss:2011mq,Comelli:2011zm},
\begin{equation}\label{eq:flrw-metric-ansatz}
        	\dd s_{\rm g}^2 = -\dd t^2 + a^2 \left( \frac{\dd r^2}{1-k r^2}+ r^2\dd \Omega^2\right)\,,\qquad
	\dd s_{\rm f}^2 = -X^2\dd t^2 + b^2 \left( \frac{\dd r^2}{1-kr^2}+r^2 \dd \Omega^2\right)\,,
\end{equation}
where $a$ and $b$ are the scale factors of $\gmn$ and $\fmn$, respectively, while $X$ is the lapse of $\fmn$.
These metric functions depend on timte $t$ only.
Further, $k$ is the spatial curvature, which must be common to both metrics~\cite{Nersisyan:2015oha}.
In this coordinate system, $k<0$, $k=0$, $k>0$ describes a closed, flat, or open universe, resp.
Let us define the Hubble rates and the scale factor ratio as
\begin{flalign}
	H = \frac{\dot a}{a}\,,\ \ H_{\rm f} = \frac{\dot b}{X b}\,, \ \ y=\frac{b}{a}\,,
\end{flalign}
where dot denotes derivative with respect to cosmic time $t$.

On the bidiagonal FLRW ansatz~\eqref{eq:flrw-metric-ansatz}, the Bianchi constraint~\eqref{eq:bianchi-constraint} simplifies to
\begin{flalign}
	(\dot b - X \dot a)(\beta_1 + 2\beta_2 y + \beta_3 y^2) = 0\,.
\end{flalign}
This equation has two branches of solution.
We use the so-called dynamical branch solution with $X=\dot b/\dot a$ for the remainder of this paper\footnote{On the
other branch of solutions, referred to as algebraic, the scale factor ratio is forced to be constant, $y={\rm const}$.
Since this solution is found to be pathological~\cite{Comelli:2012db,Cusin:2015tmf}, we focus on the dynamical branch solution.}.
Then the Hubble rates are related as $H = y H_{\rm f}$.

We take the matter source to be a perfect fluid with stress-energy tensor
\begin{flalign}
	T^{\mu\nu} = (\rho + p) u^\mu u^\nu + p\, g^{\mu\nu}
\end{flalign}
with energy density $\rho$, pressure $p$ and $u^\mu$ the $4$-velocity of the fluid.
We split energy density and pressure into a non-relativistic and a relativistic part as $\rho=\rho_{\rm m} + \rho_{\rm r}$ and $p=p_{\rm m} + p_{\rm r}$, resp.
The subscript $\rm m$ refers to non-relativistic matter, while the subscript $\rm r$ refers to radiation.
The conservation of energy-momentum~(\ref{eq:energy-conservation}) leads to the continuity equation
\begin{flalign}
	\dot \rho_{i} = -3 H (1+w_i) \rho_{i}\,,\quad i={\rm m, r}\,,
\end{flalign}
with $w_i = p_i / \rho_i$ the equation of state.
Therefore, the energy density of matter and radiation evolve with the scale factor $a$ of the physical metric as in standard cosmology.
To be explicit, the continuity equation is solved by
\begin{flalign}
	\rho_i = \rho_{i,0} a^{-3(1+w_i)}\,,
\end{flalign}
unless $w_i = -1$, with the integration constants chosen such that $\rho_{i,0}=\rho_i(1)$, where $a=1$ corresponds to the present time.
Radiation has an equation of state of $w_{\rm r}=1/3$ while for matter $w_{\rm m}=0$.

The time-time component of the two modified Einstein equations~(\ref{eq:mod-einstein}) become
\begin{equation}\label{eq:friedmann}
	3 H^2 + 3\frac{k}{a^2} = \frac{1}{m_{\rm g}^2}\left( \rho_{\rm de} + \rho_{\rm m} + \rho_{\rm r}\right)\,,\qquad
	3 H^2 + 3\frac{k}{a^2} = \frac{1}{m_{\rm g}^2}\rho_{\rm pot}\,.
\end{equation}
Here we defined the energy density of the dynamical dark energy induced by the
bimetric potential as
\begin{equation}\label{eq:dark-energy-density}
	\frac{\rho_{\rm de}}{ m_{\rm g}^2} = \beta_0 + 3\beta_1 y + 3\beta_2 y^2 + \beta_3 y^3
\end{equation}
with the time-dependent function $y$.
The Friedmann equation of $\fmn$ is sourced by potential energy only with energy density
\begin{flalign}
	\frac{\rho_{\rm pot}}{m_{\rm g}^2} =\frac{1}{\alpha^2}\left( \frac{\beta_1}{y} + 3\beta_2 + 3\beta_3 y + \beta_4 y^2\right)\,.
\end{flalign}
Interpreting the effect of spatial curvature as an energy source, we can define the energy density
\begin{flalign}
	\rho_{\rm k} = -3m_{\rm g}^2k a^{-2}\,.
\end{flalign}
Then, the Friedmann~\cref{eq:friedmann} of $\gmn$ can be written as
\begin{flalign}\label{eq:friedmann-final}
	3H^2 = \frac{1}{m_{\rm g}^2}\left(\rho_{\rm de} + \rho_{\rm k} + \rho_{\rm m} + \rho_{\rm r}\right)\,.
\end{flalign}

It remains to determine the time-evolution of $\rho_{\rm de}$, which in general cannot be written as a polynomial in $1/a$.
Subtracting the the Friedmann equations in~\eqref{eq:friedmann} yields
\begin{flalign}\label{eq:quartic-pol-y}
	\alpha^2\beta_3 y^4 + (3\alpha^2\beta_2-\beta_4)y^3 + 3(\alpha^2\beta_1-\beta_3)y^2+ \left(\alpha^2\beta_0-3\beta_2+\alpha^2\frac{\rho_{\rm m}+ \rho_{\rm r}}{m_{\rm g}^2}\right)y - \beta_1  = 0\,.
\end{flalign}
This equation determines the time evolution of $y$ in terms of the time evolution
of the matter energy densities.
It is a quartic polynomial in $y$, hence there are up to four real-valued solutions
with different time evolutions of $y$.
Following~\cite{vonStrauss:2011mq,Konnig:2013gxa}, we can distinguish three classes of solutions by inspecting the early-time behavior, for which the energy densities classically diverge:
\begin{itemize}
\item
\textit{Finite branch.}
The first possible early-time behavior is that $y$ approaches zero so as to compensate the diverging value of the energy density in the linear piece of the polynomial~\eqref{eq:quartic-pol-y}.
Hence, during early times the scale factor ratio can be approximated as $y\sim m_{\rm g}^2\beta_1/(\alpha^2\rho)$.
Existence of the finite branch requires $\beta_1>0$.
The scale factor ratio monotonically increases in time.
In the asymptotic future, the energy densities vanish and $y$ approaches a constant value $c$, which corresponds to the lowest-lying, strictly positive root of~\cref{eq:vacuum-solution}~\cite{Luben:2020xll}.
For positive matter and radiation energy densities, this branch of solution satisfies the cosmological stability bound, a generalization of the Higuchi bound to FLRW spacetime~\cite{Fasiello:2013woa,Konnig:2015lfa}.

\item
\textit{Infinite branch.}
The alternative is that $y$ is large as the energy densities are large.
Then, at early times, the evolution can be approximated as $y^3\sim -\rho/(m_{\rm g}^2\beta_3)$.
This reveals that the infinite branch exists only if $\beta_3<0$.
The scale factor ratio monotonically decreases in time.
In the asymptotic future, when the energy densities vanish, $y$ approaches the highest lying, strictly positive root of~\cref{eq:vacuum-solution}~\cite{Luben:2020xll}.
This branch, however, inevitably violates the cosmological stability bound and is therefore not a physical solution~\cite{Fasiello:2013woa,Konnig:2015lfa}.

\end{itemize}
Due to the high degree of the polynomial~\eqref{eq:quartic-pol-y}, there are further \textit{exotic branches} of solutions, which however do not give rise to a viable expansion history~\cite{Konnig:2015lfa}.
Therefore, we pick out the finite branch solution in~\cref{sec:statistical-analysis}.
The scale factor ratio then satisfies $0<y<c$~\cite{Konnig:2015lfa,Luben:2020xll}.

Let us pause for a moment and summarise.
The bimetric potential contributes a dynamical dark energy to the Friedmann equation in the form of $\rho_{\rm de}$, which depends on the scale factor ratio $y$.
Since $y$ approaches a constant value $c$ in the asymptotic future, $\rho_{\rm de}$ approximates the effect of a cosmological constant at late times.
We have identified two possible time evolutions for $y$ yielding a consistent expansion history.
Out of these, only the finite branch solution satisfies the cosmological stability bound.
Therefore, $y$ decreases when looking back in time, implying that also $\rho_{\rm de}$ decreases.
In other words, the (self-)interactions of the massive and massless spin-2 field are dynamically increasing as time proceeds.
Hence, the (self-)interaction energy is necessarily phantom or negative~\cite{Konnig:2015lfa,Luben:2019yyx}.

\subsection{Physical parametrisation}
\label{sec:into-physical-parametrisation}

The solutions presented thus far are parametrised in terms of the Planck mass ratio $\alpha$ and the interaction parameters $\beta_n$.
Vacuum solutions are labeled by $c$ that are the roots of~\cref{eq:vacuum-solution}.
The bimetric action and hence the equations of motion are invariant under a rescaling of these parameters implying that one of them is redundant.

In~\cite{Luben:2020xll} a parametrisation that is invariant under this rescaling is proposed and worked out.
It relies on the following physical parameters: the mass $\mFP$ of the spin-2 field, its coupling to matter $\bar\alpha$, and the effective cosmological constant $\Lambda$.
The hence baptised physical parametrisation allows to unambiguously combine various theoretical and observational bounds on the parameter space of bimetric theory.

Let us give more detail on the physical parametrisation and on the already inferred theoretical consistency bounds.
Since there are up to four real-valued roots $c$ of~\cref{eq:vacuum-solution}, the relation between the physical and interaction parameters is not unique.
However, a vacuum solution and the cosmic expansion history are well-defined only under the following conditions:
\begin{itemize}
\item
\textit{Consistent de Sitter vacuum.}
Without loss of generality we restrict ourselves to roots with $c>0$\footnote{Solutions
with $c<0$ can be mapped onto solutions with $c>0$ because the equations of motion are invariant under the combined rescaling $c\rightarrow-c$, $\beta_n\rightarrow (-1)^n\beta_n$.}.
We impose that the mass of the spin-2 field is positive, $\mFP>0$, and we restrict ourselves to de Sitter vacua, $\Lambda>0$.
For a massive spin-$2$ field propagating in de Sitter space, unitarity forbids
the mass to be arbitrarily small.
Instead, the Higuchi bound~\cite{Higuchi:1986py}
\begin{flalign}\label{eq:higuchi-bound}
	3\mFP^2 > 2\Lambda
\end{flalign}
has to be satisfied.
If the bound is violated, the helicity-$0$ mode of the massive spin-$2$ field
has a negative norm (Higuchi ghost).

\item
\textit{Consistent asymptotic future.}
The cosmic expansion has to approach a consistent de Sitter vacuum in the asymptotic future.
Since only the finite branch solution to the Friedmann equations is physical, the scale factor ratio satisfies $0<y<c$.
Hence, the lowest-lying strictly positive root of~\cref{eq:vacuum-solution} as the unique physical vacuum solution.
Identifying the asymptotic future of the cosmic expansion history with the physical vacuum solution implies further, model-dependent conditions on the bimetric parameters.

\item
\textit{Consistent cosmic expansion history.}
The consistency of the early universe for a cosmic expansion history on the finite branch requires~\cite{Konnig:2013gxa}
\begin{flalign}\label{eq:pos-b1}
	\beta_1>0\,,
\end{flalign}
as can be seen from~\cref{eq:friedmann}.
Otherwise $H^2\leq0$ during early times.

\end{itemize}
These restrictions imply a unique relation between the physical and interaction parameters.
The details for all bimetric models with up to three non-vanishing interaction parameters are worked out in Ref.~\cite{Luben:2020xll}.
For completeness, we summarise the explicit expressions in~\cref{tab:viable-cosmo-cond}.
Using the physical parametrisation, in~\cref{sec:statistical-analysis} we compute bounds from cosmological observations.

\section{Bounds from cosmological observations}
\label{sec:statistical-analysis}

We now compare bimetric theory to cosmological observables.
We restrict ourselves to background observables that constrain the Hubble rate at different redshifts.
In particular, we use Supernovae type 1a, Baryon Acoustic Oscillations and the first peak of the oscillations in the Cosmic Microwave Background, which we discuss in greater detail in~\cref{sec:data}.

Measurements of perturbative quantities are not included because of the aforementioned gradient instability in the scalar sector~\cite{Comelli:2012db,Khosravi:2012rk,Berg:2012kn,Sakakihara:2012iq,Konnig:2014dna,Comelli:2014bqa,DeFelice:2014nja,Solomon:2014dua,Konnig:2014xva,Lagos:2014lca,Konnig:2015lfa,Akrami:2015qga}.
The scale at which the gradient instability kicks in is set by the spin-2 mass~\cite{Luben:2019yyx}.
Taking nonlinearities into account possibly stabilises the perturbations at early times~\cite{Mortsell:2015exa,Aoki:2015xqa,Hogas:2019ywm}.
However, a framework to study cosmological phenomena on the perturbative level within bimetric theory is still pending.

\subsection{Parametrisation used for model fitting}

We have already explained how to get the time-evolution of the scale factor ratio $y$, the time evolution of $\rho_{\rm de}$ and hence the time evolution of $H$.
To ease these computations, it is convenient to introduce the energy density parameters
\begin{flalign}
	\Omega_{i} = \frac{\rho_{i}}{3 m_{\rm g}^2 H^2} \,, \quad i =\{{\rm de, k, m, b, r}, \gamma\}\,.
\end{flalign}
The Friedmann equation~\eqref{eq:friedmann-final} can then be written in the compact form
\begin{equation}
	\Omega_{\rm de}+\Omega_{\rm k}+\Omega_{\rm m}+\Omega_{\rm r} = 1\,.
\end{equation}
Evaluating this equation at present times yields the following relation between the parameters as
\begin{equation}\label{eq:friedmann-today}
	\Omega_{\rm de0}+\Omega_{\rm k0}+\Omega_{\rm m0}+\Omega_{\rm r0} = 1\,.
\end{equation}
We choose to express the parameter $\Omega_{\rm m0}$ in terms of the other parameters in our statistical analysis.
In order to compute $\Omega_{\rm de}$, we introduce the following constant parameters:
\begin{equation}
B_n = \frac{\alpha^{-n} \beta_n}{ 3 H_0^2 }\,,\
\Omega_{\rm FP}= \frac{ \mFP^2 }{ 3 H_0^2 }\,, \ 
\Omega_{\Lambda}= \frac{ \Lambda }{ 3 H_0^2 }\,.
\end{equation}
The relations between $B_n$ and the physical parameters $\bar\alpha$, $\OFP$, and $\OL$ are presented in~\cref{sec:details-physical-parametrisation} as derived in~\cite{Luben:2020xll}.
Note that $B_n$ is rescaling-invariant and hence observable.
Further, in analogy to the definition of $\bar\alpha$, we define the rescaled scale factor ratio as $\bar y = \alpha y$.
To be explicit, the energy density parameter of the induced dynamical dark energy is given as
\begin{equation}\label{eq:omega-de}
E^2\Omega_{\rm de} = B_0 + 3B_1 \bar y + 3 B_2 \bar y^2 + B_3 \bar y^3\,,
\end{equation}
where $E=H/H_0$ is the normalised Hubble rate, as follows from~\cref{eq:dark-energy-density}.
The scale factor ratio $\bar y$ is determined by the quartic polynomial~\eqref{eq:quartic-pol-y}, which in terms of the energy density parameters reads
\begin{equation}
B_3 \bar y^4 + (3B_2-B_4)\bar y^3 + 3(B_1-B_3)\bar y^2 + (B_0-3B_2 + E^2\Omega_{\rm m} + E^2\Omega_{\rm r} )\bar y - B_1 =0\,.
\end{equation}
As last ingredient, we need to determine $\Omega_{\rm de0}$ as initial condition.
In order to do so, we determine the value of the scale factor ratio at present time $\bar y_0$ using the $\fmn$-Friedmann equation~\eqref{eq:friedmann} evaluated today:
\begin{equation}
0 = B_1 + (3B_2 + \Omega_{\rm k0} -1)\bar y_0 + 3B_3 \bar y_0^2 + B_4 \bar y_0^3 
\end{equation}
This polynomial has up to three real-valued roots, out of which we pick the one that satisfies $0 < \bar y_0 < \bar \alpha$ in order to ensure that we are on the finite branch.
Plugging the result into~\cref{eq:omega-de} evaluated today, determines $\Omega_{\rm de0}$, which in turn determines $\Omega_{\rm m0}$ via~\cref{eq:friedmann-today}.

\subsection{Cosmological data}\label{sec:data}
We now present the cosmological observables that are used in our statistical analysis.
We use three different types of observations: Supernovae type 1a (SN1a), Baryon Acoustic Oscillations (BAOs) and Cosmic Microwave Background (CMB).
Before discussing them in detail, let us first introduce some useful quantities to enlighten the notation.

We start with the definition of the following integral:
\begin{flalign}
\mathcal{I}_k(z) =
\begin{cases}
\frac{1}{\sqrt{|\Omega_{\rm k0}|}}\sinh\left(\sqrt{|\Omega_{\rm k0}|}\int_0^z \frac{\dd z^\prime}{E(z^\prime)}\right) & ,\ k<0 \\
\int_0^z \frac{\dd z^\prime}{E(z^\prime)}& ,\ k=0 \\
\frac{1}{\sqrt{|\Omega_{\rm k0}|}}\sin\left(\sqrt{|\Omega_{\rm k0}|}\int_0^z \frac{\dd z^\prime}{E(z^\prime)}\right) & ,\ k>0
\end{cases},
\end{flalign}
where $E(z)=H(z)/H_0$. We use this quantity to write the various distance indicators, such as the luminosity distance $d_{\rm L} $ and the angular diameter distance $d_{\rm A} $:
\begin{flalign}
d_{\rm L}(z) = \frac{c}{H_0} (1+z) \mathcal{I}_k(z)\,, \qquad
d_{\rm A}(z) = \frac{c}{H_0} \frac{1}{1+z}  \mathcal{I}_k(z) \,.
\end{flalign}
Another important quantity is the comoving sound horizon, defined as 
\begin{flalign}
r_{\rm s}(z)=\frac{c}{\sqrt{3}}\int_{z}^\infty\frac{dz^\prime}{H(z^\prime)\sqrt{1+(3\Omega_{\rm b0}/4\Omega_{\rm \gamma 0})/(1+z^\prime)  }}\,.
\end{flalign}
 Here $\Omega_{\rm b0}$ and $\Omega_{\rm \gamma 0}$ are the baryons and photons density parameters. $\Omega_{\rm \gamma 0}$ is fixed by the temperature $T_{\rm CMB}$ of the CMB~\cite{Chen_2019}:
\begin{flalign}
\frac{3}{4\Omega_{\rm \gamma 0} h^2} = 31500(T_{\rm CMB}/2.7\text{K})^{-4},\qquad T_{\rm CMB}=2.7255\text{K}\,.
\end{flalign}
Here, $h = H_0 / 100\, {\rm km}\,{\rm s}^{-1}\,{\rm Mpc}^{-1}$ is the normalised Hubble rate today.
The photons density parameter is related to the total radiation density parameter via
\begin{flalign}
\Omega_{\rm r0} = \Omega_{\rm \gamma 0}(1+0.2271 N_{\text{eff}})\,.
\end{flalign}
$N_{\text{eff}}$ is the effective number of neutrino species, that we fix to the standard value $N_{\text{eff}}=3.046$ for our analysis.
We now proceed introducing separately the different observables.

\subsubsection{Supernovae type 1a}\label{sec:SN1a}

Supernovae of type 1a serve as standard candles whose luminosities can be inferred independently of their redshift. This allows to build up the local distance ladder. We use the catalogue of the Joint Light-curve Analysis (JLA) that contains 740 SN1a events as reported in Ref.~\cite{Betoule:2014frx}.

The  observed quantity in the case of Supernovae is the apparent magnitude $m$
\begin{flalign}\label{eq:luminosity-distance}
	m &= M + 25 + 5 \log_{10} d_{\rm L}
	= \mathcal M + 5\log_{10} D_{\rm L}\,,
\end{flalign}
where $M$ is the absolute magnitude,  $D_{\rm L} = H_0 d_{\rm L}$  and $\mathcal M = M - 5\log_{10} H_0 + 25$.
In this formula $\mathcal M$ is just an additive constant and appears as a nuisance parameter. This implies that the absolute magnitude $M$ and $H_0$ are degenerate parameters. 

In order to compare the luminosity distance to the observed apparent magnitude $m_{\rm obs}$,
we have to callibrate the supernovae as~\cite{Betoule:2014frx}
\begin{flalign}
	m = m_{\rm obs} - \Delta_{\rm M} + \alpha X_1 - \beta C\,,
\end{flalign}
where $m$ is related to the luminosity distance as in Eq.~(\ref{eq:luminosity-distance}).
Here, $X_1$ and $C$ are the light-curve parameters of the supernova.
For the nuisance parameters $\alpha$, $\beta$ and $\Delta_{\rm M}$ we use the best-fit values
$\alpha=0.140\pm0.006$ and $\beta=3.139\pm0.072$ as obtained in Ref.~\cite{Betoule:2014frx}.
The parameter $\Delta_{\rm M}$ is a correction to the absolute magnitude $M$ of the supernova
that depends on the stellar mass $M_\ast$ of the host galaxy of the supernova as
\begin{flalign}
	\Delta_{\rm M} =
	\begin{cases}
	 -0.060 \pm 0.012 &,\ {\rm if}\, M_\ast < 10^{10} M_\odot \\
		0 &,\ {\rm otherwise.}
	\end{cases}
\end{flalign}
Marginalising over $\mathcal M$, the log-likelihood for the SN-data is given by
\begin{equation}
	-2 \log\mathcal L_{\rm SN} (\theta) = S_2 - \frac{S_1^2}{S_0}\,,
\end{equation}
where we have defined
\begin{equation}
	S_0 = \sum_{ij} (C^{-1}_{\rm SN})_{ij}\,, \qquad
	S_1 = \sum_{ij} (C^{-1}_{\rm SN})_{ij} y_i\,,\qquad
	S_2 = \sum_{ij} (C^{-1}_{\rm SN})_{ij} y_i y_j\,,
\end{equation}
in terms of $y = m - 5\log_{10} D_{\rm L}$.
The covariance matrix $C_{\rm SN}$ is given in Ref.~\cite{Betoule:2014frx}. The errors
on $\alpha$, $\beta$ and $\Delta_M$ are added in quadrature.

\subsubsection{CMB}\label{sec:CMB}

CMB observations constrain background cosmology through the measurement of two different distance ratios~\cite{Komatsu:2008hk}. The first is the ratio between the angular diameter distance $d_{\rm A}(z_\ast)$ and the sound horizon $r_{\rm s}(z_\ast)$ at decoupling epoch $z_\ast$. This ratio is measured through the location of the first peak of the CMB power spectrum:
\begin{flalign}
\label{eq:lA}
l_{\rm A} = (1+z_\ast)\frac{\pi d_{\rm A}(z_\ast)}{r_{\rm s}(z_\ast)}\,.
\end{flalign}
The second is the angular distance divided by the Hubble horizon at the decoupling epoch, $d_{\rm A}(z_\ast) H(z_\ast)/c$. This last information is usually implemented in the shift parameter, defined as:
\begin{flalign}
\label{eq:shift}
R(z_\ast) = H_0\sqrt{\Omega_{\rm m0}} (1+z_\ast) d_{\rm A}(z_\ast)/c\,,
\end{flalign}
which differs from $d_{\rm A}(z_\ast) H(z_\ast)/c$ by a factor of $\sqrt{1+z_\ast}$.
This parameter is obtained assuming
\begin{flalign}
\label{eq:approx}
H^2(z_\ast) = H^2_0\Omega_{\rm m0}(1+z_\ast)^3,
\end{flalign}
which correspond to neglect all contributions (other than $\Omega_{\rm m0}$) to the evolution history at the time of decoupling $H(z_\ast)$.
We use that the early-universe is not altered in bimetric theory compared to general relativity as our working assumption.
This is justified because on the finite branch the energy density arising from the bimetric potential does not contribute to the Hubble rate at large redshifts.

For our statistical analysis we follow the standard procedure~\cite{Zhai_2020}, implementing CMB data in the three distance priors $(l_A,\text{ }R,\text{ }\Omega_{\rm b0} h^2)$.
We use the latest Planck values~\cite{Chen_2019}
\begin{flalign}
l_{\rm A} (z_\ast) =301.471^{+0.089}_{-0.090}\,,\qquad R(z_\ast)=1.7502 \pm 0.0046\,,\qquad \Omega_{\rm b0} h^2=0.02236 \pm 0.00015
\end{flalign}
In our analysis $R(z_\ast)$ and $l_A(z_\ast)$ are computed from \cref{eq:shift} and \cref{eq:lA}, where the decoupling epoch $z_\ast$ is obtained via the fitting function~\cite{Hu:1995en}:
\begin{equation}
\begin{split}
	z_\ast &= 1048 ( 1 + 0.00124\, (\Omega_{\rm b0}h^2)^{-0.738})(1+g_1 (\Omega_{\rm m0}h^2)^{g_2})\,,\\
	g_1 & = \frac{ 0.0783\, (\Omega_{\rm b0}h^2)^{-0.238}}{ 1+39.5\, (\Omega_{\rm b0}h^2)^{0.763} }\,, \qquad
	g_2  = \frac{0.560}{1+21.1\, (\Omega_{\rm b0}h^2)^{1.81}}.
\end{split}
\end{equation}
Defining the CMB data vector as
\begin{flalign}
	X_{\rm CMB}^{\rm T} =
	\left(\begin{matrix}
	l_{\rm A} - 301.471,\text{ }  R(z_\ast) -1.7502,\text{ }  \Omega_{\rm b0} h^2 - 0.02236
	\end{matrix}\right)
\end{flalign}
the likelihood for the CMB is written as:
\begin{flalign}
-2 \log \mathcal{L}_{\text{CMB}} = X_{\rm CMB}^{\rm T} C_{\rm CMB}^{-1} X_{\rm CMB}\,.
\end{flalign}
where the covariance matrix $C_{\rm CMB}$ can be found in~\cite{Chen_2019}.

\subsubsection{BAO}

Barionic Acoustic Oscillations (BAOs) are another important tool to constrain background cosmology. We follow a procedure similar to~\cite{Lindner:2020eez}, using measurements at 10 different redshifts $z=0.106,\text{ } 0.15,\text{ } 0.38,\text{ } 0.51,\text{ } 0.61,\text{ } 0.72,\text{ } 0.978,\text{ } 1.23,\text{ } 1.526,\text{ } 1.944$. The data that we consider are taken from the surveys \textit{6dFGS}~\cite{Beutler_2011},  \textit{SDSS MGS}~\cite{Ross_2015}, \textit{BOSS DR12}~\cite{Alam_2017}, \textit{BOSS DR14}~\cite{Bautista_2018}, \textit{eBOSS QSO}~\cite{Zhao_2018} and are summarised, e.g. in table 2 of~\cite{Lindner:2020eez}.

The relevant length scale in the case of BAOs is the sound horizon at the drag epoch $r_{\rm s}(z_{\rm d})$.
The drag epoch can be derived with the following fitting formula~\cite{Hu:1995en}:
\begin{align}
\begin{split}
	 	&z_{\rm d} =\frac{1291 (\Omega_{\rm m0} h^2)^{0.251}}{1+0.659(\Omega_{\rm m0} h^2)^{0.828}}[1+b_1(\Omega_{\rm b0} h^2)^{b_2}]\,,\\
	 	&b_1=0.313(\Omega_{\rm m0} h^2)^{-0.419}[1+0.607(\Omega_{\rm m0} h^2)^{0.674}]\,,\qquad
	 	b_2=0.238(\Omega_{\rm m0} h^2)^{0.223}
\end{split}
\end{align}
The other relevant quantities in our study are the effective distance measure $d_{\rm V}$ and the redshift-weighted comoving distance $d_{\rm M}$:
\begin{equation}
	d_{\rm V}(z)=\Biggl[(1+z)^2d_{\rm A}^2(z)\frac{cz}{H(z)}\Biggr]^{1/3}\,,\qquad
	d_{\rm M}(z)=(1+z)\, d_{\rm A}(z).
\end{equation}
The likelihood for the statistical analysis is derived using the covariance matrices taken from the original references~\cite{Beutler_2011,Ross_2015,Alam_2017,Bautista_2018,Zhao_2018}.

\subsection{Statistical analysis and results}
\label{sec:cosmo-constraints}

Using the aforementioned data sets, we perform a statistical analysis via MCMC sampling for all models with up to three non-vanishing interaction and vacuum energy parameters $\beta_n$.
The $\Lambda$CDM-model corresponds to the $\beta_0$-model in this sense.
We classify models according to their number of free parameters $\beta_n$: models with one, two or three free parameters $\beta_n$ are referred to as one-, two- or three-parameter models, respectively.
In terms of the physical parameters, that means that one, two or all three of the parameters $\ba$, $\Omega_{\rm FP}$ and $\Omega_\Lambda$ are independent.
The best-fit values and the 1-dimensional ($1\sigma$) error intervals for all free and derived parameters are reported in~\cref{tab:results1,tab:results2}.

In order to estimate the goodness-of-fit for each model, we compute the Bayesian Information Criterion (BIC)~\cite{Schwarz:1978tpv},
\begin{equation}
	{\rm BIC} = \chi^2 + k \ln N
\end{equation}
where $\chi^2=-2\ln \mathcal L$ is evaluated at maximum likelihood, $k$ is the number of model parameters and $N$ is the number of data points.
For our combined data set SN+CMB+BAO we have $N=740+3+10=753$.
The number of model parameters is given by $k=3+\hat k$ with $\hat k$ the number of independent physical parameters as will become clear from the next paragraph.
We will use the BIC of the $\Lambda\rm CDM$-model as reference.
The difference $\Delta\rm BIC$ allows to estimate the preference of one model over another as~\cite{Kass:1995loi,Lindner:2020eez}:
strong support ($\Delta\rm BIC<-12$), favorable ($\Delta\rm BIC<-6$), inconclusive ($\Delta\rm BIC<6$), disfavored ($\Delta\rm BIC<12$), strongly disfavored ($\Delta\rm BIC\geq12$).
The corresponding $\Delta\rm BIC$ are reported in~\cref{tab:results1,tab:results2} as well.

\textit{Independent parameters and priors.}
We use the theoretical bounds of~\cite{Luben:2020xll} that ensure a consistent cosmology, which we summarise in~\cref{tab:viable-cosmo-cond}, as flat priors.
For the parameter $\Omega_\Lambda$ we use $[0,1]$ as flat prior.
For the other physical parameters $\ba$ and $\OFP$ we use a log-scale to resolve large regions of the parameter space.
We again use flat priors with $\log_{10}(\ba)\in[-100,100]$ and $\log_{10}(\Omega_{\rm FP})\in[-2,122]$ if applicable\footnote{The
$\Lambda$CDM-model does not contain these parameters.
For the two-parameter models, $\ba$ and $\OFP$ are not independent.
For the $\beta_0\beta_1$-model we scan over the parameter $\log_{10}(\ba)$, while we scan over $\log_{10}(\OFP)$ for the $\beta_1\beta_{2,3,4}$-models.}.
The lower bound on $\OFP$ is chosen because of the Higuchi bound, while the upper bound corresponds to the Planck-scale.
In addition to the physical parameters, the models further depend on the cosmological parameters.
We choose to scan over $H_0$, $\Omega_{\rm k0}$ and $\Omega_{\rm b0}$ as free parameters with flat priors $[45{\rm\frac{km/s}{Mpc}},85{\rm\frac{km/s}{Mpc}}]$, $[0,1]$ and $[0,1]$, respectively.
Note that we analytically marginalise over $H_0$ in the case of supernovae.
The other energy density parameters $\Omega_{\rm de0}$ and $\Omega_{\rm m0}$ are computed from~\cref{eq:omega-de,eq:friedmann-today}.
We use $[-1,1]$ and $[0,1]$ as flat priors on these parameters, respectively.

\begin{table}
\centering
{\def\arraystretch{1.2}
\begin{tabular}{ l | l l | l l }
Model & $\beta_0$ & $\beta_1$ & $\beta_0\beta_1$ &  $\beta_1\beta_n$  \\
\hline\hline
$\bar\alpha$ &  & $3^{-1/2}$ &$<0.2$ & $<0.016$ \\
$\mFP$ [eV] &  & $(1.82\pm0.02)\times 10^{-32}$ & $(1.33\pm0.02)\times 10^{-32}$ & $>4.24\times10^{-31}$ \\
$\Lambda$ [$10^{-64}{\rm eV}^2$] & $1.76\pm0.05$ & $2.5\pm0.1$ & $1.8\pm0.1$&$1.76^{+0.1}_{-0.09}$  \\
\hline
$H_0$ [$\frac{{\rm km} / {\rm s}}{{\rm Mpc}}$] & $68.8^{+1.4}_{-1.2}$ & $73.1^{+1.6}_{-1.4}$ &$68.8\pm0.2$ & $69.0\pm1.0$\\
$\Omega_\Lambda$ & $0.69\pm0.01$ & $0.86\pm0.01$ &$0.69^{+0.03}_{-0.02}$&$0.69\pm0.01$\\
$\Omega_{\rm m0}$ & $0.31\pm0.01$ & $0.27\pm0.01$ & $0.31\pm0.02$ & $0.31\pm0.01$ \\
$\Omega_{\rm de0}$ & $0.69\pm0.01$ & $0.73\pm0.01$ & $0.69\pm0.02$ & $0.69\pm0.01$ \\
$\Omega_{\rm k0}$ & $0.002^{+0.004}_{-0.003}$& $-0.007^{+0.004}_{-0.003}$ & $0.002^{+0.005}_{-0.001}$&$0.002\pm0.004$ \\
$\Omega_{\rm b0}$ & $0.0224\pm0.0001$ & $0.0223\pm0.0003$ & $0.0224\pm0.0004$& $0.0224\pm0.0003$ \\
\hline
$\chi^2$ & $694.7$ & $726.0$ &$694.7$ & $694.7$ \\
$\Delta{\rm BIC}$ & $721$ & $+31$ &$+6.6$  &$+6.6$ \\
\hline\hline
\end{tabular}}
\caption{The best-fit values and errors at 68\% c.l. from SN+CMB+BAO for the one- and two-parameter models are shown.
To compute the values of $\mFP$ and $\Lambda$ from $\OFP$ and $\OL$, we use the obtained value of $H_0$ for each model.
The $\beta_0$-model is the $\Lambda$CDM-model and hence does not contain the parameters $\ba$ and $\mFP$.
The last row contains the parameter values for the $\beta_1\beta_2$-, $\beta_1\beta_3$- and $\beta_1\beta_4$-model as indicated by the subscript $n$, for which the statistical analysis yielded the exact same results as expected.}
\label{tab:results1}
\end{table}

\begin{table}
\centering
{\def\arraystretch{1.2}
\begin{tabular}{ l | l l l }
Model & $\beta_0\beta_1\beta_4$ & $\beta_0\beta_1\beta_3$ & $\beta_0\beta_1\beta_2$ \\
\hline\hline
$\bar\alpha$ & $<0.14$ & $<0.03$ & $<0.06$  \\
$\mFP$ [eV] & $>1.01\times10^{-32}$ & $>1.17\times 10^{-32}$ & $>1.08\times10^{-32}$  \\
$\Lambda\, [10^{-64}{\rm eV}^2]$ & $1.77^{+0.10}_{-0.09}$ & $1.77\pm0.09$ & $1.77^{+0.10}_{-0.09}$\\
\hline
$H_0$ [$\frac{{\rm km} / {\rm s}}{{\rm Mpc}}$] & $68.8^{+1.5}_{-1.3}$ & $68.9^{+1.4}_{-1.3}$ & $68.9^{+1.4}_{-1.3}$  \\
$\Omega_\Lambda$ & $0.69^{+0.02}_{-0.01}$ & $0.69^{+0.02}_{-0.01}$ & $0.69^{+0.03}_{-0.01}$  \\
$\Omega_{\rm m0}$ & $0.31\pm0.01$ & $0.30\pm0.01$ & $0.31\pm0.01$  \\
$\Omega_{\rm de0}$ & $0.69\pm0.01$ & $0.69\pm0.01$ & $0.69\pm0.01$  \\
$\Omega_{\rm k0}$ & $0.002\pm0.004$ & $0.002^{+0.004}_{-0.003}$ & $0.002^{+0.006}_{-0.001}$  \\
$\Omega_{\rm b0}$ & $0.0224\pm0.0003$ & $0.0224\pm0.0003$ & $0.0224\pm0.0003$  \\
\hline
$\chi^2$ & $694.7$ & $694.7$ & $694.7$ \\
$\Delta{\rm BIC}$ & $+11.6$ & $+11.6$ & $+11.6$  \\
\hline\hline
\end{tabular}
\vspace{1cm}

\begin{tabular}{ l | l l l }
	Model  & $\beta_1\beta_2\beta_3$ & $\beta_1\beta_2\beta_4$ & $\beta_1\beta_3\beta_4$ \\
	\hline\hline
	$\bar\alpha$ & $<0.004$ & $<0.01$ & $<0.002$  \\
	$\mFP$ [eV] & $>1.10\times10^{-32}$ & $>1.15\times 10^{-32}$ & $>1.10\times10^{-32}$  \\
	$\Lambda\, [10^{-64}{\rm eV}^2]$ & $1.78\pm ^{+0.09}_{-0.1}$ & $1.77\pm0.09$ & $1.77^{+0.10}_{-0.09}$\\
	\hline
	$H_0$ [$\frac{{\rm km} / {\rm s}}{{\rm Mpc}}$] & $68.9^{+1.4}_{-1.3}$ & $69.0^{+1.3}_{-1.2}$ & $68.9^{+1.4}_{-1.3}$  \\
	$\Omega_\Lambda$ & $0.69\pm0.01$ & $0.69^\pm0.01$ & $0.69\pm0.01$  \\
	$\Omega_{\rm m0}$ & $0.31\pm0.01$ & $0.30\pm0.01$ & $0.31\pm0.01$  \\
	$\Omega_{\rm de0}$ & $0.69\pm0.01$ & $0.69\pm0.01$ & $0.69\pm0.01$  \\
	$\Omega_{\rm k0}$ & $0.002\pm0.004$ & $0.0023^{+0.0003}_{-0.004}$ & $0.002\pm0.003$  \\
	$\Omega_{\rm b0}$ & $0.0224\pm0.0003$ & $0.0224\pm0.0003$ & $0.0224\pm0.0003$  \\
	\hline
	$\chi^2$ & $694.7$ & $694.7$ & $694.7$ \\
	$\Delta{\rm BIC}$ & $+11.6$ & $+11.6$ & $+11.6$  \\
	\hline\hline
\end{tabular}}

\caption{The best-fit values and errors at 68\% c.l. from SN+CMB+BAO for the three-parameter models are shown.
To compute the values of $\mFP$ and $\Lambda$ from $\OFP$ and $\OL$, we use the obtained value of $H_0$ for each model.}
\label{tab:results2}
\end{table}

\textit{Reference model and consistency of data sets.}
Let us start with a discussion of the results for the $\Lambda$CDM-model as reference.
The 1- and 2-dimensional posterior distributions for the parameters $H_0$, $\Omega_{\rm m0}$ and $\Omega_\Lambda=\Omega_{\rm de0}$ are presented in the left panel of~\cref{fig:posterior_Om-H0-Ode_B0_B1}.
While supernovae do not constrain $H_0$ due to the degeneracy with the absolute magnitude, also CMB and BAOs do not yield robust constraints individually.
However, this geometric degeneracy is broken when combining all three sets of data.
We then obtain the value $H_0=(68.8\pm1.2)\rm{km/s/Mpc}$, which is in good agreement with current constraints~\cite{Betoule:2014frx,Aubourg:2014yra,Aghanim:2018eyx}.
Our value is slightly but not significantly larger, which can be understood as an artifact of using the distance priors~\cite{Chen_2019}.
Combining the data sets stabilises spatial curvature to the value $\Omega_{\rm k0} = 0.002\pm0.003$ such that the universe is spatially flat within $68\%$ c.l.
Also the other parameters as reported in~\cref{tab:results1} are in good agreement with current constraints.
We conclude that our analysis is robust.
At the best-fit point we have $\chi^2 = 694.7$ leading to ${\rm BIC} = 721.2$.


\subsubsection{$\beta_1$-model}

\begin{figure}
\includegraphics[scale=0.5]{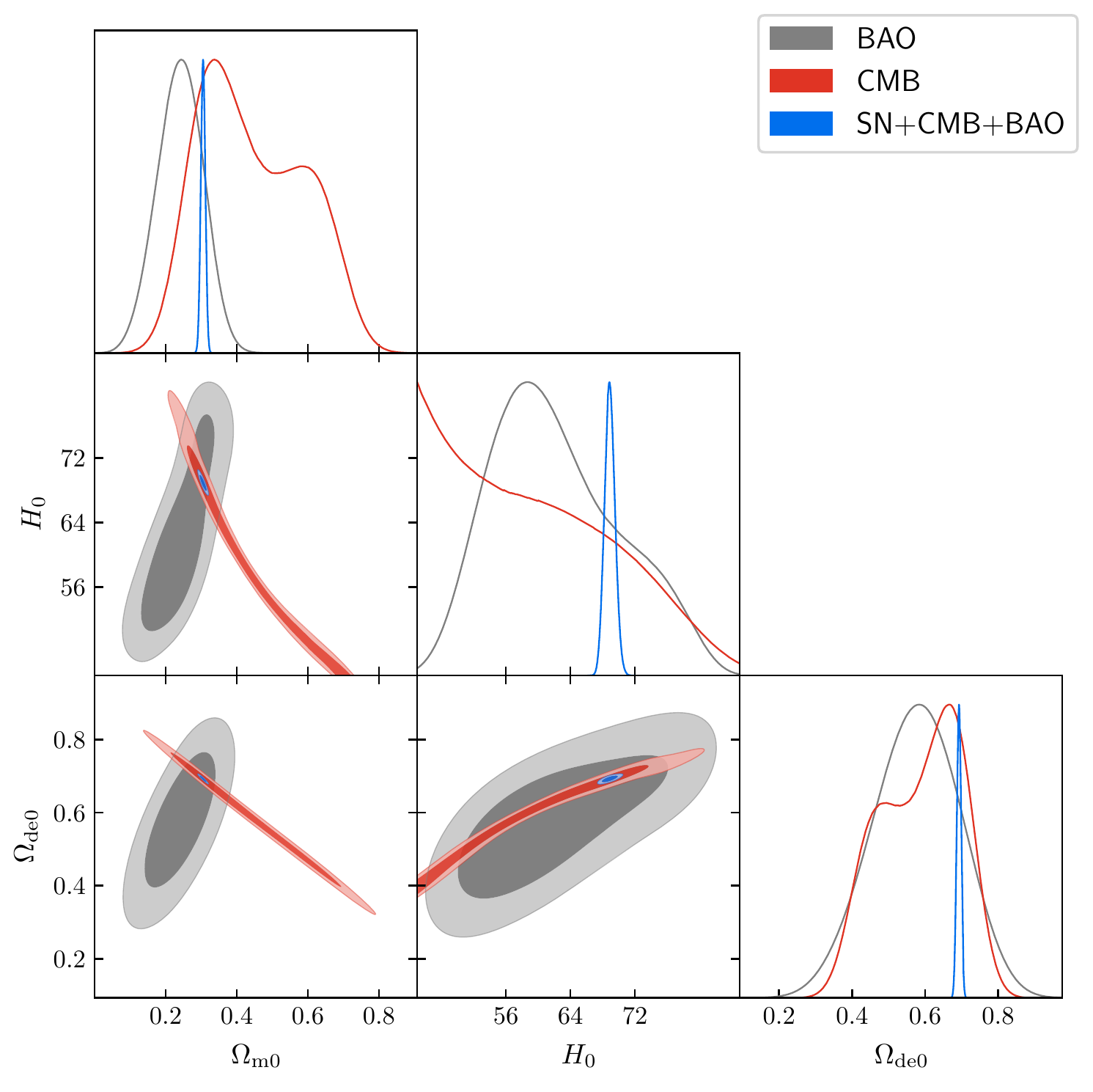}
\includegraphics[scale=0.5]{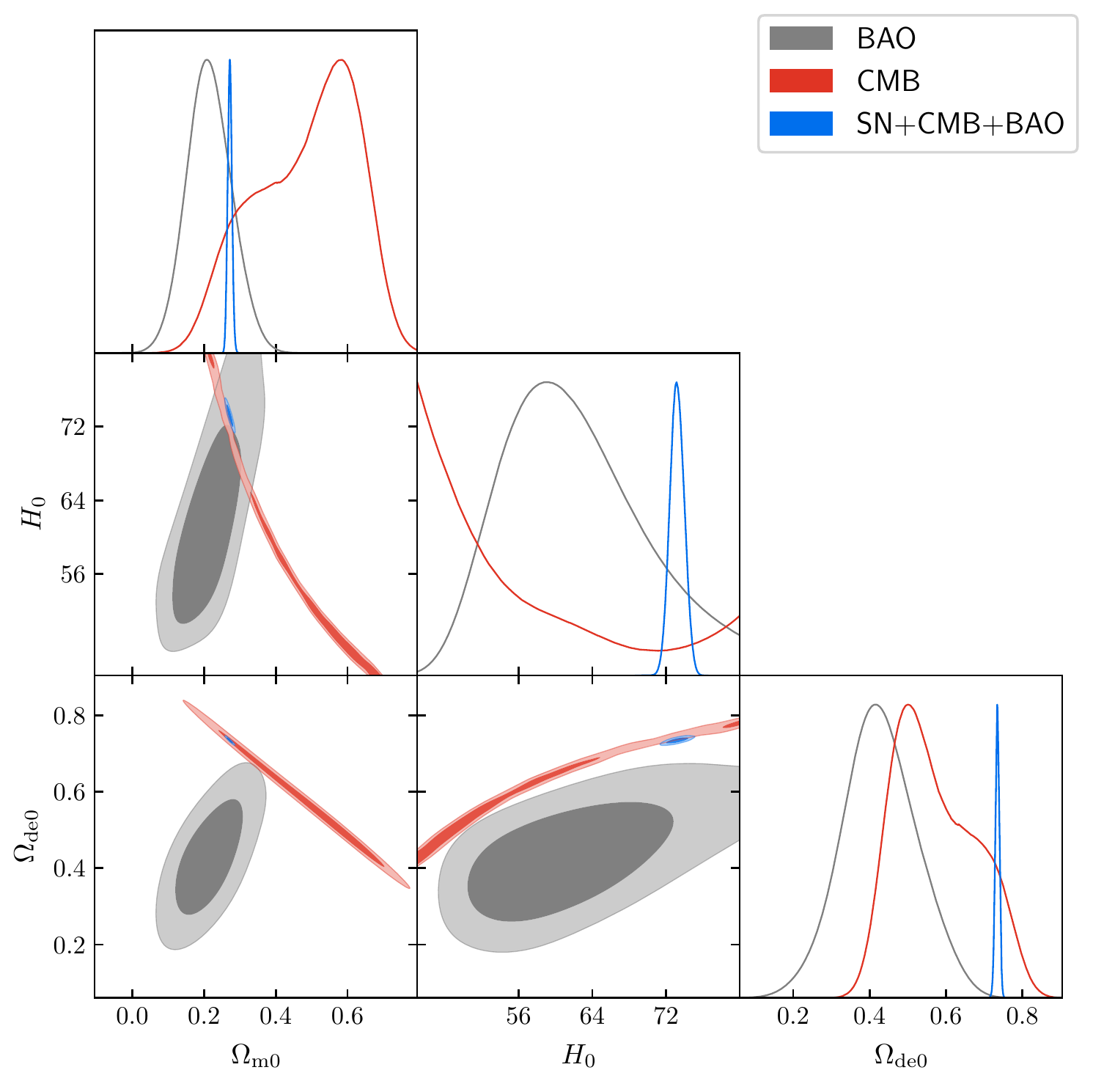}
\caption{Marginalised posterior distribution of the cosmological parameters in the $\Lambda\rm CDM$-model (left) and the $\beta_1$-model (right).
The contour lines correspond to $68\%$ and $95\%$ c.l.
The results reflect a tension between the early- and late-time data sets for the $\beta_1$-model.}
\label{fig:posterior_Om-H0-Ode_B0_B1}
\end{figure}

As discussed before, the $\beta_1$-model is the only bimetric one-parameter model that can possibly give rise to a viable expansion history.
The posterior distributions are presented in the right panel of~\cref{fig:posterior_Om-H0-Ode_B0_B1}.
Also this model is subject to the geometric degeneracy such that both $H_0$ and $\Omega_{\rm k0}$ are unconstrained by CMB and BAO individually.
Combining all three data sets stabilises these parameters.
Data then favors a slightly spatially curved universe with $\Omega_{\rm k0} = -0.007^{+0.004}_{-0.003}$.
The Hubble rate today is constrained to the remarkable high value of $H_0=73.1^{+1.6}_{-1.4}{\rm km/s/Mpc}$, which is in perfect agreement with local determinations of $H_0$~\cite{DiValentino:2020vnx}.
We emphasise that we obtained this result without local prior on $H_0$.
The value of the spin-2 mass is constrained to be small with $\mFP=(1.82\pm0.02)\times 10^{-32}\,{\rm eV}$, which is close to the Higuchi bound.

However, the cosmological early- and late-time measurements are incompatible within this model, as can be seen already from the right panel of~\cref{fig:posterior_Om-H0-Ode_B0_B1}.
Quantitatively, $\Delta{\rm BIC}=+31$ indicates that this model is indeed strongly disfavored by cosmological background data.
Therefore, this model does not solve the $H_0$-tension, despite its large favored value.
We conclude that the bimetric $\beta_1$-model is statistically ruled out.
Nonetheless, we summarise the obtained best-fit values for the other parameters in~\cref{tab:results1}.

\subsubsection{Two-parameter models}
\begin{figure}
	\includegraphics[scale=0.5]{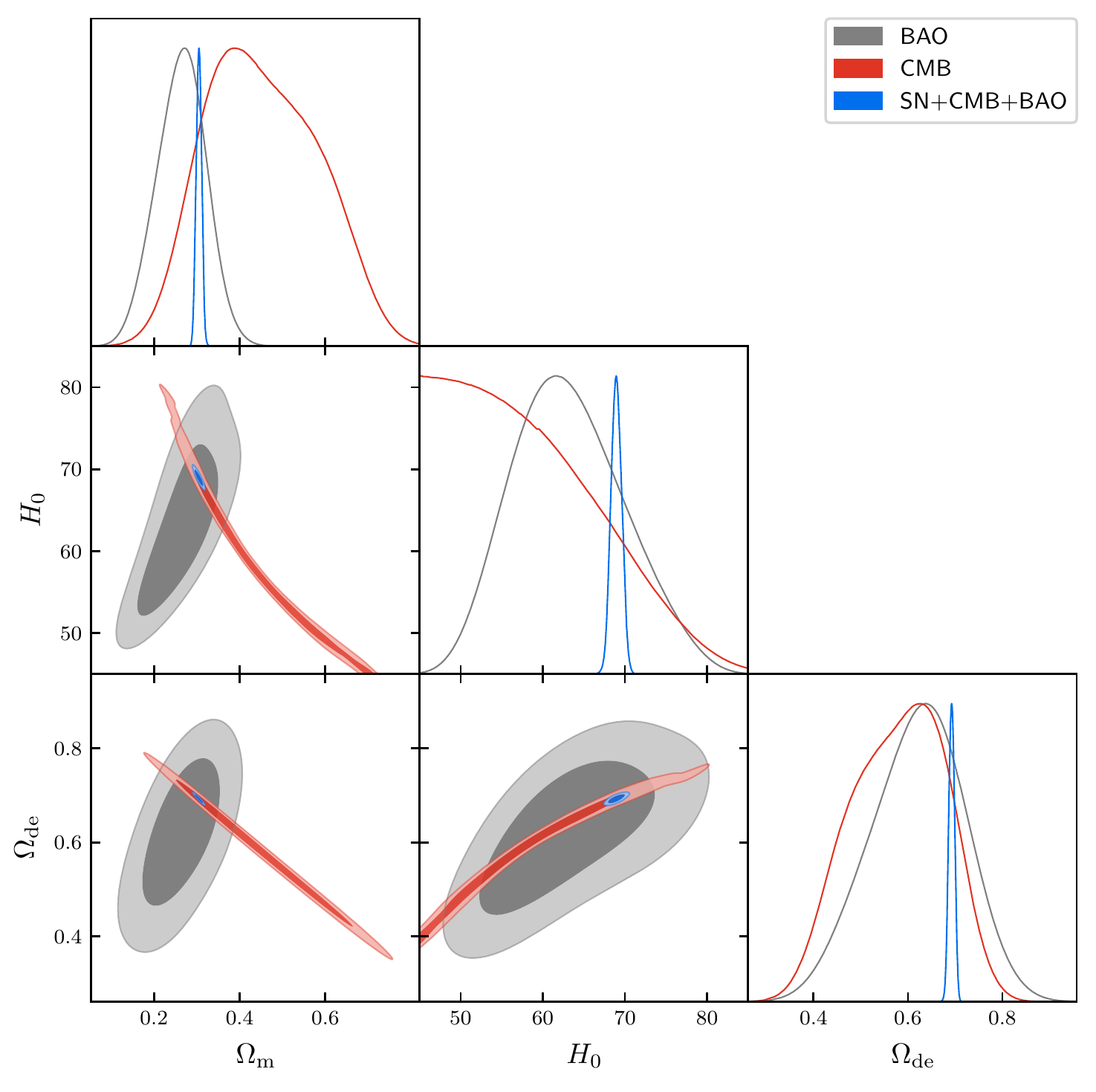}
	\includegraphics[scale=0.5]{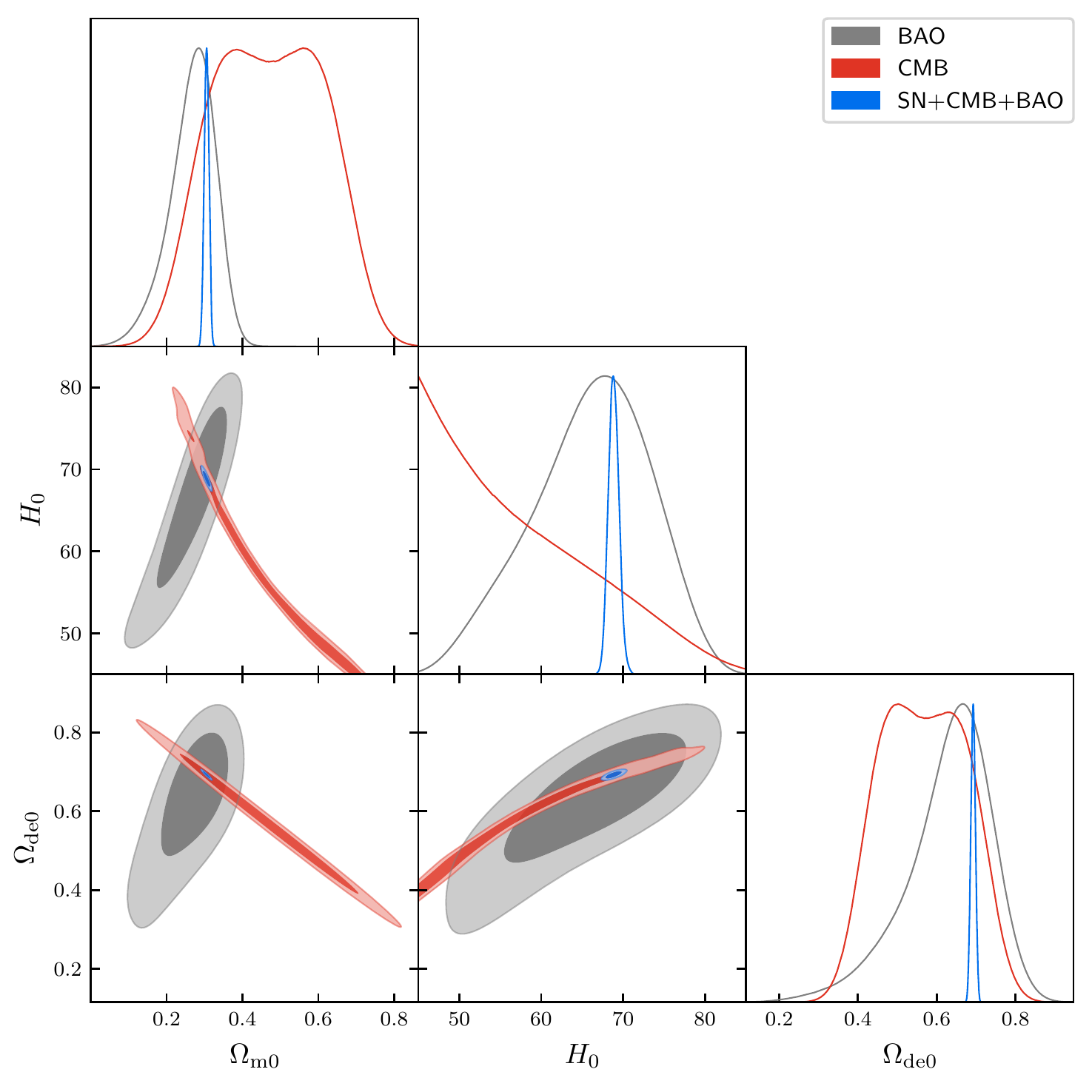}
	\caption{Marginalised posterior distribution of the cosmological parameters in the $\beta_0\beta_1$-model (left) and the $\beta_1\beta_3$-model (right).
	The contour lines correspond to $68\%$ and $95\%$ c.l.}
	\label{fig:posterior2params}
\end{figure}

As already discussed, consistency of the early universe on the finite branch requires $\beta_1>0$. This means that there are four possibilities for the two-parameter models: the $\beta_0\beta_1$ model and the $\beta_1\beta_n$ models with $n=2,3,4$. From our analysis it emerges that the results for all the $\beta_1\beta_n$ models are very similar and statistically compatible. For this reason, we explicitly show only the results for the $\beta_1\beta_3$ model and we refer to it as $\beta_1\beta_n$ for the rest of this section. In \cref{fig:posterior2params} we plot the 1- and 2- dimensional marginalised posterior distribution for these models, where we can see that the data sets are compatible. This is true also for SN data, that are not shown explicitly in this figure. We conclude that the data sets can be combined in the statistical analysis.

In ~\cref{tab:results1} we summarise the results of the combined analysis. The main difference between the $\beta_0\beta_1$ model and the $\beta_0\beta_n$ models are the values of $\bar{\alpha}$ and $\mFP$. In both cases, $\bar{\alpha}$ is constrained only from above with a maximum value of $0.2$ for the $\beta_0\beta_1$ model and $0.016$ for the $\beta_1\beta_n$ models. In the $\beta_0\beta_1$ model the value of $\mFP$ is constrained to be $(1.33\pm0.02)\times 10^{-32}\, {\rm eV}$, which lies close to the Higuchi bound. In the $\beta_1\beta_n$ models, however, the value of the Fierz-Pauli mass is constrained only weakly from below with $\mFP>4.24\times10^{-31}\,{\rm eV}$.
We will discuss the constrains on $\bar{\alpha}$ and $\mFP$ in more detail in~\cref{sec:cosmological-vs-local}, where we will compare the results of the statistical analysis with local tests of gravity.

Despite these differences, the other parameters for the $\beta_0\beta_1$ and $\beta_1\beta_n$ models are very similar with each other and compatible with $\Lambda\rm CDM$. The only exception is spatial curvature, that seems to be slightly positive for the $\beta_0\beta_1$ model. However, this model still prefers an almost flat universe with $\Omega_{\rm k0}=0.002^{+0.005}_{-0.001}$, which remains very close to $0$ at $1\sigma$ and compatible with the $\Lambda\rm CDM$ value. Moreover, the value of $\chi^2$ in both models coincides with $\Lambda\rm CDM$. This makes the $\beta_0\beta_1$ and $\beta_1\beta_n$ models slightly disfavored with a $\Delta{\rm BIC}=+6.6$, since these models have more free parameters with respect to $\Lambda\rm CDM$.

\subsubsection{Three-parameter models}

\begin{figure}
\includegraphics[scale=0.5]{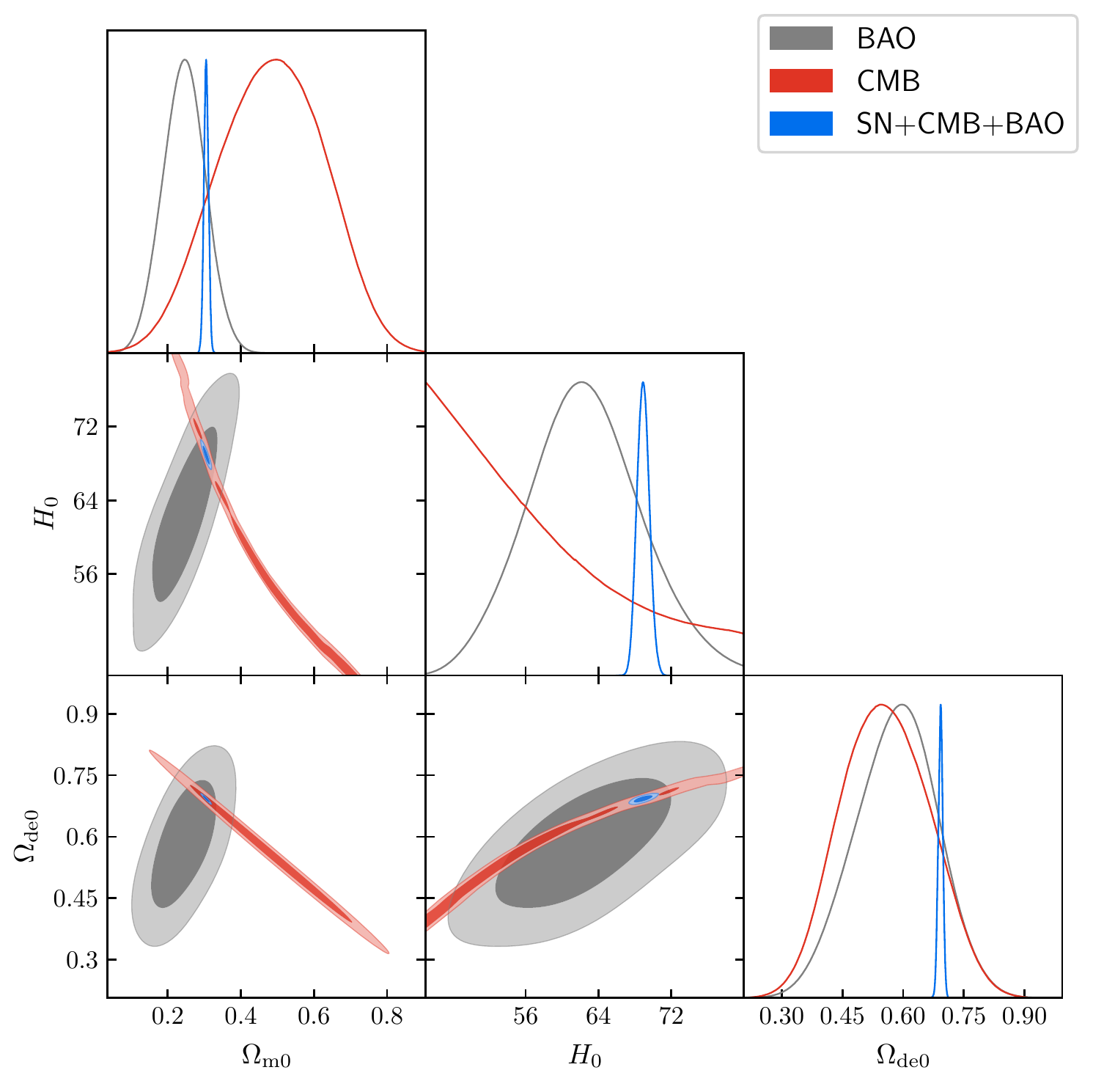}
\includegraphics[scale=0.5]{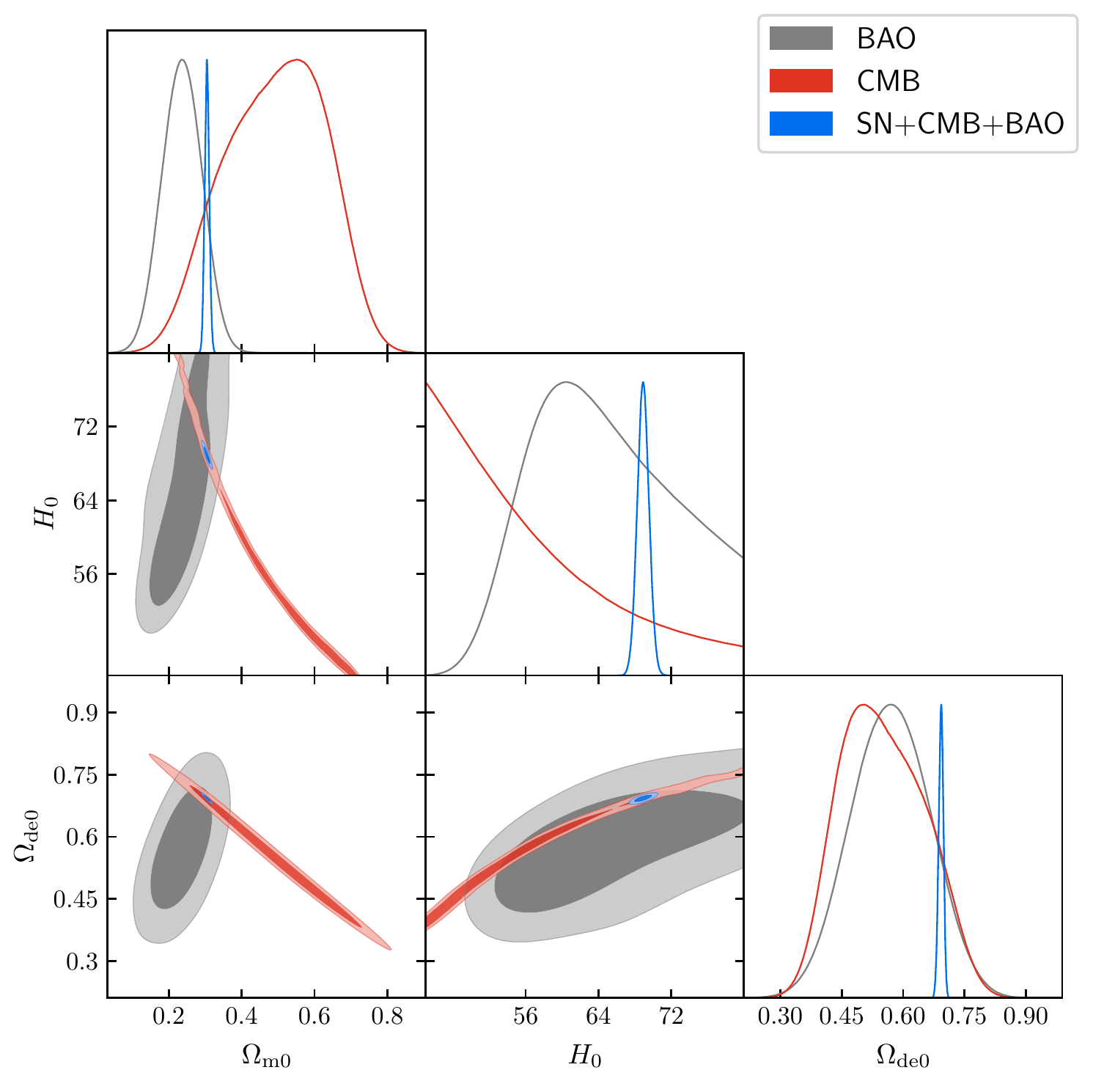}
\caption{Marginalised posterior distribution of the cosmological parameters in the $\beta_0\beta_1\beta_2$-model (left) and the $\beta_0\beta_1\beta_3$-model (right).
The contour lines correspond to 99\% and 95\% c.l.}
\label{fig:posterior_Om-H0-Ode_B012_B013}
\end{figure}

\begin{figure}
\includegraphics[scale=0.5]{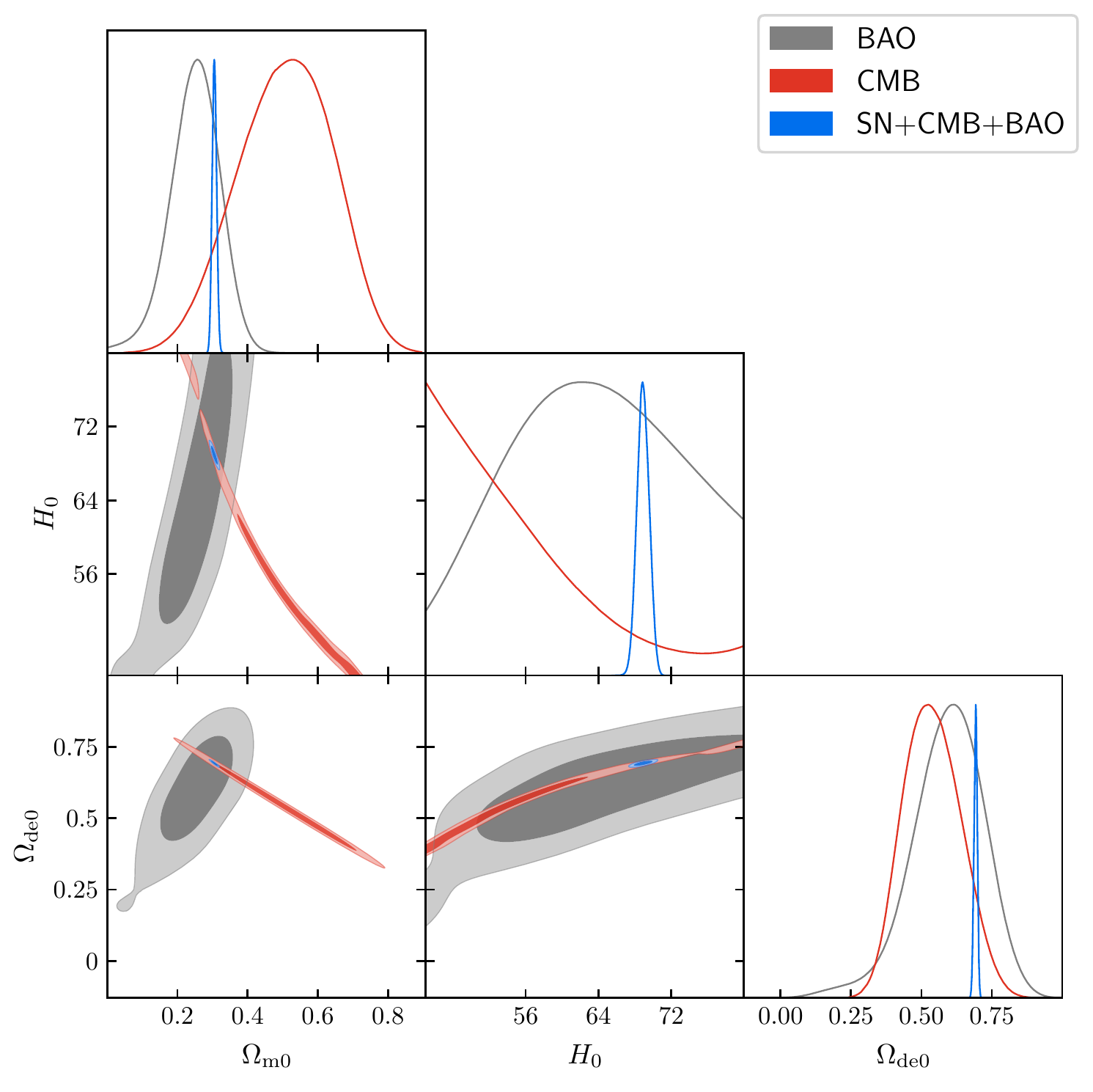}
\includegraphics[scale=0.5]{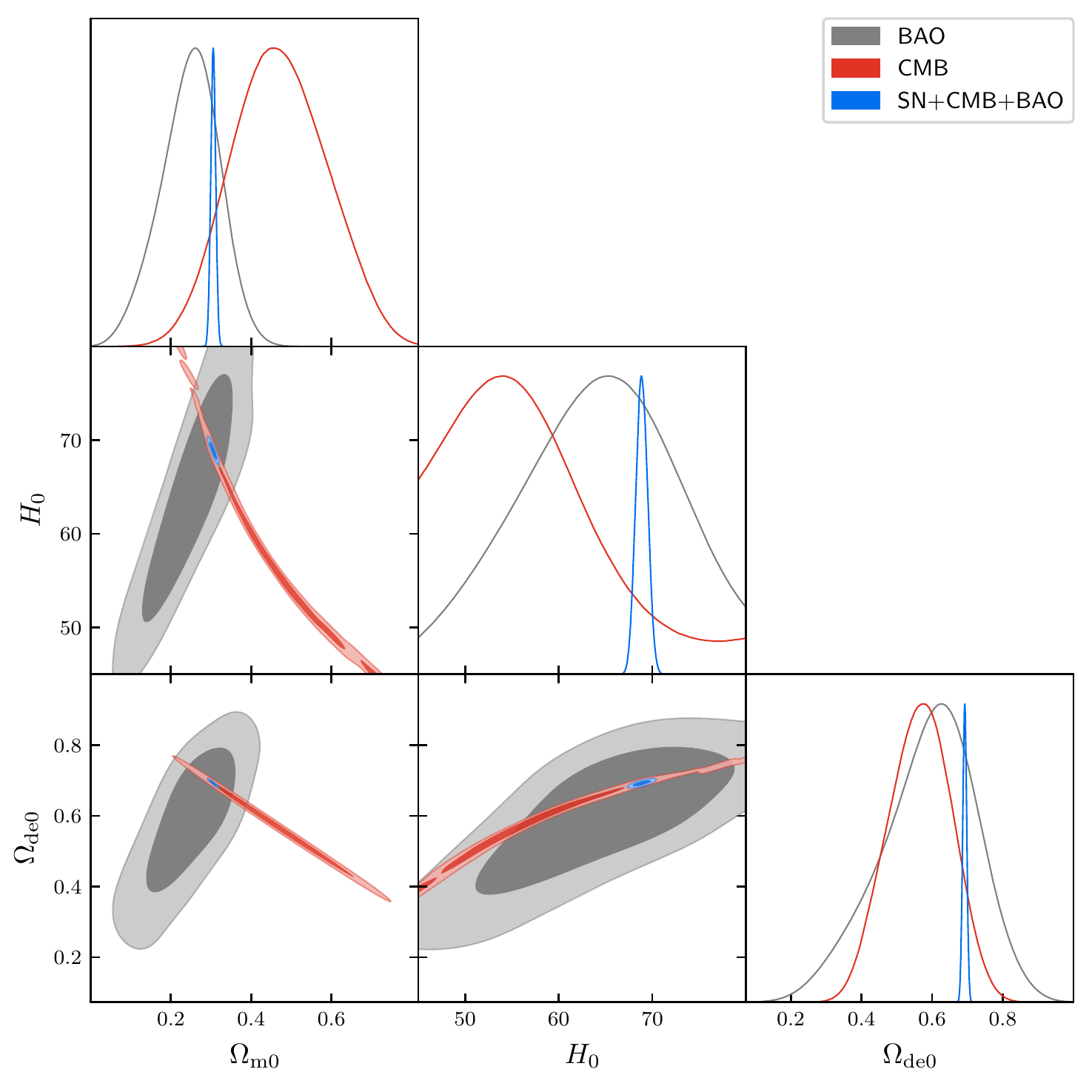}
\caption{Marginalised posterior distribution of the cosmological parameters in the $\beta_0\beta_1\beta_4$-model (left) and the $\beta_1\beta_2\beta_3$-model (right)}
\label{fig:posterior_Om-H0-Ode_B014_B123}
\end{figure}

\begin{figure}
	\includegraphics[scale=0.5]{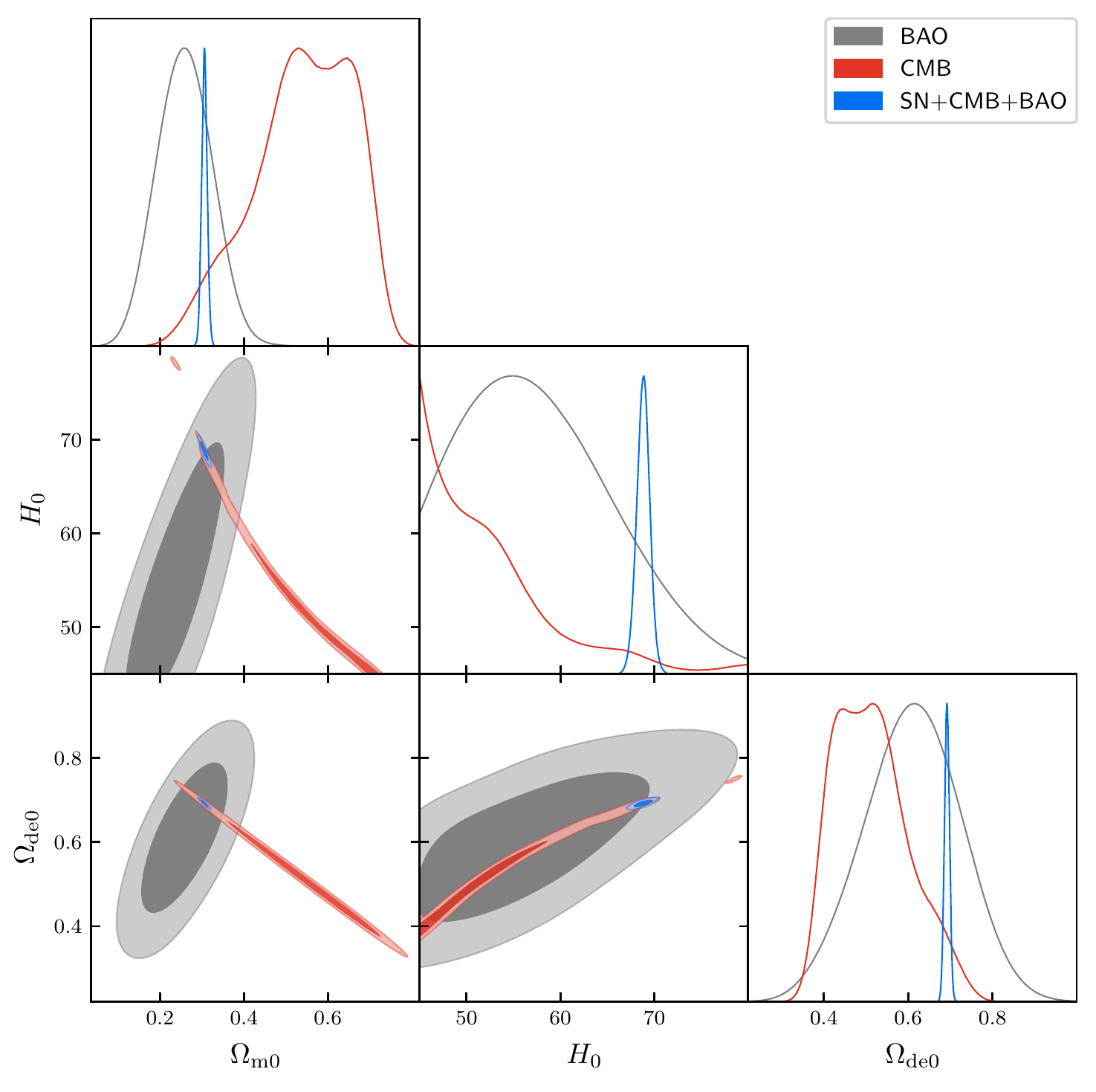}
	\includegraphics[scale=0.5]{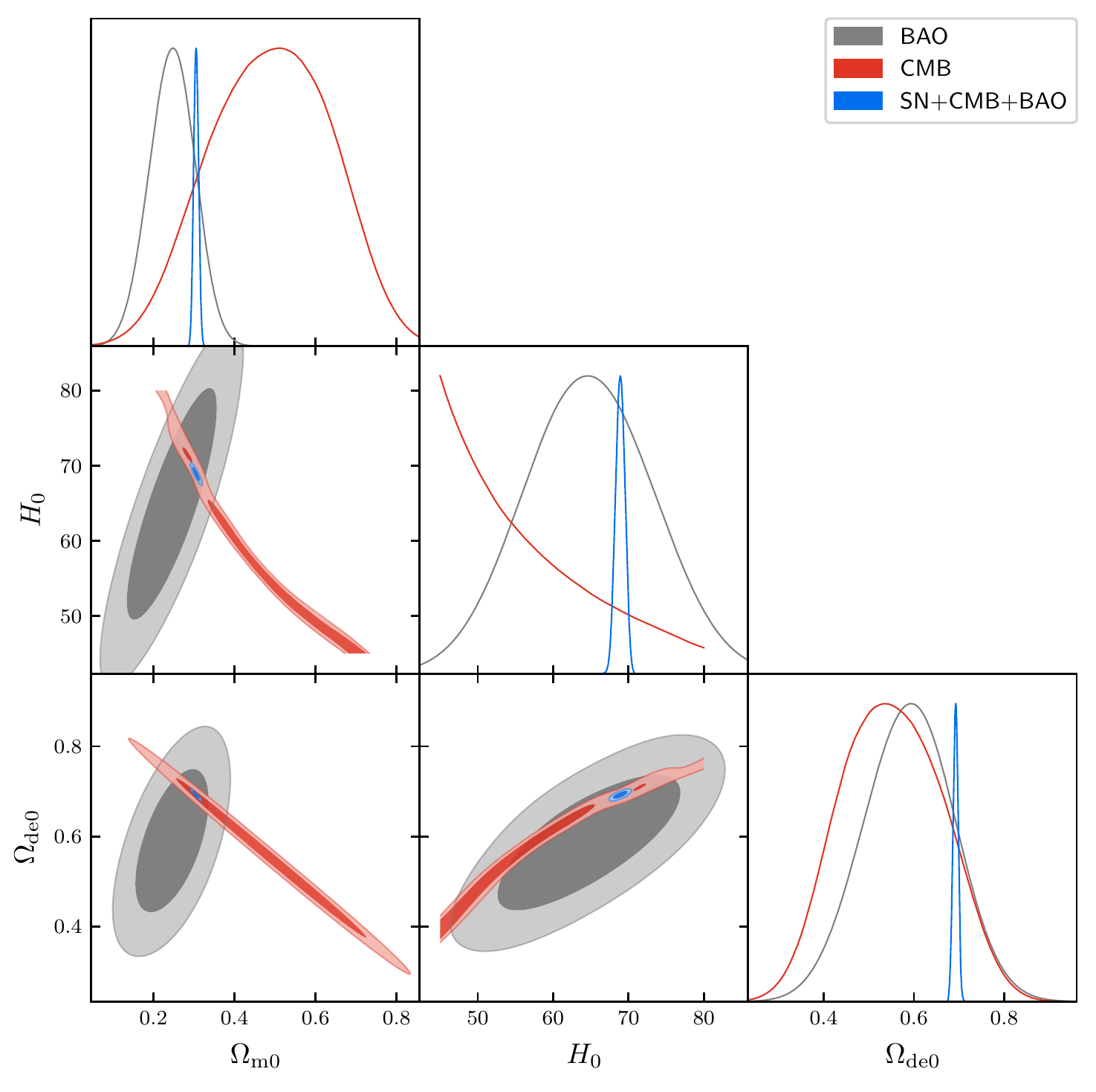}
	\caption{Marginalised posterior distribution of the cosmological parameters in the $\beta_1\beta_2\beta_4$-model (left) and the $\beta_1\beta_3\beta_4$-model (right)}
	\label{fig:posterior_Om-H0-Ode_B124_B134}
\end{figure}

We finish by discussing models with three free parameters. We will see that the parameter constraints are similar across these models. Data sets are compatible for all the three parameter models, as we can see in \cref{fig:posterior_Om-H0-Ode_B012_B013,fig:posterior_Om-H0-Ode_B014_B123,fig:posterior_Om-H0-Ode_B124_B134}. Therefore, combining the data sets for these models is trustworthy\footnote{This is true also for SN data, even if they are not shown explicitly in \cref{fig:posterior_Om-H0-Ode_B012_B013,fig:posterior_Om-H0-Ode_B014_B123,fig:posterior_Om-H0-Ode_B124_B134}.}.

$\bullet$ \textit{Models with $\beta_4=0$.}
We first discuss the three-parameter models without vacuum energy in the $f$-sector.
We present the 1- and 2- dimensional marginalised posterior distribution in the left and right panel of~\cref{fig:posterior_Om-H0-Ode_B012_B013} and the left panel of~\cref{fig:posterior_Om-H0-Ode_B014_B123} for the $\beta_0\beta_1\beta_2$-, $\beta_0\beta_1\beta_3$- and $\beta_0\beta_1\beta_4$-model, respectively. The parameter constraints are reported in~\cref{tab:results2} and agree with the $\Lambda\rm CDM$ values within $1\sigma$ and are all quite similar across these models. The only exception is that the $\beta_0\beta_1\beta_2$-model seems to prefer a slightly curved universe at $1\sigma$. However, the value of the curvature is very small and compatible with $\Lambda\rm CDM$ at $1\sigma$ and therefore we do not consider it as a statistically relevant difference.
The bimetric parameters $\ba$ and $\mFP$ are constrained only weakly. We will discuss more about the constraints on these parameters in the $\bar{\alpha}-m_{\rm FP}$ plane when comparing with local tests in~\cref{sec:cosmological-vs-local}.
We find that these models are consistent with the current data because $\chi^2=694.7$ as in the $\Lambda$CDM-model.
However, the BIC is higher due to the two additional free parameters indicating that these models are statistically disfavored.

$\bullet$ \textit{Models with $\beta_0=0$.} The result for this models is very similar to models with $\beta_4=0$. The only exception is that $\bar\alpha$ is more constrained of  about one order of magnitude, as it can be seen in~\cref{tab:results2}. We show the 1- and 2- dimensional marginalised posterior distribution in the right panel of~\cref{fig:posterior_Om-H0-Ode_B014_B123} for the $\beta_1\beta_3\beta_4$-model and in~\cref{fig:posterior_Om-H0-Ode_B124_B134} for the $\beta_1\beta_2\beta_4$- and $\beta_1\beta_3\beta_4$-model. Again, these models have the same $\chi^2$ of the $\Lambda$CDM-model, which makes them statistically disfavored with a higher BIC.

\section{Bounds from local tests of gravity}
\label{sec:local-bounds}

Having identified those regions of the parameter space that are in agreement with current observations in background cosmology, we now turn to local tests of gravity.
In~\cref{sec:local-tests-review} we give a brief review on local tests of gravity and present the existing constraints on the parameters $\ba$ and $\mFP$.
However, the existing bounds were derived without taking into account the Vainshtein screening mechanism.
In regions where Vainshtein screening is active, the bounds on $\ba$ and $\mFP$ are not trustworthy.
In~\cref{sec:vainshtein-screening} we identify the regions of the parameter space that give rise to Vainshtein screening in spherically symmetric systems.
We finally confront cosmological with local bounds in~\cref{sec:cosmological-vs-local} in parameter regions without Vainshtein screening.

\subsection{Local tests of gravity}
\label{sec:local-tests-review}

In this section we briefly summarise local tests of gravity.
For details and reviews we refer to Ref.~\cite{Fischbach:1999bc,Adelberger:2003zx,Adelberger:2009zz,Will:2014kxa,Murata:2014nra}.
Extra degrees of freedom (such as the massive spin-2 field of bimetric theory) modify the Newtonian gravitational potential as compared to GR.
This is usually attributed to a \textit{fifth force}.
If the force mediator is massive, it contributes a Yukawa potential to the gravitational potential felt by a massive test body.
Following standard notation in the literature, the gravitational potential for a static and spherically-symmetric configuration around a massive source can be parametrised as
\begin{equation}\label{eq:lin-grav-pot}
	V(r) = -\frac{1}{m_{\rm P}} \left(\frac{1}{r} + \xi\frac{e^{-r/\lambda}}{r}\right)\,,
\end{equation}
with $m_{\rm P}$ the Planck mass.
This is referred to as Yukawa parametrisation.
The first term corresponds to the usual potential arising from massless gravitons.
The second term arises from a massive force mediator with Compton wavelength $\lambda$ and coupling to matter $\xi$.
While in the limit $\xi \ll 1$ the fifth force is suppressed, it is most prominent on length scales $r\sim \lambda$.
As we will see in~\cref{sec:vainshtein-screening}, these parameters are related to the bimetric parameters as
\begin{equation}\label{eq:Yukawa-parameters}
	\xi = \frac{4\ba^2}{3}\,, \qquad \lambda = \mFP^{-1}\,.
\end{equation}

\begin{figure}
\centering
\includegraphics[width=0.8\textwidth]{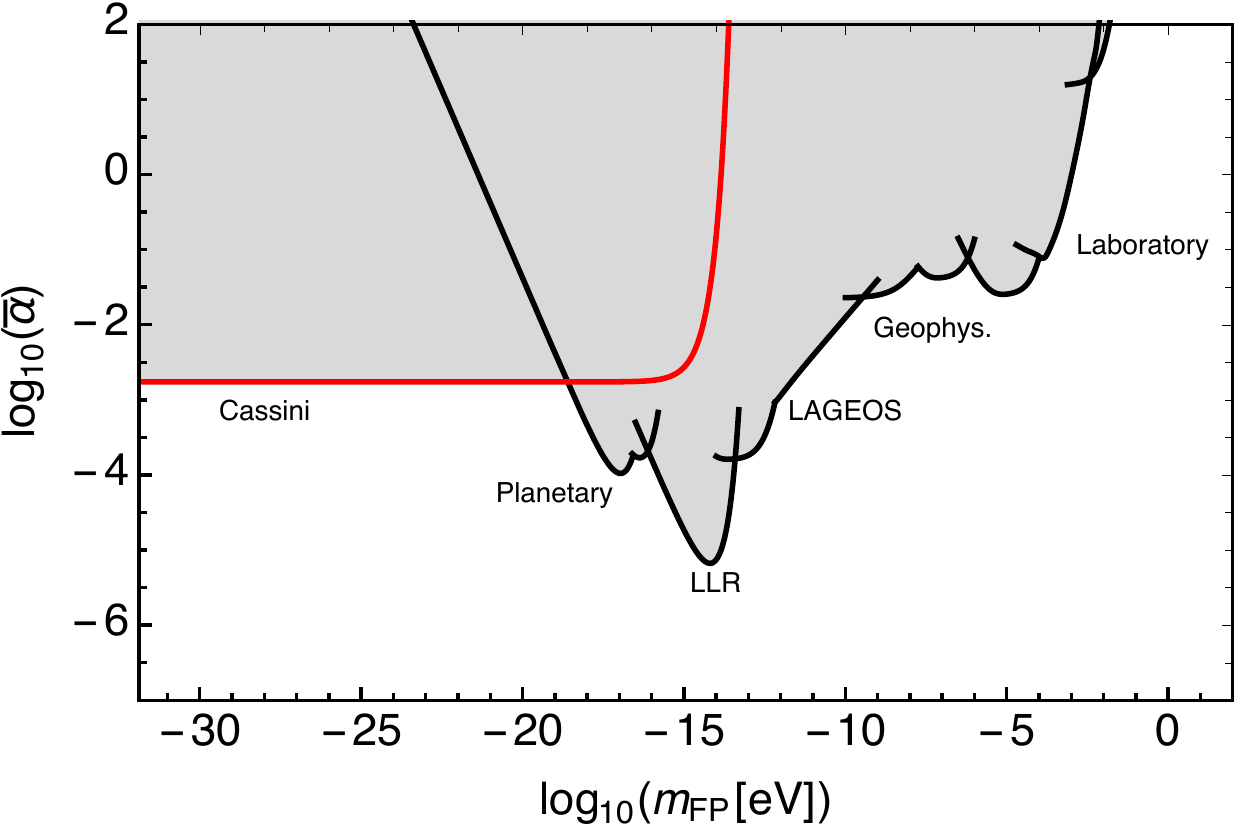}
\caption{Bounds on the bimetric parameters $\ba$ and $\mFP$ from local tests of gravity.
The black lines result from tests of the Newtonian gravitational potential as reported in~\cite{Talmadge:1988qz,Fischbach:1999bc,Adelberger:2003zx,Hoyle:2004cw,Smullin:2005iv,Adelberger:2009zz}.
Tests of the scalar curvature yield the red line~\cite{Bertotti:2003rm,Hohmann:2017uxe}.
The gray-shaded region is excluded by $95\%$ c.l.
Note that the Vainshtein mechanism is not taken into account in the derivation of these bounds.}
\label{fig:local-tests}
\end{figure}

In~\cref{fig:local-tests} we present the constraints from local tests of gravity on the bimetric parameters $\ba$ and $\mFP$ as indicated by the black lines.
The gray-shaded region is excluded by 95\% c.l.
To compute the constraints, we follow the procedure as described in~\cite{Fischbach:1999bc,Adelberger:2003zx,Adelberger:2009zz}.
The laboratory constraints correspond to the Irvine~\cite{Hoskins:1985tn}\footnote{Note that the bound from~\cite{Spero:1980zz} is not included, yet.}, E\"ot-Wash~\cite{Hoyle:2004cw} and Stanford~\cite{Smullin:2005iv} experiments.
The geophysical constraints correspond to the lake~\cite{Cornaz:1994yp,Hubler:1995pd} and tower~\cite{Thomas:1989,Kammeraad:1990bj,Romaides:1996uh} experiments.
The constraints from measurements with the LAGEOS satellite are reported in~\cite{Smith:1985,Rapp:1987,Dickey:1994zz}.
The constraint from Lunar-Laser-Ranging (LLR) results from lunar precession~\cite{Adelberger:2003zx}.
The planetary constraints are reported in~\cite{Talmadge:1988qz}.

The aforementioned tests of the Newtonian potential provide only weak constraints on modifications if the mass of the fifth force mediator is small.
In this region of the parameter space, tests of the scalar curvature, i.e. the deflection and delay of light induced by matter, provide powerful constraints.
We use the gravitational slip parameter $\gamma$, which is unity in GR, $\gamma=1$.
Within the context of bimetric theory, the parameter has been calculated to be~\cite{Hohmann:2017uxe}
\begin{equation}
	\gamma = \frac{3+2\ba^2 e^{-\mFP r}}{3+4\ba^2 e^{-\mFP r}}\,.
\end{equation}
The most precise value has been obtained within the solar system.
The measurement of the time delay of radar signals sent between the Earth and the Cassini spacecraft on its way to Saturn yields $\gamma-1 = (2.1\pm2.3)\times 10^{-5}$~\cite{Bertotti:2003rm}.
The radio signals were passing by the sun at a distance of $1.6$ solar radii, i.e. $r\simeq7.44\times 10^{-3}\,{\rm AU}\simeq 1.31\times 10^{-15}\,{\rm eV}^{-1}$.
Using these values defines the red line in~\cref{fig:local-tests}.
The region above that line is excluded by 95\% c.l.
For small spin-2 masses, the coupling to matter is constraint to be $\ba\lesssim 1.7\times 10^{-3} $.
A collection of bounds on $\gamma$ from time delay and deflection of light form galaxy halo to solar system scales can be found in~\cite{Will:2014kxa}, which however are less constraining than the bound used here.

Summarising, the most stringent constraint comes from LLR with $\ba < 8.9\times10^{-6}$ at $95\%$ c.l. for $\mFP\simeq 6.5\times 10^{-15}\,{\rm eV}$.
On the other hand, for $\mFP\gtrsim 10^{-2}\,{\rm eV}$ the bounds on $\ba$ are weak.
This, however, is not the end of the story in the framework of bimetric theory due to the Vainshtein screening mechanism as we discuss in the next section.

\subsection{Vainshtein screening}
\label{sec:vainshtein-screening}

Local tests of gravity put tight constraints on modifications of the gravitational potential.
Historically, the linear theory of Fierz and Pauli~\cite{Fierz:1939ix} was discarded because it did not pass basic solar system tests.
This is due to the so-called vDVZ discontinuity~\cite{vanDam:1970vg,Zakharov:1970cc}: the helicity-0 mode of the massive graviton couples to the trace of the energy-momentum tensor even in the limit $\mFP\rightarrow 0$.
Vainshtein argued that including nonlinear interaction terms for the massive spin-2 field restores GR in that limit thus curing the problem~\cite{Vainshtein:1972sx}.
This is indeed the case as explicitly demonstrated in~\cite{Babichev:2009jt,Babichev:2009us,Babichev:2010jd}.
For a review on the Vainshtein mechanism we refer to~\cite{Babichev:2013usa}.

Bimetric theory incorporates the Vainshtein mechanism as explicitly demonstrated for static and spherically symmetric systems~\cite{Volkov:2012wp,Babichev:2013pfa,Enander:2013kza,Enander:2015kda}.
We leave the technical details to~\cref{sec:Vainshtein-derivation}, where we repeat and extend the analysis by relaxing the assumption of asymptotic flatness to allow for a non-vanishing cosmological constant.
Here we limit ourselves to discussing the general features.
The radius below which nonlinearities have to be taken into account is referred to as Vainshtein radius  $r_{\rm V} = (r_{\rm S} / \mFP^2)^{1/3}$ with $r_{\rm S}$ the Schwarzschild radius of the matter source.
For larger radii, $r\gg r_{\rm V}$, the linear approximation is valid.
The resulting gravitational potential for small and large radii can be written as~\cite{Babichev:2013pfa,Enander:2013kza}
\begin{flalign}\label{eq:gravitational-potential}
    V(r) =
    \begin{cases}
        - \frac{1}{m_{\rm g}^2} \frac{1}{r}   & r \ll r_V \\
         - \frac{1}{m_{\rm P}^2} \left( \frac{1}{r} + \frac{4\bar\alpha^2}{3}\frac{e^{ -\mFP r}}{r}\right)  & r \gg r_V
    \end{cases}\,,
\end{flalign}
where $m_{\rm P}^2 = (1+\ba^2)m_{\rm g}^2$ is the unscreened Planck mass.
This justifies our identification~(\ref{eq:Yukawa-parameters}).
On the other hand, since $r_{\rm V}$ depends on the spin-2 mass and the mass of the central source, the constraints on $\ba$ and $\mFP$ summarised in~\cref{fig:local-tests} are not trustworthy.

However, Vainshtein screening exists only in a subregion of the parameter space~\cite{Babichev:2013pfa,Enander:2015kda}, as we review in~\cref{sec:Vainshtein-derivation}.
In this section, we will identify this subregion in terms of $\ba$ and $\mFP$.
The Vainshtein mechanism relies on the following three necessary conditions:
\begin{itemize}
\item
\textit{Consistent screening regime.}
The existence of a nonlinear solution that restores GR in spacetime regions close to the matter source requires
\begin{equation}\label{eq:pos-b3}
	\frac{c^3\beta_3}{c\beta_1 + 2c^2\beta_2 +c^3 \beta_3} > 1\,.
\end{equation}
Since the denominator is positive if $\mFP^2>0$, this bound implies that $\beta_3$ must be strictly positive\footnote{Recall that the infinite branch solution in the context of background cosmology is well-defined only for $\beta_3<0$. Hence, requiring of a working Vainshtein mechanism rules out the infinite branch.}, $\beta_3>0$.
That means, any bimetric model with $\beta_3=0$ does not have a working Vainshtein mechanism (for some related caveats, see~\cite{Babichev:2013pfa}).

\item
\textit{Consistent asymptotics.}
The nonlinear equations must give rise to a solution that matches the linearised solution for $r\gg r_{\rm V}$.
This requires
\begin{equation}\label{eq:consistent-asymptotics}
	\frac{c^2\beta_2+c^3\beta_3}{c\beta_1 + 2c^2\beta_2 +c^3 \beta_3} < \sqrt{\frac{c^3\beta_3}{c\beta_1 + 2c^2\beta_2 +c^3 \beta_3}}
\end{equation}
to be satisfied.
Combined with~\cref{eq:pos-b3} it follows that $\beta_2$ must be strictly negative, $\beta_2<0$.
Bimetric models with $\beta_2=0$ do not give rise to Vainshtein screening.

\item
\textit{Consistent Vainshtein-Yukawa solution.}
The nonlinear solution realising Vainshtein screening has to be smoothly connected to the linear solution realising the Yukawa-type fifth force without branch cuts.
The existence of such a well-defined Vainshtein-Yukawa solution imposes another restriction on the parameters.
The explicit expression is lengthy so that we shift its presentation to~\cref{eq:Vainshtein-Yukawa-cond}.

\end{itemize}
Following~\cite{Luben:2020xll}, we express these conditions in terms of the physical parameters $\bar\alpha$, $\mFP$, and $\Lambda$ in this section.
The relations between the interaction and physical parameters are collected in~\cref{sec:details-physical-parametrisation}.
It suffices to study models with all interaction parameters $\beta_1$, $\beta_2$ and $\beta_3$ being free as only those models can possibly give rise to a viable cosmic expansion history while incorporating the local Vainshtein mechanism.

\begin{figure}
\centering
\includegraphics[width=0.8\textwidth]{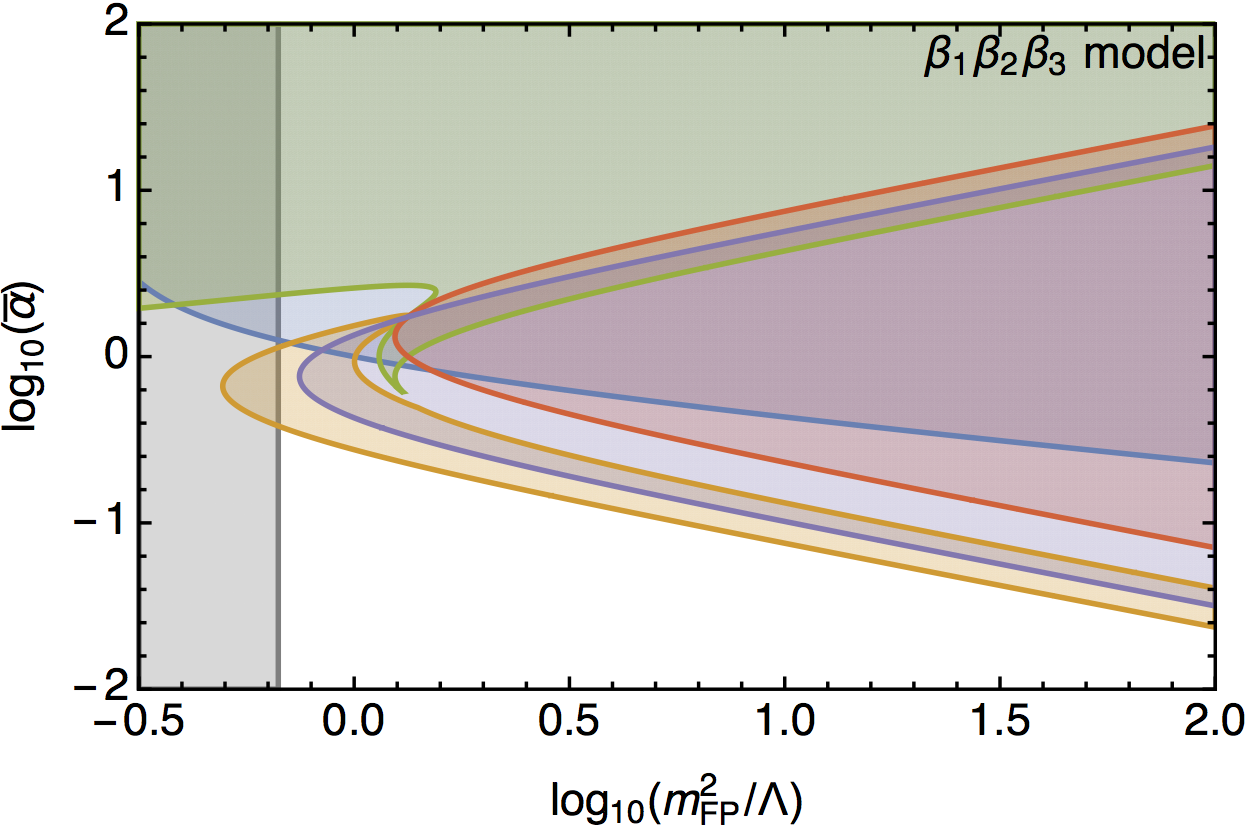}
\caption{The physical parameter space of the $\beta_1\beta_2\beta_3$ model is shown.
The model gives rise to a consistent cosmology only outside the
blue-shaded region, which indicates where the vacuum point is not well-defined,
and the red-shaded region, in which $\beta_1$ is non-positive.
The Vainshtein mechanism works only outside
the purple-shaded region, which does not give rise to a consistent screening regime,
the green-shaded region, which indicates where the Vainshtein solution does not have the right asymptotics,
and the yellow-shaded region, in which there is no Vainshtein-Yukawa solution.
Finally, the Higuchi bound is violated in the gray-shaded region.
Summarising, this bimetric model gives rise to a viable expansion history while incorporating a working Vainshtein mechanism
in the region that is left white.}
\label{fig:b123-param-space}
\end{figure}

Cosmologically viable one- and two-parameter models are not able to incorporate Vainshtein screening.
The $\beta_1\beta_2\beta_3$-model is the only three-parameter model with a possibly viable background cosmology and a working Vainshtein mechanism.
In~\cref{fig:b123-param-space} we collect all the aforementioned theoretical bounds on the physical parameter space of that model.
The expressions are too lengthy to display them here explicitly.
The blue- and red-shaded regions are excluded as these do not give rise to a viable
expansion history.
The purple-shaded region violates the bound in~\cref{eq:pos-b3}.
In the green-shaded region the bound in~\cref{eq:consistent-asymptotics} is violated.
The Vainshtein-Yukawa solution does not exist in the yellow-shaded region, which represents the most stringent bound.
To summarise, we expand the most-stringent bound on the physical parameters
for $\mFP^2\gg\Lambda$ to find
\begin{flalign}
	16 \bar\alpha^2\mFP^2 \lesssim \Lambda\,.
\end{flalign}
If this approximate bound is satisfied, the $\beta_1\beta_2\beta_3$-model incorporates the local Vainshtein mechanism.
As can be seen from~\cref{fig:b123-param-space}, the parameter region giving rise to a viable cosmic expansion history is only slightly larger.

Before moving on, let us point out the following caveats.
The equivalence principle is violated in two-body systems due to nonlinearities~\cite{Hiramatsu:2012xj}.
The assumption of spherical symmetry entering the derivation of the gravitational potential might not be appropriate to describe, e.g. the earth-moon system.
Further, Vainshtein screening is weaker or even completely absent in systems with cylindrical or planar symmetry, respectively, as demonstrated in the context of galileons~\cite{Bloomfield:2014zfa}.
This has to be taken into account for gravity tests without spherical symmetry.

\subsection{Local vs. cosmological bounds}
\label{sec:cosmological-vs-local}

We are finally in the position to compare the constraints from local tests of gravity, shown in~\cref{fig:local-tests}, to the parameter regions that give rise to Vainshtein screening.
Further, we compare the local constraints to the constraints that we inferred from background cosmology in~\cref{sec:cosmo-constraints}.
Since only the $\beta_1\beta_2\beta_3$-model incorporates Vainshtein screening, the local bounds are directly applicable to all the other models considered in this paper.
The discussed observational bounds in this section always refer to 95\% confidence level.

\begin{figure}
	\begin{center}		
		\includegraphics[width=0.49\textwidth]{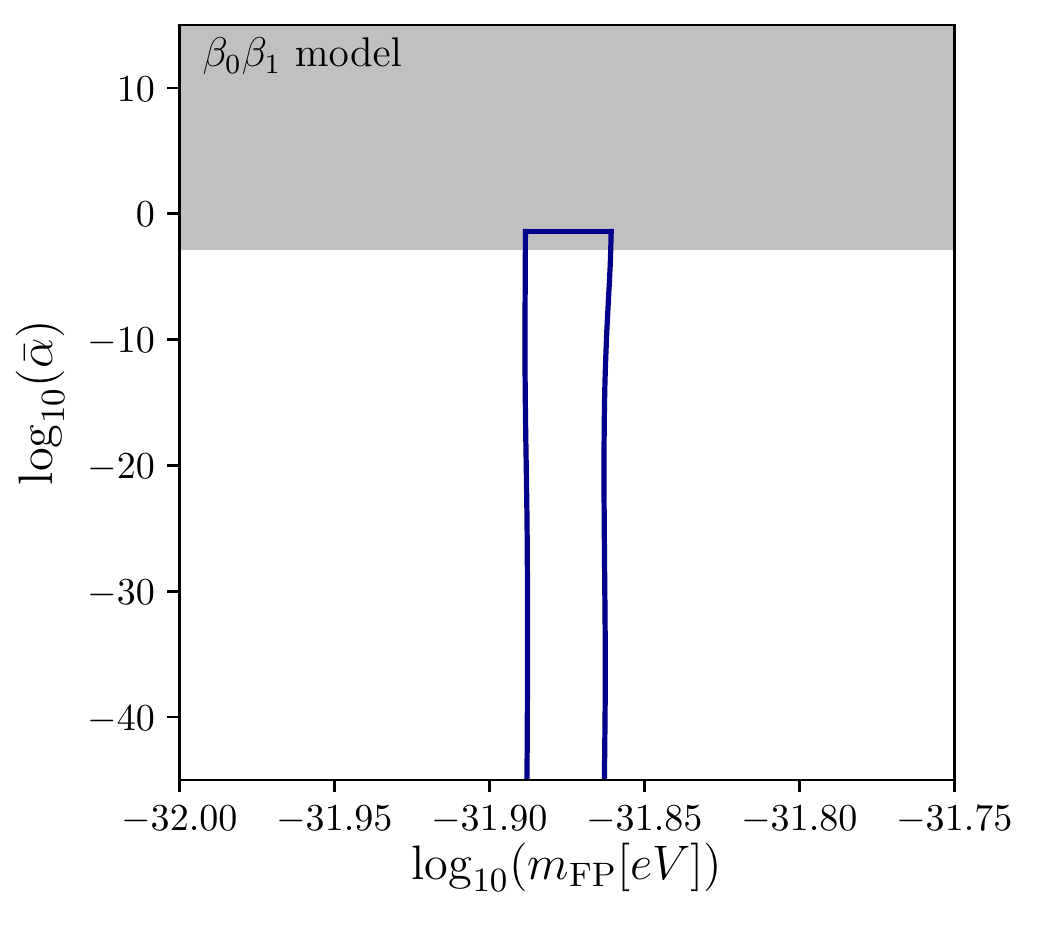}
		\includegraphics[width=0.453\textwidth]{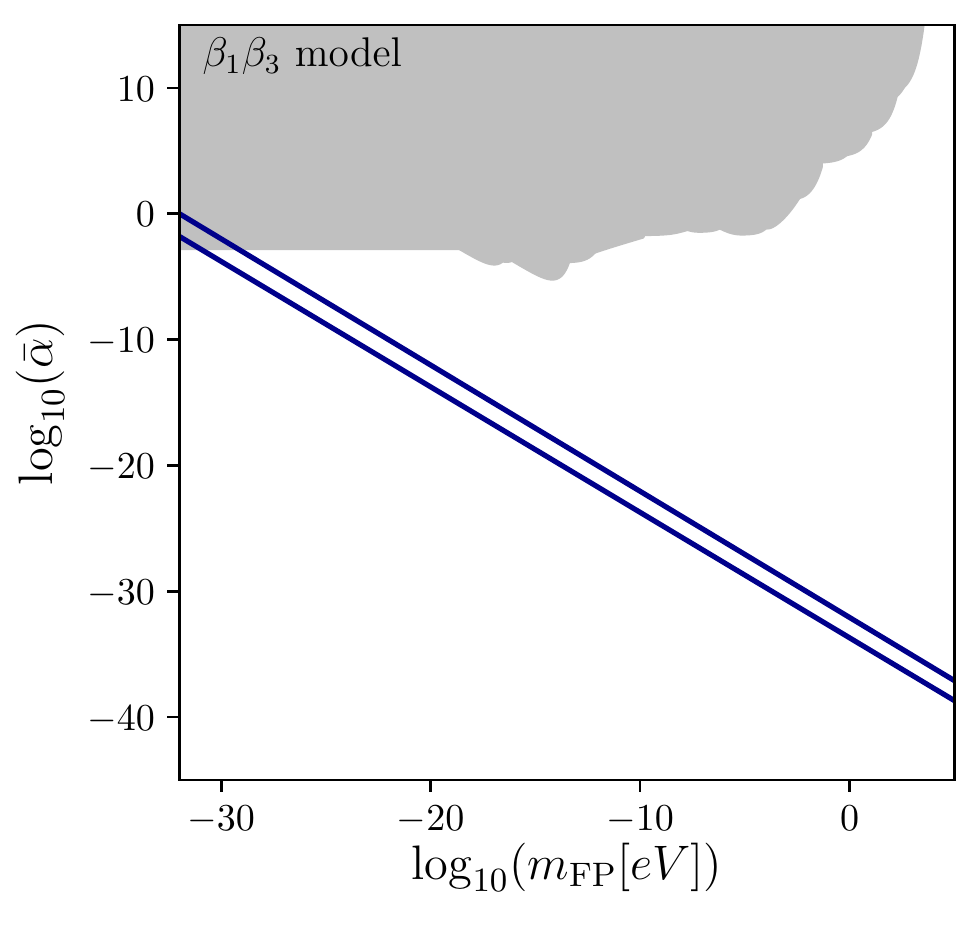}
		\caption{Cosmological constraints compared to local tests for the $\beta_0\beta_1$-model (left) and for the $\beta_1\beta_3$-model (right). The blue lines delimit the $2\sigma$ regions from the statistical analysis. The gray region is excluded by local tests at $2\sigma$.
			\label{fig:comparison2}}
	\end{center}
\end{figure}

\begin{figure}
	\centering
	\includegraphics[width=0.49\textwidth]{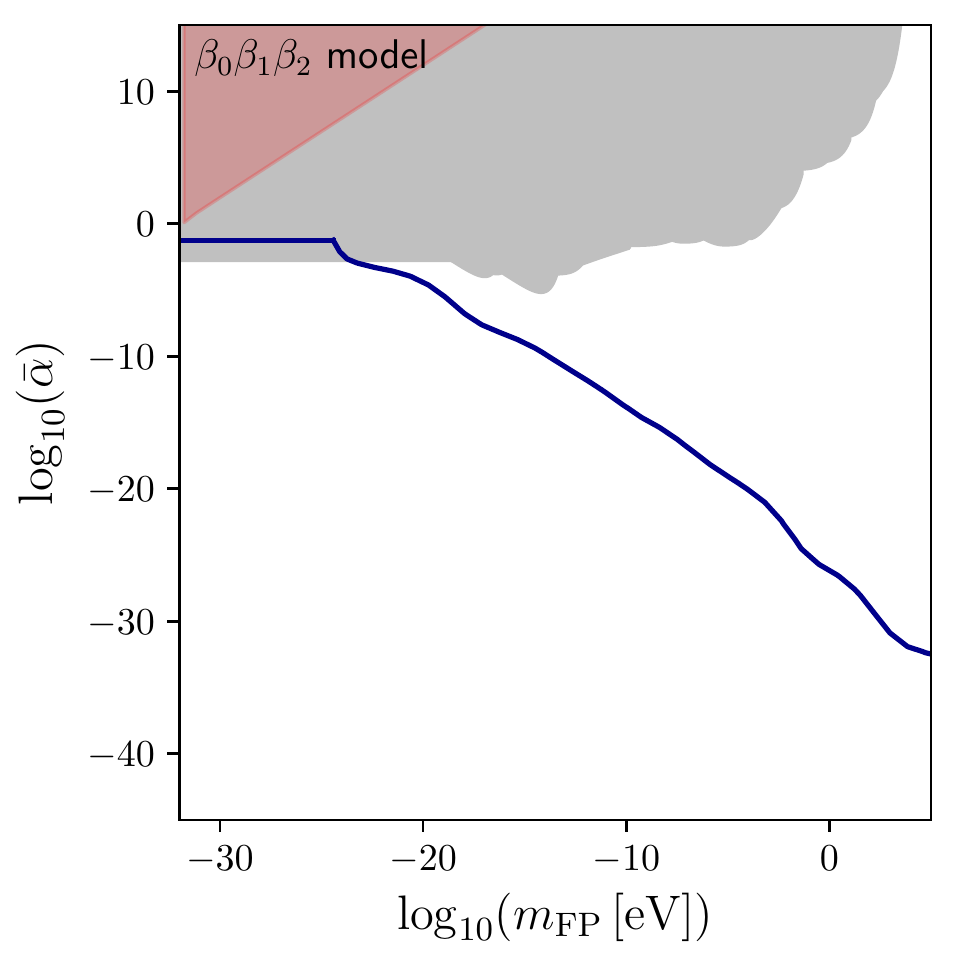}
	\includegraphics[width=0.49\textwidth]{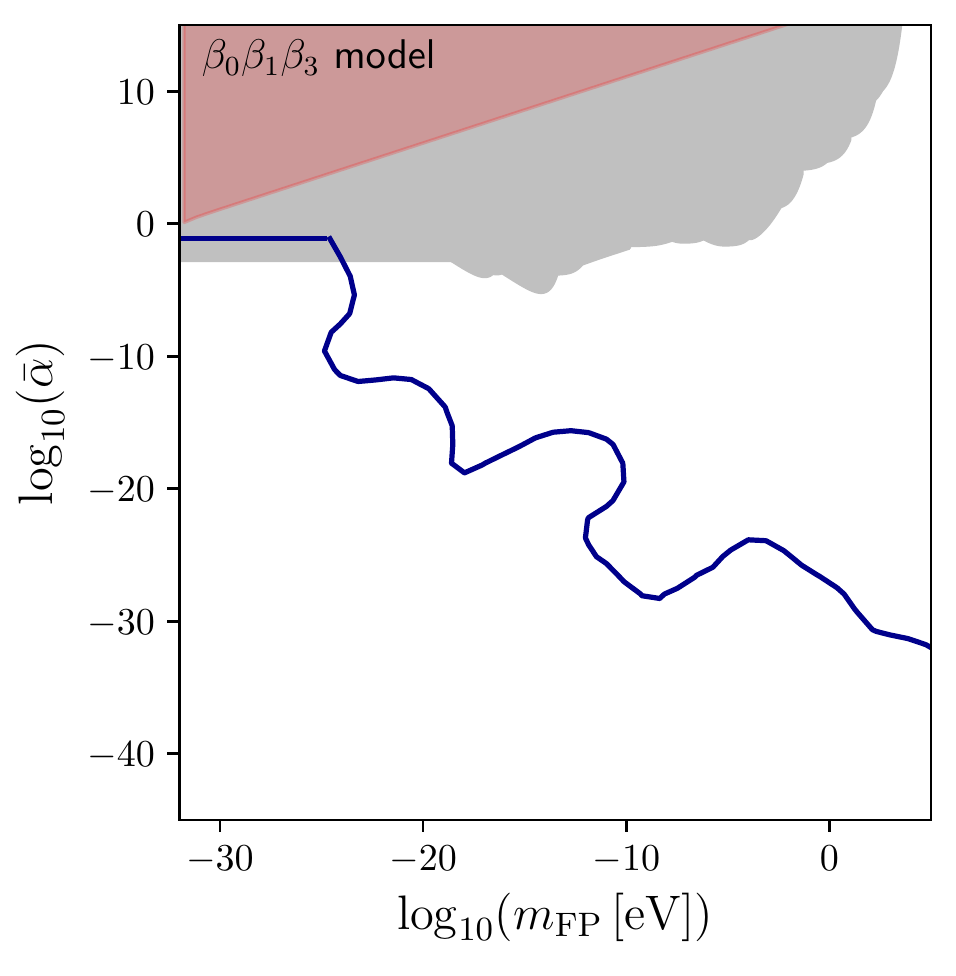}
	\caption{The combined bounds from local and cosmological tests for the $\beta_0\beta_1\beta_2$- (left) and $\beta_0\beta_1\beta_3$-model (right).
		The gray-shaded region is excluded by local tests of gravity at $95\%$ c.l.
		The red-shaded region is excluded by theoretical constraints.
		The region above the blue line is excluded by our bounds from cosmology at $95\%$ c.l.}
	\label{fig:comparison_B012-B013}
\end{figure}

\begin{figure}
	\centering
	\includegraphics[width=0.453\textwidth]{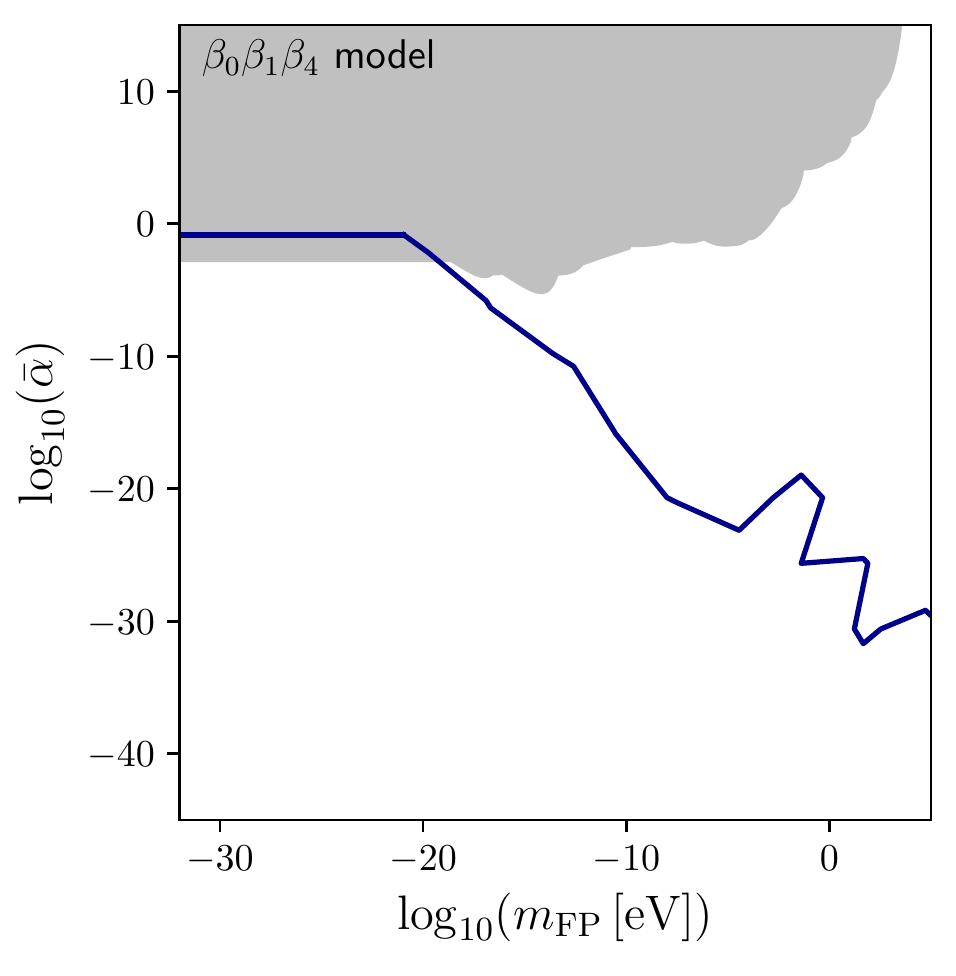}
	\includegraphics[width=0.49\textwidth]{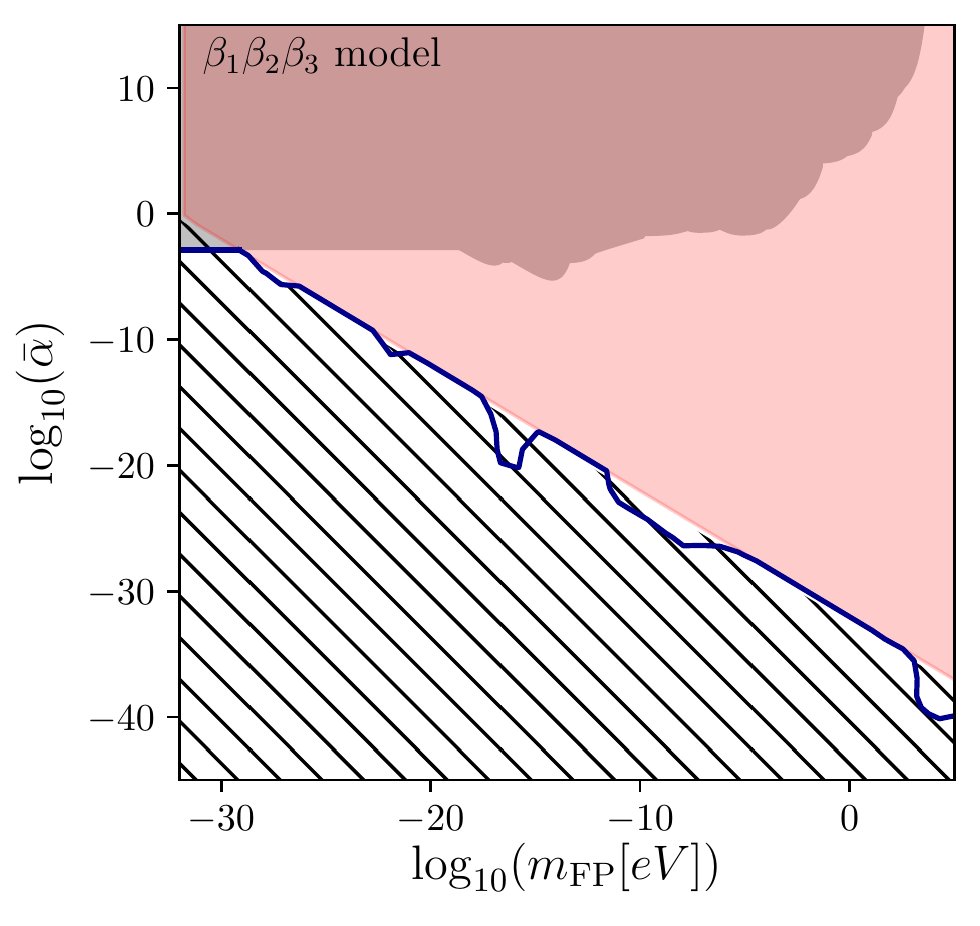}
	\caption{The combined bounds from local and cosmological tests for the $\beta_0\beta_1\beta_4$- and $\beta_1\beta_2\beta_3$-model.
		The gray-shaded region is excluded by local tests of gravity at $95\%$ c.l.
		The red-shaded region is excluded by theoretical constraints.
		The region above the blue line is excluded by our bounds from cosmology at $95\%$ c.l.
		The hatched region in the right panel represents where the Vainshtein mechanism is working.}
	\label{fig:comparison_B014-B123}
\end{figure}

\begin{figure}
	\centering
	\includegraphics[width=0.49\textwidth]{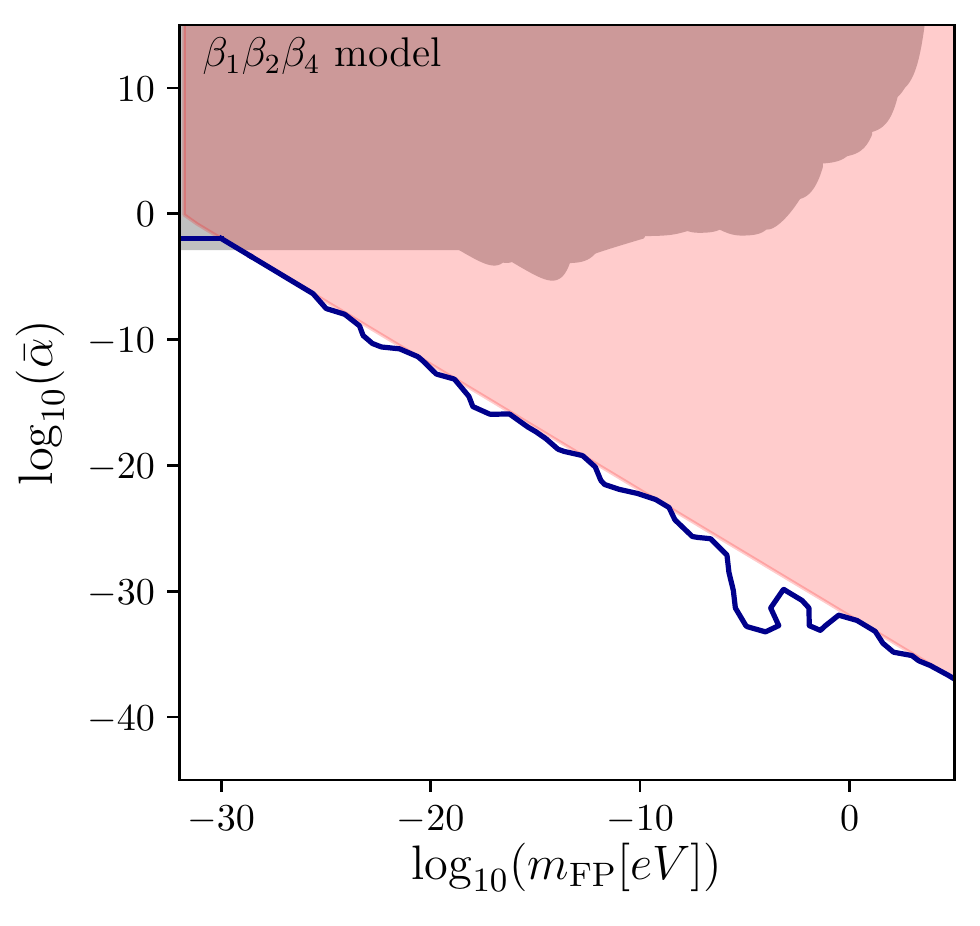}
	\includegraphics[width=0.49\textwidth]{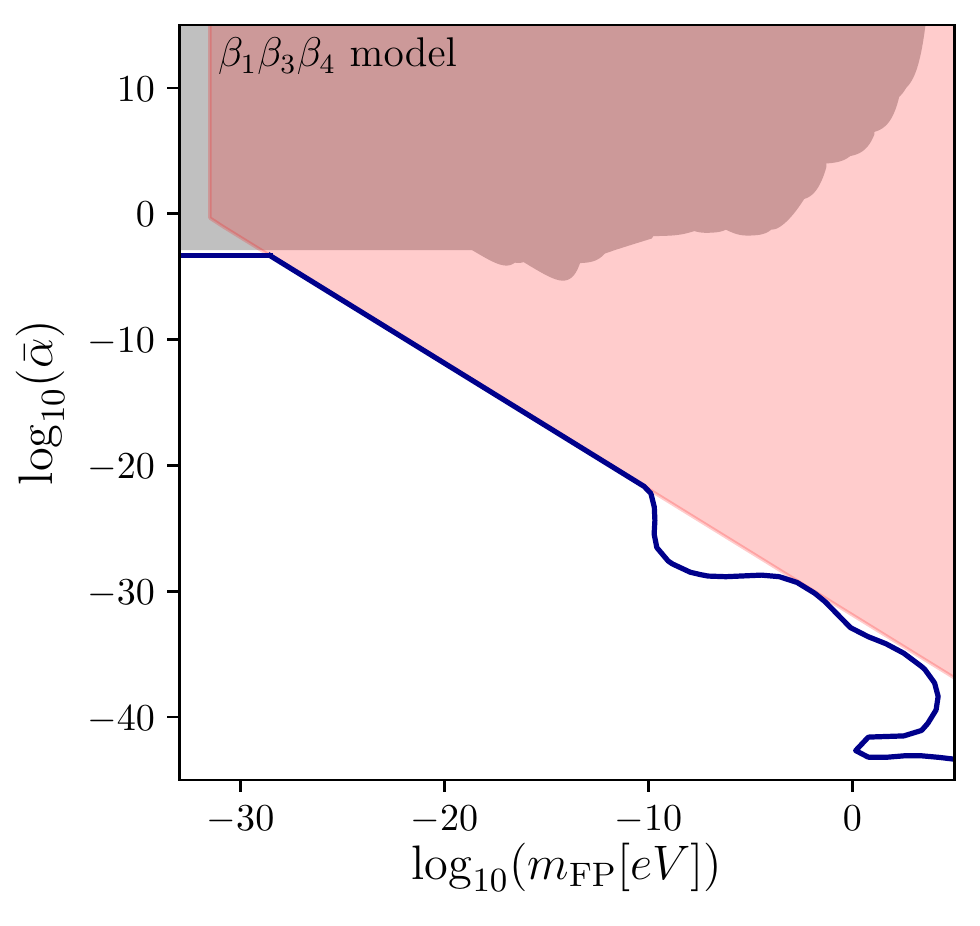}
	\caption{The combined bounds from local and cosmological tests for the $\beta_1\beta_2\beta_4$- and $\beta_1\beta_3\beta_4$-model.
		The region above the gray line is excluded by local tests of gravity at $95\%$ c.l.
		The red-shaded region is excluded by theoretical constraints.
		The region above the blue line is excluded by our bounds from cosmology at $95\%$ c.l.}
	\label{fig:comparison_B124-B134}
\end{figure}

Let us start discussing the $\beta_1$-model.
As mentioned before, the cosmological data sets are not compatible, strongly disfavoring this model.
If we nonetheless combine the data sets, the spin-2 mass is constrained to $\mFP = (1.82\pm 0.02)\times 10^{-32}\,{\rm eV}$.
The coupling to matter is fixed to the value $\ba = 1/\sqrt{3}\simeq 0.58$, which is incompatible with the observational bound from Cassini $\ba \lesssim 1.7\times 10^{-3}$ in that mass regime.
Since this model does not give rise to Vainshtein screening, we conclude that the $\beta_1$-model is completely ruled out by the combined analysis of cosmological and local data.

Next let us discuss the two-parameter models, which do not give rise to Vainshtein screening as well.
This implies that the bounds from local tests of gravity are indeed trustworthy.
In~\cref{fig:comparison2} we compare the bounds from local and cosmological tests for the two parameter models $\beta_0\beta_1$ and $\beta_1\beta_3$.
The blue line defines the boundary of the region that is consistent with cosmological data.
The gray-shaded region is excluded by local tests of gravity.
For the $\beta_0\beta_1$-model, cosmological data constraints the spin-2 mass to $\mFP=(1.33\pm0.02)\times 10^{-32}\, {\rm eV}$ while Cassini constrains the coupling to matter to $\ba \lesssim 1.7\times 10^{-3}$, which is slightly more stringent than the cosmological bound.
For the $\beta_1\beta_n$-model, the physical parameters are correlated as can be seen in the right panel of~\cref{fig:comparison2}.
Indeed, they are approximately related as $\ba^2 \mFP^2 \simeq \frac{12}{n}\Lambda$ for $n=\{2,3,4\}$ (see~\cref{sec:physical-para-2})~\cite{Luben:2020xll}.
As before, Cassini provides the most stringent constraint on $\ba$.
Due to the parameter correlation, this implies a more stringent constraint on the spin-2 mass than inferred from cosmology of $\mFP \gtrsim 1.85\times 10^{-30}\,{\rm eV}$ at 95\% c.l.
Therefore, all these two-parameter models are in agreement with observational constraints from cosmology and local tests of gravity, but driven to their GR-limits.

In~\cref{fig:comparison_B012-B013,fig:comparison_B014-B123,fig:comparison_B124-B134} we combine theoretical, cosmological and local bounds in single plots for all the three parameter models.
In all these plots, the regions above the blue lines are excluded at 95\% c.l. by our cosmological bounds inferred in~\cref{sec:data}.
The red-shaded regions in the plots are excluded by theoretical constraints that ensure a consistent cosmic expansion history, which we used as hard priors in the cosmological analysis.
The region excluded by local bounds at 95\% c.l. (if the Vainshtein mechanism is not taken into account) is shown as a gray-shaded region.
For large spin-2 masses, the bounds from cosmology are the most stringent ones.
For smaller spin-2 masses, the most stringent bound is provided by tests of the scalar curvature with the Cassini spacecraft.
Summarising, without taking Vainshtein screening into account, all three-parameter models  are consistent with both local tests of gravity and background cosmology.
However, since the coupling of the massive spin-2 field to matter is forced to be small, i.e. $\ba\lesssim 1.7\times 10^{-3}$, all these models are essentially forced into their GR-limits.
Although local tests of gravity allow for large couplings to matter if the spin-2 mass is large, $\mFP \gtrsim 10^{-2}\,{\rm eV}$, cosmological data excludes this parameter region.
Therefore, deviations from GR are substantially suppressed for these models on all scales.

Let us discuss further about the cosmological bounds on $\ba$ and $\mFP$.
From inspecting~\cref{fig:comparison_B012-B013,fig:comparison_B014-B123,fig:comparison_B124-B134} we notice that the 95\% contour line falls into two regimes. The first is the small spin-2 mass regime. Here,  below some critical value that depends on the model, the observational constraints becomes independent of $\mFP$. This is particularly evident for the models with $\beta_0\neq 0$.
In this regime, the coupling to matter $\ba$ is bounded from above as reported in~\cref{tab:results1,tab:results2}.
The second regime is above the critical value for $\mFP$, where the contour line can be approximated as:
\begin{equation}\label{eq:fitting}
	\bar{\alpha}^{c_1}\cdot {m_{\rm FP}}\leq10^{-c_2}\text{ }\text{eV}.
\end{equation}
We give the explicit value of the coefficients $c_1$ and $c_2$ in~\cref{tab:coefficients}, as the result of a polynomial fit. This behavior can be seen in~\cref{fig:comparison_B012-B013,fig:comparison_B014-B123,fig:comparison_B124-B134}, that however show only a portion of the $\bar{\alpha}-m_{\rm FP}$ plane scanned by the cosmological analysis\footnote{The wiggles in the contour lines of~\cref{fig:comparison_B012-B013,fig:comparison_B014-B123,fig:comparison_B124-B134} are a consequence of zooming in on a portion of the full scanned space, and are interpreted as artifacts of the Markov chains having finite size.}. Indeed, the approximation \eqref{eq:fitting} is valid up to the Planck scale, i.e. beyond the zoomed-in regions in the plots\footnote{The limits on $\bar{\alpha}$ and $m_{\rm FP}$ used in our statistical analysis can be found in \cref{sec:cosmo-constraints}.}.  

\begin{table}
	\centering
	{\def\arraystretch{1.2}
		\begin{tabular}{ l | l | l}
			\hline\hline
			Model & $c_1$ & $c_2$\\
			\hline
			$\beta_0\beta_1\beta_4$ & $0.97\pm0.01$ & $23.6\pm0.4$ \\
			$\beta_0\beta_1\beta_3$ & $1.117\pm0.006$ & $32.9\pm0.2$ \\
			$\beta_0\beta_1\beta_2$ & $0.964\pm0.006$ & $24.6\pm0.2$ \\
			$\beta_1\beta_2\beta_3$ & $1.047\pm0.007$ & $35.5\pm0.2$\\
			$\beta_1\beta_2\beta_4$ & $1.021\pm0.006$ & $32.7\pm0.2$\\
			$\beta_1\beta_3\beta_4$ & $0.959\pm0.008$ & $32.7\pm0.3$\\
			\hline\hline
	\end{tabular}}
	\caption{The value of numerical coefficients for the fitting formula \eqref{eq:fitting}, obtained with a polynomial fit. We also show $1\sigma$ errors in the determination of the coefficients.}
	\label{tab:coefficients}
\end{table}

For the $\beta_1\beta_2\beta_3$-model, we need to take into account the Vainshtein mechanism when including local tests of gravity.
The parameter region, which supports Vainshtein screening, is depicted as a hatched area in the right panel of~\cref{fig:comparison_B014-B123}.
As we have seen before, the region incorporating Vainshtein screening is almost exactly complementary to the region that is excluded by our theoretical bounds.
In particular, the entire parameter region that is consistent with cosmological data gives rise to Vainshtein screening.
From the right panel of~\cref{fig:comparison_B014-B123} we see that there is a small overlap between the region excluded by local tests and the region where the Vainshtein mechanism is active.
In this interesting region, which is characterized by a potentially large value of $\ba$, local tests are not trustworthy.
Unfortunately this small region seems to be excluded by cosmological data at 95\% c.l. in our statistical analysis.
However, this is probably caused by prior domination.
Indeed, this is a very small region of the full parameter space explored by the statistical analysis and it is close to the hard prior given by the theoretical constraints.
To further investigate this issue, one should perform another statistical analysis focusing on this small region of parameter space.

For the models with $\beta_0=0$, the $95\%$ contour lies very close to the theoretical priors except for a few wiggles.
We interpret the wiggles as an artifact of the finite size of the Markov chains and conclude that the used data sets are not constraining these parameters.
Instead, the statistically favored regions almost exactly coincide with the parameter regions allowed by our theoretical priors.
To stabilise the value of $\ba$ and $\mFP$, we would need to include other data sets, which however is beyond the scope of the present analysis.

On the other hand, models with $\beta_4=0$ are subject to rather weak theoretical priors.
Nonetheless, the $95\%$ lines show the same tendency as for the other models.
Hence, the data sets constrain these parameters substantially beyond the prior knowledge, but is not able to stabilise their values.
To understand why the upper-right region in the $\ba-\mFP$-plane is excluded also for these models, we note the following.
Expanding the expression for $\beta_0$ in the limit $\mFP^2\gg\Lambda$, we find that $\beta_0\simeq - \frac{12}{n} \ba^2 \mFP^2$ for the $\beta_0\beta_1\beta_n$-model with $n=\{2,3,4\}$.
This allows to conclude that data disfavors large negative values of $\beta_0$.
This means that, in order to be cosmologically viable, $\rho_{\rm de}$ is allowed to change its sign only at sufficiently large redshifts.

\section{Conclusions and outlook}
\label{sec:conclude}

Massive spin-2 fields might play a crucial role in our universe.
Their (self-)interaction energy can drive the observed cosmic accelerated expansion at late times.
Furthermore, a massive spin-2 field serves as ideal dark matter candidate as it interacts with ordinary matter only gravitationally~\cite{Babichev:2016hir,Babichev:2016bxi,Chu:2017msm}.
Recently, another intriguing aspect has been explored within a completely different context.
As argued in~\cite{Geng:2020qvw}, massive spin-2 states could be essential in solving the black hole information paradoxon.

In this paper, we have further studied the observational viability of bimetric theory.
We confronted all models with up to three free interaction parameters $\beta_n$ with measurements of SN1a, BAOs and CMB.
We find that all two and three parameter models are in perfect agreement with these data sets.
Compared to the $\Lambda$CDM model, bimetric models are statistically disfavored due to the larger number of free parameters.
However, the self-accelerating solutions with $\beta_0=0$ appear more appealing from a theoretical perspective as these models do not suffer the cosmological constant problems~\cite{Weinberg:1988cp,Martin:2012bt}.

The best-fit values for the cosmological parameters do not significantly differ from the best-fit values of the $\Lambda$CDM model.
We have also studied spatial curvature within bimetric theory, and
our results show that a spatially flat universe is always preferred.

The only exception from the previous summary is the $\beta_1$-model, which represents the simplest bimetric model that gives rise to self-accelerating solutions on the finite branch.
Our statistical analysis shows that the aforementioned data sets are inconsistent leading to a substantially higher $\chi^2$ at the best fit point.
We therefore conclude that the $\beta_1$-model is statistically disfavored.
This conclusion is further strengthened when simultaneously confronting this model with local tests of gravity.
Cosmological data selects a spin-2 mass of $\mFP = (1.82\pm 0.02)\times 10^{-32}\,{\rm eV}$, for which local tests of gravity demand a coupling to matter of $\ba\lesssim 1.7\times 10^{-3}$ at 95\% c.l.
Within the $\beta_1$-model, this parameter is fixed to be $\ba=3^{-1/2}$.
Therefore, this model is in severe conflict with observations.

We also analysed the other bimetric models with up to three free parameters, for which we compared the bounds from background cosmology to the ones obtained from local tests of gravity.
All models are consistent with both types of observations even if the Vainshtein mechanism is not active.
However, the coupling of the massive spin-2 field to ordinary matter is forced to be small, $\ba\lesssim 10^{-3}$ (see~\cref{sec:local-tests-review} and~\cref{tab:results1,tab:results2} for the precise values).
For large spin-2 masses, the cosmological observations force the coupling to be even smaller.
This suppresses the deviations from GR on all scales for these models.

For the $\beta_1\beta_2\beta_3$-model, the Vainshtein mechanism removes the constraint on $\ba$ and $\mFP$ implied by local tests for sufficiently small values of the spin-2 mass. In this region, where $\ba$ can be large, the theory is not in its GR limit. As we mentioned in \cref{sec:cosmological-vs-local}, the fact that this small region is excluded by our cosmological analysis might be a numerical effect, and it would be interesting to further investigate this issue in another analysis.
Further, more general bimetric models, i.e. those with four or all five parameters $\beta_n$ being free, might not be forced into their GR-limits.
In any case, even if the restricted models studied in this paper turn out to be forced into their GR-limits, their theoretical motivation would persist: self-accelerating models with $\beta_0=0$ have an effective cosmological constant that is technical natural in the sense of t'Hooft~\cite{tHooft:1979rat}.

The data sets used in this paper are only weekly constraining the physical parameter space of $\ba$ and $\mFP$.
To stabilise their values and effectively constrain bimetric theories, a possibility would be taking into account further measurements.
Cosmological perturbations around the FLRW background solution are expected to have constraining power as these probe different redshifts and scales.
However, in bimetric theory linear scalar perturbations are plagued by gradient instabilities at early times~\cite{Comelli:2012db,Konnig:2014dna,Konnig:2014xva,Lagos:2014lca,DeFelice:2014nja,Konnig:2015lfa}.
The instability sets in when the Hubble rate is of the order of the spin-2 mass, $H\simeq \mFP$~\cite{Luben:2019yyx}.
Nonlinear terms become as important as the linear term rendering perturbation theory invalid.
Indeed, taking nonlinearities into account might stabilise the solution in analogy to the Vainshtein mechanism~\cite{Mortsell:2015exa,Aoki:2015xqa,Hogas:2019ywm,Luben:2019yyx}.
Structure formation and nonlinear cosmological perturbations are therefore main open issues within bimetric theory, which should be addressed in the future.

\section*{Acknowledgements}

We thank Marcus H\"og\r{a}s and Edvard M\"ortsell for interesting discussions and useful comments on the manuscript.
We are grateful to Thomas Hahn, Martin Kerscher and Kerstin Paech for their kind help regarding the statistical analysis and numerical implementation.
We further acknowledge the usage of the {\tt Getdist} package, which we we extensively used for the statistical analysis~\cite{Lewis:2019xzd}.
The work of ML is supported by a grant from the Max Planck Society.
The work of AC and JW is supported by the Excellence Cluster ORIGINS which is funded by the Deutsche Forschungsgemeinschaft (DFG, German Research Foundation) under Germany's Excellence Strategy - EXC-2094 - 390783311.

\appendix

\section{Details on the physical parametrisation}
\label{sec:details-physical-parametrisation}
In order to provide an overview over the relation between the physical and interaction
parameters, we summarise the results of Ref.~\cite{Luben:2020xll} in this appendix.
We immediately work in terms of the cosmological parameters that we use for parameter fitting.

In the $\beta_1$-model, there is only one free physical parameter, in terms of
which the interaction parameter is given as
\begin{equation}
B_1=\frac{1}{\sqrt{3}}\Omega_{\Lambda}=\frac{\sqrt{3}}{4}\OFP\,,
\end{equation}
while the mixing angle is fixed to be $\bar\alpha=1/\sqrt{3}$.

\subsection{Two-parameter models}
\label{sec:physical-para-2}

Next, we turn to the two-parameter models with $\beta_1\ne0$.
We choose to express the interaction parameters on terms of $\OFP$ and $\OL$, while
the mixing angle $\bar\alpha$ is a derived parameter. However, one can equivalently
eliminate one of the physical parameters by $\bar\alpha$.
For the $\beta_0\beta_1$-model, the relation between the parameters is
\begin{equation}
	B_0 = -3 \OFP + 4\OL\,, \ B_1=\sqrt{(\OFP-\OL)\OL}\,,
\end{equation}
in terms of which the mixing angle is determined to be $\bar\alpha = \sqrt{\OFP/\OL-1}$.

For the $\beta_1\beta_2$-model, the relation is given by
\begin{gather}
\begin{split}
	B_1 & = \frac{1}{2}  \sqrt{\frac{3\OFP - 2\OL - \Delta_{12}}{3\OL}}\left( 3\OFP - 3\OL + \Delta_{12} \right)\,,\\
	B_2 & = - \frac{1}{6} \left( 3\OFP - 5\OL + \Delta_{12} \right)
\end{split}
\end{gather}
where we defined $\Delta_{12}=\sqrt{9\OFP^2 -12\OFP \OL + \OL^2}$.
The mixing angle is given by $\bar\alpha^2=(3\OFP - 2\OL - \Delta_{12})/(3\OL)$.
To avoid numerical instabilities, it is convenient to expand the expressions for $\OFP\gg\OL$.
In this limit, the parameter relation reads
\begin{flalign}
	\bar\alpha \simeq \sqrt{\frac{\OL}{6\OFP}}\,,\qquad
	B_1 \simeq \sqrt{\frac{3\OFP \OL}{2}}\,, \qquad
	B_2 \simeq -\OFP\,.
\end{flalign}

Next, we turn to the $\beta_1\beta_3$-model, where the relations read
\begin{flalign}\label{eq:param-relations-b13}
	B_1 & = \frac{1}{4}\sqrt{ \frac{2 \OFP-\OL - 2\Delta_{13} }{\OL} }
	\sqrt{ 3\OFP - 2\OL + 3 \Delta_{13} }\,,\\
	B_3 & = -\sqrt{ \frac{2 \OFP-\OL - 2\Delta_{13} }{\OL}}
	\left( 4\OFP^2 - 7\OFP\OL + 2\OL^2 + \Delta_{13}(4\OFP-5\OL) \right)\,.
\end{flalign}
with $\Delta_{13}=\sqrt{(\OFP-\OL)\OFP}$.
The mixing angle is given by $\bar\alpha^2=(2 \OFP-\OL - 2\Delta_{13})/(\OL)$.
Again, to avoid numerical instabilities, we expand the expressions for $\OFP\gg\OL$
yielding
\begin{flalign}
	\bar\alpha \simeq \frac{1}{2}\sqrt{\frac{\OL}{\OFP}}\,,\qquad
	B_1 \simeq \frac{3}{4}\sqrt{\OFP \OL}\,, \qquad
	B_3 \simeq - \sqrt{\frac{\OFP^3}{\OL}}\,.
\end{flalign}

Finally, in the $\beta_1\beta_4$-model the parameter relations read
\begin{flalign}
	B_1 & = \frac{1}{3} \sqrt{(3\OFP-\OL)\OL}\,,\\
	B_4 & =  - \frac{9\OFP^2 - 15\OFP\OL + 4\OL^2}{3\OL}
\end{flalign}
while the mixing angle reads $\bar\alpha^2 = \OL/(3\OFP-\OL)$.

\begin{table}
\centering
{\def\arraystretch{1.2}
\begin{tabular}{ l | l | l}
\hline\hline
Model & Viable cosmology & Vainshtein screening\\
\hline
$\beta_0\beta_1$ & $\OFP>\OL$ & -- \\
\hline
$\beta_1\beta_2$ & $3\OFP>(2+\sqrt{3})\OL$ & -- \\
\hline
$\beta_1\beta_3$ & $\OFP>\OL$ & -- \\
\hline
$\beta_0\beta_1\beta_2$ & $3\OFP > 2(1+\bar\alpha^2)\OL$ & -- \\
\hline
$\beta_0\beta_1\beta_3$ & $4\OFP > (1+\bar\alpha^2)^2\OL$ & -- \\
\hline
$\beta_1\beta_2\beta_3$ & $4\bar\alpha^4\OFP < (1+\bar\alpha^2)^2\OL$ & $16\ba^2 \OFP \lesssim \OL$\\
 & $6\bar\alpha^2\OFP < (3+4\bar\alpha^2+\bar\alpha^4)\OL$ \\
 \hline
$\beta_1\beta_2\beta_4$ & $3\bar\alpha^2 \OFP < 2(1+\bar\alpha^2)\OL$ & -- \\
 & $3\bar\alpha^4\OFP < (1+\bar\alpha^2)^2\OL$ & \\
 \hline
$\beta_1\beta_3\beta_4$ & $\ba^2\OFP < (1+\ba^2)\OL$ & -- \\
\hline\hline
\end{tabular}}
\caption{This table summarises the theoretical constraints on all bimetric models
with up to three non-vanishing interaction parameters.
The cosmology constraints were derived in Ref.~\cite{Luben:2020xll} and follow from
requiring a consistent cosmic expansion history.
We only state the most stringent constraints other than the Higuchi bound, $3\OFP>2\OL$.
Only the $\beta_1\beta_2\beta_3$-model has a working Vainshtein mechanism.
We report the approximation of the most stringent bound.
}
\label{tab:viable-cosmo-cond}
\end{table}

\subsection{Three-parameter models}

We continue by stating the parameter relations for the three-parameter models with $\beta_1\ne0$.
For the $\beta_0\beta_1\beta_2$-model, the relations read
\begin{gather}
\begin{split}
	B_0 &= - \frac{ 6\bar\alpha^2 \OFP - (1+4\bar\alpha^2 + 3\bar\alpha^4)\OL}{1+\bar\alpha^2}\,,\\
	B_1 &= \bar\alpha\left( \frac{3\OFP}{1+\bar\alpha^2} - 2\OL \right)\,\\
	B_2 &= - \frac{\OFP}{1+\bar\alpha^2} + \OL\,.
\end{split}
\end{gather}
In the $\beta_0\beta_1\beta_3$-model, the physical and interaction parameters are related as
\begin{gather}
\begin{split}
	B_0 &= - \frac{4\bar\alpha^2\OFP - (1+2\bar\alpha^2+\bar\alpha^4)\OL}{1+\bar\alpha^2}\,,\\
	B_1 &= \frac{\bar\alpha}{2}\left(\frac{3 \OFP}{1+\bar\alpha^2} - \OL \right)\,,\\
	B_3 &= - \frac{1}{2\bar\alpha}\left( \frac{\OFP}{1+\bar\alpha^2} - \OL \right)\,.
\end{split}
\end{gather}
Next, the $\beta_0\beta_1\beta_4$-model gives rise the the following relations,
\begin{equation}
	B_0 = - \frac{3\ba^2\OFP}{1+\bar\alpha^2} + \OL\,,\qquad
	B_1 = \frac{\bar\alpha \OFP}{1+\bar\alpha^2}\,,\qquad
	B_4 = -\frac{1}{\bar\alpha^2}\left( \frac{\OFP}{1+\bar\alpha^2} - \OL \right)\,.
\end{equation}
Turning to models with $\beta_0=0$, the parameter relations
for the $\beta_1\beta_2\beta_3$-model read
\begin{gather}
\begin{split}
	B_1 &= \frac{- 6\bar\alpha^2\OFP + (3 + 4\bar\alpha^2 + \bar\alpha^4)\OL}{4\bar\alpha(1+\bar\alpha^2)}\,,\\
	B_2 &= \frac{4\bar\alpha^2\OFP - (1 + 2\bar\alpha^2 + \bar\alpha^4)\OL}{2\bar\alpha^2(1+\bar\alpha^2)}\,,\\
	B_3 &= - \frac{6\bar\alpha^2\OFP - (1+4\bar\alpha^2 + 3\bar\alpha^4)\OL }{4\bar\alpha^3(1+\bar\alpha^2)}\,.
\end{split}
\end{gather}
Next, the $\beta_1\beta_2\beta_4$-model has the following parameter relations,
\begin{gather}
\begin{split}
	B_1 &= \frac{-3\bar\alpha^2 \OFP + 2(1+\bar\alpha^2)\OL}{3\bar\alpha(1+\bar\alpha^2)}\,,\\
	B_2 &= \frac{ 3\bar\alpha^2 \OFP - (1+\bar\alpha^2) \OL }{ 3\bar\alpha^2 (1+\bar\alpha^2) }\,,\\
	B_4 &= - \frac{6\bar\alpha^2 \OFP - (1+4\bar\alpha^2 + 3\bar\alpha^4)\OL}{3\bar\alpha^4(1+\bar\alpha^2)}\,.
\end{split}
\end{gather}
Finally, in the $\beta_1\beta_3\beta_4$-model the interaction and physical parameters are related as
\begin{gather}
\begin{split}
	B_1 &= \frac{-\bar\alpha^2\OFP + (1+\bar\alpha^2)\OL}{2\bar\alpha(1+\bar\alpha^2)}\,,\\
	B_3 &= \frac{3\bar\alpha^2 \OFP - (1+\bar\alpha^2)\OL}{2\bar\alpha^3(1+\bar\alpha^2)}\,,\\
	B_4 &= \frac{-4\bar\alpha^2 \OFP + (1+2\bar\alpha^2+\bar\alpha^4)\OL}{\bar\alpha^4(1+\bar\alpha^2)}\,.
\end{split}
\end{gather}

\section{Vainshtein screening in static, spherically symmetric systems}
\label{sec:Vainshtein-derivation}
In this appendix, we will solve the bimetric equations of motion for the
static, spherically symmetric, and bidiagonal ansatz.
This allows to study the Vainshtein mechanism in these systems,
which restores General Relativity on small length scales.
However, a solution that incorporates the Vainshtein mechanism
does not exist for every set of bimetric parameters.
Instead, demanding existence of such a solution implies conditions
on the bimetric parameters.
The situation was studied in detail in Refs.~\cite{Babichev:2013pfa,Enander:2015kda}
however with flat asymptotics.
We aim at generalizing the analysis allowing for a non-zero cosmological
constant~\cite{Platscher:2016adw} and compute the conditions for the
Vainshtein mechanism to work in this more general setup.
As we will see, our results reduce to the ones of Ref.~\cite{Babichev:2013pfa,Enander:2015kda}
in the limit of vanishing cosmological constant.

Assuming staticity and spherical symmetry, the metrics $\gmn$ and $\fmn$ can
be written in the bidiagonal case as~\cite{Babichev:2013pfa}
\begin{gather}
\begin{split}
	\dd s_{\rm g}^2 & = - e^{-\nu_{\rm g}}\dd t^2 + e^{\lambda_{\rm g}}\dd r^2 + r^2 \dd\Omega^2\,,\\
	\dd s_{\rm f}^2 & = - e^{-\nu_{\rm f}}\dd t^2 + e^{\lambda_{\rm f}}(r+r\mu)'^2\dd r^2 + (r+r\mu)^2 \dd\Omega^2\,.
\end{split}
\end{gather}
The metric functions $\nu_{\rm g,f}$, $\lambda_{\rm g,f}$, and $\mu$ depend on radius $r$ only.
The field $\mu$ can be thought of as a St\"uckelberg field that restores diffeomorphisms
in the $r$-dimension.

First, let us introduce the following short-hand notation,
\begin{flalign}
	\beta = \frac{1+\bar\alpha^2}{\bar\alpha^2 \mFP^2}c^2(\beta_2 + \beta_3 c)\,,\qquad \gamma = \frac{1+\bar\alpha^2}{\bar\alpha^2 \mFP^2} c^3\beta_3\,.
\end{flalign}
To simplify the equation of motion, we assume that the gravitational fields and their
derivatives are small,
$\{\lambda_{\rm g,f},\nu_{\rm g,f}\}\ll1$ and $\{r\lambda'_{\rm g,f},r\nu'_{\rm g,f}\}\ll1$,
but we keep all nonlinearities in $\mu$.
This results in the following set of Einstein equation for $\gmn$,
\begin{gather}
\begin{split}\label{eq:Einstein-Vainshtein-g}
	\frac{(r\lambda_{\rm g})'}{r^2} & =
	\Lambda + \frac{\rho(r)}{m_{\rm g}^2} +
	\frac{\bar\alpha^2\mFP^2}{1+\bar\alpha^2}\left(\frac{1}{2}(\lambda_{\rm f}-\lambda_{\rm g}) + \frac{1}{r^2} \Big[ r^3\left(\mu + \beta \mu^2 + \frac{\gamma}{3}\mu^3\right)\Big]'\right)\,\\
	\frac{\lambda_{\rm g}}{r^2}-\frac{\nu'_{\rm g}}{r}&= \Lambda + \frac{\bar\alpha^2\mFP^2}{1+\bar\alpha^2}\left(\frac{1}{2}(\nu_{\rm f}-\nu_{\rm g}) + 2\mu+\beta \mu^2\right)\,,\\
	-\frac{\lambda'_{\rm g}}{2r} + \frac{(r\nu_{\rm g}')'}{2r} &=\Lambda + \frac{\bar\alpha^2\mFP^2}{1+\bar\alpha^2}\left( \frac{1}{2}(\lambda_{\rm f}-\lambda_{\rm g}+\nu_{\rm f}-\nu_{\rm g}) +\frac{1}{r}\Big[r^2\Big(\mu + \frac{\beta}{2} \mu^2\Big)\Big]'\right)\,.
\end{split}
\end{gather}
With the same approximations, the Einstein equations of $\fmn$ can be arranged to
\begin{gather}
\begin{split}\label{eq:Einstein-Vainshtein-f}
	\frac{\left((r+r\mu)\lambda_{\rm f}\right)'}{r^2} &=
	\Lambda- \frac{\mFP^2}{1+\bar\alpha^2}\Bigg(\frac{1}{2}(\lambda_{\rm f}-\lambda_{\rm g}) + \frac{1}{r^2} \Big[r^3\Big( \mu +(1+\beta)\mu^2+\frac{1+\beta+\gamma}{3}\mu^3 \Big) \Big]' \Bigg)\,,\\
	(r+r\mu)'\frac{\lambda_{\rm f}}{r^2}-(1+\mu)\frac{\nu'_{\rm f}}{r} &= \Lambda - \frac{\mFP^2}{1+\bar\alpha^2}\Big(\frac{1}{2}(\nu_{\rm f}-\nu_{\rm g}) + 2\mu + (1+\beta)\mu^2\Big)(r+r\mu)'  \,,\\
	\frac{\lambda_{\rm f}}{2r} - \frac{1}{2r}\left(\frac{(r+r\mu)\nu_{\rm f}'}{(r+r\mu)'}\right)'&=\Lambda - \frac{\mFP^2}{1+\bar\alpha^2}
	\Bigg( \frac{1}{2}(\lambda_{\rm f}-\lambda_{\rm g}-\nu_{\rm f}-\nu_{\rm g}) +
	+\frac{1}{r}\Big[ r^2 \left(\mu + \frac{1+\beta}{2}\mu^2 \right) \Big]' \Bigg)
\end{split}
\end{gather}
Note that we do not present the $\phi\phi$-components of the Einstein equations
as these coincide with the $\theta\theta$-components.
The last ingredient is the Bianchi constraint, that simplifies to
\begin{flalign}\label{eq:Bianchi-radial-full}
	\frac{(r+r\mu)'}{r}(1+\beta \mu)(\lambda_{\rm f}-\lambda_{\rm g}) - 
	\frac{1}{2}(1+2\beta\mu+\gamma\mu^2)(\nu'_{\rm f} - (r+r\mu)'\nu'_{\rm g})=0\,.
\end{flalign}
Even in this simplified approximation, the equations are not solvable analytically.
To find analytic solutions, we will go to large radii where all metric functions
and their derivatives have to be small because we assume spacetime
to be asymptotically de Sitter while remaining far inside the de Sitter horizon $3/\sqrt{\Lambda}$.
The other limit is for small radii compared to the Compton wavelength of the
massive spin-2 field.

\subsection{Approximate solutions}

\subsubsection{Linear regime}
\label{sec:linear-regime}
First, we will solve the equations of motion  for $r\gg r_{\rm S}$ with $r_{\rm S}$ the Schwarzschild radius of a localised source to be defined later.
For large radii smaller than the de Sitter horizon, all metric functions, including $\mu$, are assumed to be small such that both metrics approach de Sitter solution asymptotically.
We expand the Einstein equations and the Bianchi constraint further in $\mu\ll1$ and $r\mu'\ll1$.
It is convenient to introduce the auxiliary functions,
\begin{flalign}
	\lambda_{\pm} = \lambda_f \pm \lambda_g\,,\ \nu_{\pm}=\nu_f \pm \nu_g
\end{flalign}
In terms of these functions, the linearised Bianchi constraint~(\ref{eq:Bianchi-radial-full}) reads,
\begin{flalign}
	2\lambda_- = r\nu'_- \,,
\end{flalign}
which is independent of the $+$-fields.
We can find linear combinations of the Einstein equations for $\gmn$ and $\fmn$
such that the $+$-fields drop out of the expression,
\begin{gather}
\begin{split}
	0& = 2r \lambda'_- + \left(2+\mFP^2 r^2 \right)\lambda_-  + 2\mFP^2 r^2 (3\mu+r\mu')\,,\\
	0& = -2\lambda_- + \mFP^2r^2 (\nu_- + 4\mu) \,,\\
	0& = r\lambda'_- - \mFP^2 r^2(\lambda_- + \nu_- + 4\mu + 2r\mu')\,,
\end{split}
\end{gather}
where we have used the Bianchi constrained already to eliminate $\nu'_-$ and $\nu''_-$
in terms of $\lambda_-$ and $\lambda'_-$.
The equations reduce to two coupled differential equations for $\lambda_-$ and $\mu$,
\begin{flalign}
	\lambda'_- = -2 \mFP^2\, r\mu\,,\qquad
	\mu'  = -\frac{(2+\mFP^2\, r^2)\lambda_- + 2\mFP^2\, r^2 \mu}{2 \mFP^2\, r^3}\,,
\end{flalign}
while $\nu_-$ is algebraically determined by the other two fields.
These equations are solved by
\begin{gather}
\begin{split}\label{eq:solution-minus-fields}
	\nu_- &= \frac{C_1 e^{-\mFP\, r}}{r}\,,\\
	\lambda_-  &=-\frac{C_1(1+\mFP\, r)e^{-\mFP\, r}}{2r}\,,\\
	\mu  &= -\frac{C_1\left(1 + \mFP^2\, r + \mFP^2 r^2 \right)e^{-\mFP\, r}}{4\mFP^2 r^3}\,,
\end{split}
\end{gather}
where $C_1$ is a constant of integration to be determined later.
We already fixed the other constant of integration by requiring that the fields
vanish in the limit $r\rightarrow\infty$
in order to ensure that the metrics are asymptotically bidiagonal and de Sitter.
Next, we turn to the other set of linearly independent equations that contain also 
the $+$-fields. These can be arranged to
\begin{gather}
\begin{split}
	0 &= (\bar\alpha^2-1)(\lambda_- + r\lambda'_-) + (1+\bar\alpha^2) (\lambda_+ + r\lambda'_+)-2(1+\bar\alpha^2)\Lambda r^2\,,\\
	0 &= (\bar\alpha^2-1)\lambda_- - (1+\bar\alpha^2)(\lambda_+ - r\nu'_+) +2(1+\bar\alpha^2)\Lambda r^2 \,,\\
	0 &= (\bar\alpha^2-1) \lambda'_- - (1+\bar\alpha^2)(\lambda'_+ - \nu'_+ - r\nu''_+ - 4\Lambda r)
\end{split}
\end{gather}
Upon using~\cref{eq:solution-minus-fields}, these equations are solved by
\begin{gather}
\begin{split}
	\nu_+ &= -\frac{2\Lambda r^2}{3}-\frac{2C_2}{r} -\frac{ C_1 e^{-\mFP\, r}}{(1+\bar\alpha^2)r}\,,\\
	\lambda_+ &= \frac{2\Lambda r^2}{3} + \frac{2C_2}{r} + \frac{C_1 (1+ \mFP\, r)e^{-\mFP\, r}}{2(1+\bar\alpha^2)r}\,,
\end{split}
\end{gather}
where $C_2$ is another constant of integration.
Again, we have already used that the functions have to be bidiagonal and de Sitter
in the limit $r\rightarrow\infty$ to fix one of the constants of integration.
Having solved the differential equations for all the fields, we can present the final solutions
for the metric fields in the linearised limit.
They are given by
\begin{gather}
\begin{split}\label{eq:linear-solutions}
	\mu &= - \frac{C_1(1+\mFP\, r+\mFP^2 r^2)e^{-\mFP\, r}}{4\mFP^2 r^3}\,,\\
	\nu_{\rm g} &= -\frac{\Lambda r^2}{3} - \frac{C_2}{r} - \frac{C_1\bar\alpha^2 e^{-\mFP\, r}}{(1+\bar\alpha^2)r}\,, \\ 
	\nu_{\rm f} &= -\frac{\Lambda r^2}{3} -\frac{C_2}{r} + \frac{C_1 e^{-\mFP\, r}}{(1+\bar\alpha^2)r}\,,\\
	\lambda_{\rm g} &= \frac{\Lambda r^2}{3} + \frac{C_2}{r} + \frac{C_1\bar\alpha^2(1+\mFP\, r)e^{-\mFP\, r}}{2(1+\bar\alpha^2)r}\,, \\
	\lambda_{\rm f} &= \frac{\Lambda r^2}{3} + \frac{C_2}{r} - \frac{C_1(1+\mFP\, r)e^{-\mFP\, r}}{2(1+\bar\alpha^2)r}\,.
\end{split}
\end{gather}
In the limit of a vanishing cosmological constant, $\Lambda=0$, our linearised solutions reduce the the results obtained in Refs.~\cite{Babichev:2013pfa,Enander:2015kda}.

Let us comment on the region of validity of these solutions.
As we will see later, $C_1\sim r_{\rm S}$.
Hence, all metric fields are small on scales $r\gg r_{\rm S}$ and inside the de Sitter horizon, $r\ll \Lambda^{-1/2}$.
This is not true however for the St\"uckelberg field $\mu$, which is small only on scales $r\gg (r_{\rm S}/\mFP^2)^{1/3}$.
In general, this scale is larger than the Schwarzschild radius of the source.
These solutions are hence not appropriate to describe the scales $r_{\rm S}\ll r \ll (r_{\rm S}/\mFP^2)^{1/3}$.
In the next section, we seek solutions that are valid on these scales.

\subsubsection{Compton regime}
Next, we want to solve the equations on scales much smaller than the Compton wavelength of the massive spin-2 field, i.e. for $r\ll\mFP ^{-1}$.
Technically that means that we omit the terms $\mFP^2r^2 \cdot\{\lambda_{\rm g,f},\nu_{\rm g,f}\}\ll 1$ in the modified Einstein equations.

Under this approximation we can integrate the $tt$-component of~\cref{eq:Einstein-Vainshtein-g} to
\begin{flalign}\label{eq:nl-sol-lg}
	\lambda_{\rm g} = \frac{r_{\rm S}}{r} + \frac{\bar\alpha^2\mFP^2}{1+\bar\alpha^2}r^2\left(\mu + \beta \mu^2 + \frac{\gamma}{3} \mu^3\right)\,,
\end{flalign}
where we defined the Schwarzschild radius
\begin{flalign}
	r_{\rm S} = \frac{1}{4\pi m_{\rm g}^2}\int_0^{R_*} \rho r^2 \dd r
\end{flalign}
of a source of mass $M$ and radius $R_*$.
The solution is valid outside the compact object
and we fixed the constants of integration by demanding regularity at the surface of
the compact object and at the origin $r=0$.
Upon using the solution for $\lambda_{\rm g}$, the $rr$-component of~\cref{eq:Einstein-Vainshtein-g} simplifies to
\begin{flalign}\label{eq:nl-sol-ng}
	r\nu'_{\rm g} = \frac{r_{\rm S}}{r}
	- \frac{\bar\alpha^2\mFP}{1+\bar\alpha^2}r^2
	\left(\mu - \frac{\gamma}{3} \mu^3 \right)\,.
\end{flalign}
The $tt$-component of~\cref{eq:Einstein-Vainshtein-f} can easily be integrated to
\begin{flalign}\label{eq:nl-sol-lf}
	\lambda_{\rm f} = - \frac{\mFP^2}{1+\bar\alpha^2}\frac{r^2}{1+\mu}
	\left(\mu+(1+\beta)\mu^2+\frac{1+\beta+\gamma}{3} \mu^3 \right)\,.
\end{flalign}
Here, we fixed the constant of integration by demanding regularity at the origin $r=0$.
With this result, the $rr$-component of~\cref{eq:Einstein-Vainshtein-g} simplifies to
\begin{flalign}
	r\nu'_{\rm f} = \frac{\mFP^2}{1+\bar\alpha^2}\frac{r^2(r+r\mu)'}{(1+\mu)^2}
	\left(\mu + 2\mu^2 + \frac{2+2\beta-\gamma}{3}\mu^3\right)\,.
\end{flalign}
We have obtained expression for all the fields that show up in the
Bianchi constraint in terms of $\mu$.
Plugging these results into the Bianchi constraint yields the following
algebraic polynomial for $\mu$:
\begin{gather}
\begin{split}\label{eq:mu-polynomial}
-3(3+3\bar\alpha^2)\mu\ \, &\\
-2 \left( 9(1+\bar\alpha^2)(1+\beta) \right)\mu^2&\\
- \left(10+9\bar\alpha^2 + 2(17+18\bar\alpha^2)\beta + 6(1+\bar\alpha^2)\beta^2 + 4(1+\bar\alpha^2)\gamma\right) \mu^3&\\
 - 2\left(1+ (7+9\bar\alpha^2)\beta + 6(1+\bar\alpha^2)\beta^2 + 4(1+\bar\alpha^2)\gamma\right)\mu^4&\\
 -\left(2(1+2\gamma)\beta + 2(1+3\bar\alpha^2)\beta^2 + 2(1+2\bar\alpha^2)\gamma - (1+\bar\alpha^2)\gamma^2 \right) \mu^5&\\
+ 2\bar\alpha^2\gamma^2\mu^6 + \bar\alpha^2\gamma^2\mu^7 & \\
= 3\left(\frac{r_{\rm V}}{r}\right)^3 (1+\bar\alpha^2) (1+\mu)^2(1-\gamma\mu^2)&\,.
\end{split}
\end{gather}
In here, we defined the Vainshtein radius
\begin{equation}
	r_{\rm V} = \left(\frac{r_{\rm S}}{\mFP^2} \right)^{1/3}
\end{equation}
that we already encountered before.
Our result coincides with the one obtained in Ref.~\cite{Enander:2015kda}.

\begin{figure}
\centering
\includegraphics[width=0.8\textwidth]{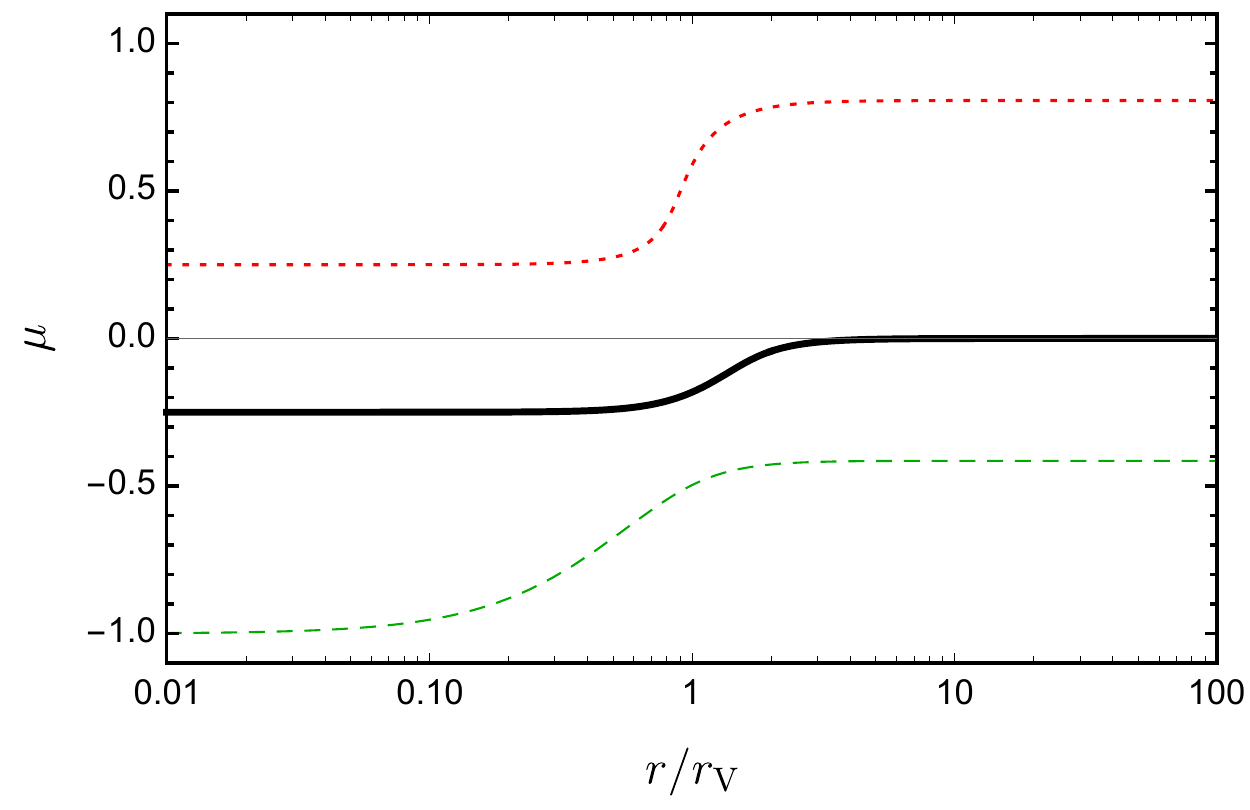}
\caption{Plot of the St\"uckelberg field $\mu$ as a function of $r/r_{\rm V}$
for the exemplary parameter values $\bar\alpha=1$, $\beta=1$, and $\gamma=16$.
The Vainshtein-Yukawa solution (solid black) smoothly evolves from zero
at large distances to a non-zero value inside the Vainshtein regime such as to restore
GR on these scales.
The other two solutions (red dotted and green dashed) also restore GR
inside the Vainshtein regime, but they do not tend to
zero asymptotically.
For the same set of parameters, there are two more solutions that are real valued
only for small radii $r$.}
\label{fig:VainshteinYukawa}
\end{figure}

Eq.~(\ref{eq:mu-polynomial}) represents a seventh order algebraic polynomial for $\mu$ and as such has up to seven solutions.
As argued in Ref.~\cite{Babichev:2013pfa} for $\gamma>0$, three
solutions are real-valued for all $r$, while two solutions are real only
up to a certain critical radius that depends on the other parameters.
The last two solutions are complex-valued.
The three everywhere real-valued solutions are shown in Fig.~\ref{fig:VainshteinYukawa}.
In the opposite case $\gamma<0$, two of the real-valued solutions become complex-valued
and the remaining real-valued solutions are neither asymptotically bidiagonal nor restore GR inside the Vainshtein regime.
The case $\gamma=0$ is special because the polynomial is only of degree five.
There is one solution that realises the Vainshtein mechanism for small
radii, which however is not asymptotically bidiagonal and it is complex-valued for very small radii.
Although there might be special situations in which the Vainshtein mechanism works
even for this case~\cite{Volkov:2012wp,Babichev:2013pfa},
we exclude $\gamma=0$ from our analysis and restrict to $\gamma>0$.

We can infer further information about the general behavior of $\mu$.
For radii much larger than the Vainshtein radius, the term $(r_{\rm V}/r)^3$ on the right hand side of~\cref{eq:mu-polynomial} is small,
which implies that the St\"uckelberg field is small as well, $\mu\ll1$.
This gives rise to three asymptotic values for $\mu$ as can be seen in Fig.~\ref{fig:VainshteinYukawa}.
One of these is given by $\mu=0$ such that we can neglect nonlinearities in $\mu$.
This solution corresponds to the one obtained when linearising the field equations as in~\cref{sec:linear-regime}.
For small radii, the term $(r_{\rm V}/r)^3$ on the right hand side is large.
This allows to identify two classes of solutions.
Either, $\mu$ on the right hand side compensates that large term.
This is the case for $\mu=-1$ or $\mu=\pm1/\sqrt{\gamma}$.
Note that the second solution exists only for $\gamma>0$.
These are the three asymptotic values that show up in Fig.~\ref{fig:VainshteinYukawa}.
The other possibility is that $\mu$ diverges as $r\rightarrow 0$.

\subsubsection{Matching the analytic solutions}
It remains to determine the constants of integration $C_1$ and $C_2$.
We do this by matching the analytic solutions in the regime where both of them are valid, i.e.
$r_{\rm V} \ll r \ll \mFP^{-1}$.
Linearising the nonlinear solutions~\cref{eq:nl-sol-lg,eq:nl-sol-lf,eq:mu-polynomial} for $r\gg r_{\rm V}$, we obtain
\begin{flalign}
	\mu = - \frac{r_{\rm S}}{3\mFP^2 r^3}\,,\qquad
	\lambda_{\rm g} = \frac{(3+2\bar\alpha^2)r_{\rm S}}{3(1+\bar\alpha^2)r}\,,\qquad
	\lambda_{\rm f} = \frac{r_{\rm S}}{3(1+\bar\alpha^2)r}\,.
\end{flalign}
Here, we linearised the St\"uckelberg field around $\mu=0$.
Expanding on the other hand the linear solutions~(\ref{eq:linear-solutions}) for $r\ll \mFP^{-1}$ yields
\begin{flalign}
	\mu = -\frac{C_1}{4\mFP^2 r^3}\,,\qquad
\lambda_{\rm g} = \frac{\bar\alpha^2 C_1+2(1+\bar\alpha^2)C_2}{2(1+\bar\alpha^2)r}\,,\qquad
\lambda_{\rm f} =-  \frac{C_1-2(1+\bar\alpha^2)C_2}{2(1+\bar\alpha^2)r}\,.
\end{flalign}
These solutions coincide, if we choose the constants of integration to be
\begin{flalign}\label{eq:int-constants}
	C_1 = \frac{4r_{\rm S}}{3}\,, \qquad C_2 = \frac{r_{\rm S}}{1+\bar\alpha^2}\,.
\end{flalign}

\subsection{Existence of Vainshtein-Yukawa branch}
\label{sec:Vainshtein-Yukawa-Condition}
So far, we have studied analytic solutions on two different scales.
For large radii, $r\gg r_{\rm V}$, the gravitational potential is a combination
of the $1/r$-law and a Yukawa-type potential:
\begin{equation}
	\nu_{\rm g} =- \frac{r_{\rm S}}{(1+\ba^2)}\left(\frac{1}{r} + \frac{4\ba^2}{3}\frac{e^{-\mFP\, r}}{r}\right)
\end{equation}
as follows from~\cref{eq:linear-solutions,eq:int-constants}.
On smaller length scales, $r_{\rm S} \ll r \ll r_{\rm V}$ the St\"uckelberg field $\mu$ is nonlinear.
The nonlinearities are such that the gravitational potential is given by
\begin{equation}
	\nu_{\rm g} = -\frac{r_{\rm S}}{r}
\end{equation}
as follows from~\cref{eq:nl-sol-ng} on a solution where $\mu$ is constant.
This is summarised in~\cref{eq:gravitational-potential}.
For the screening mechanism to work it is necessary that the solution of the St\"uckelberg field realising these two regimes exists and is real-valued for every $r$ without branch cuts.

\begin{figure}
\centering
\includegraphics[width=0.45\textwidth]{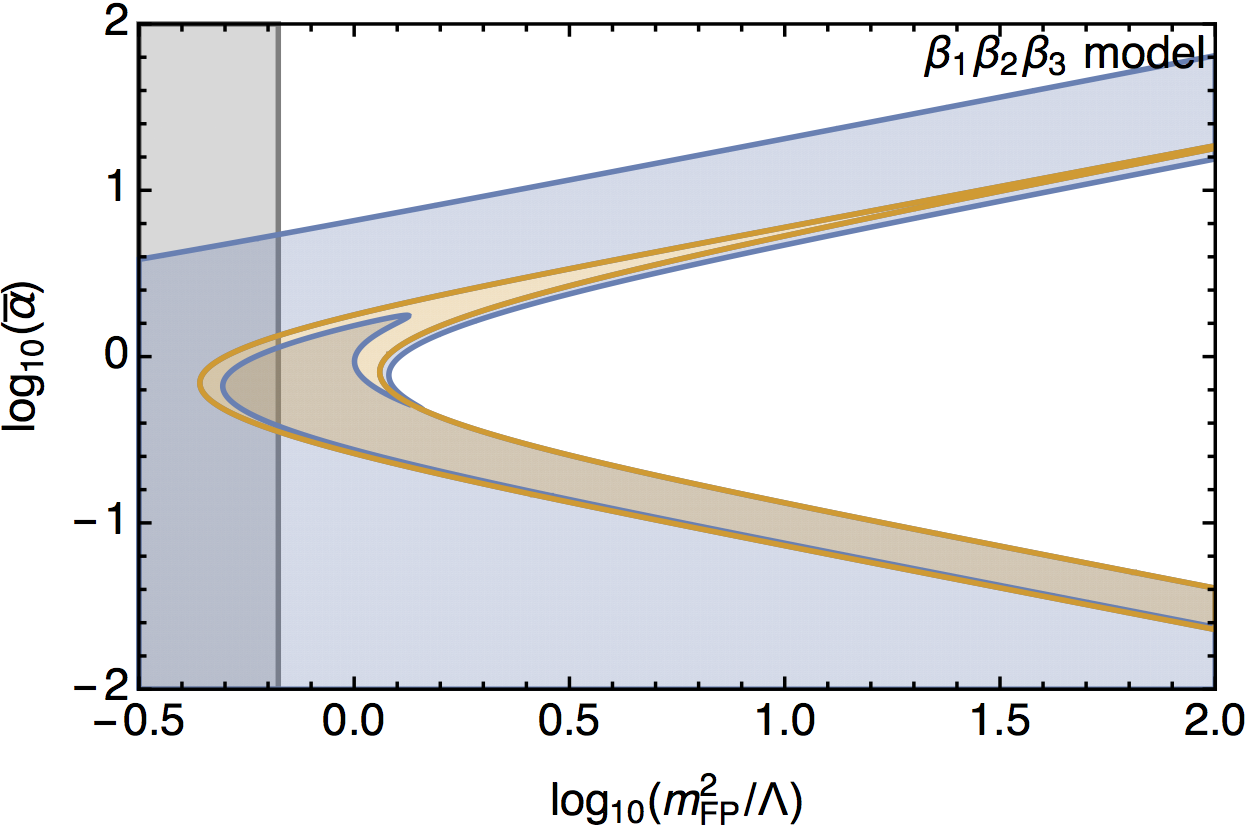}
\includegraphics[width=0.45\textwidth]{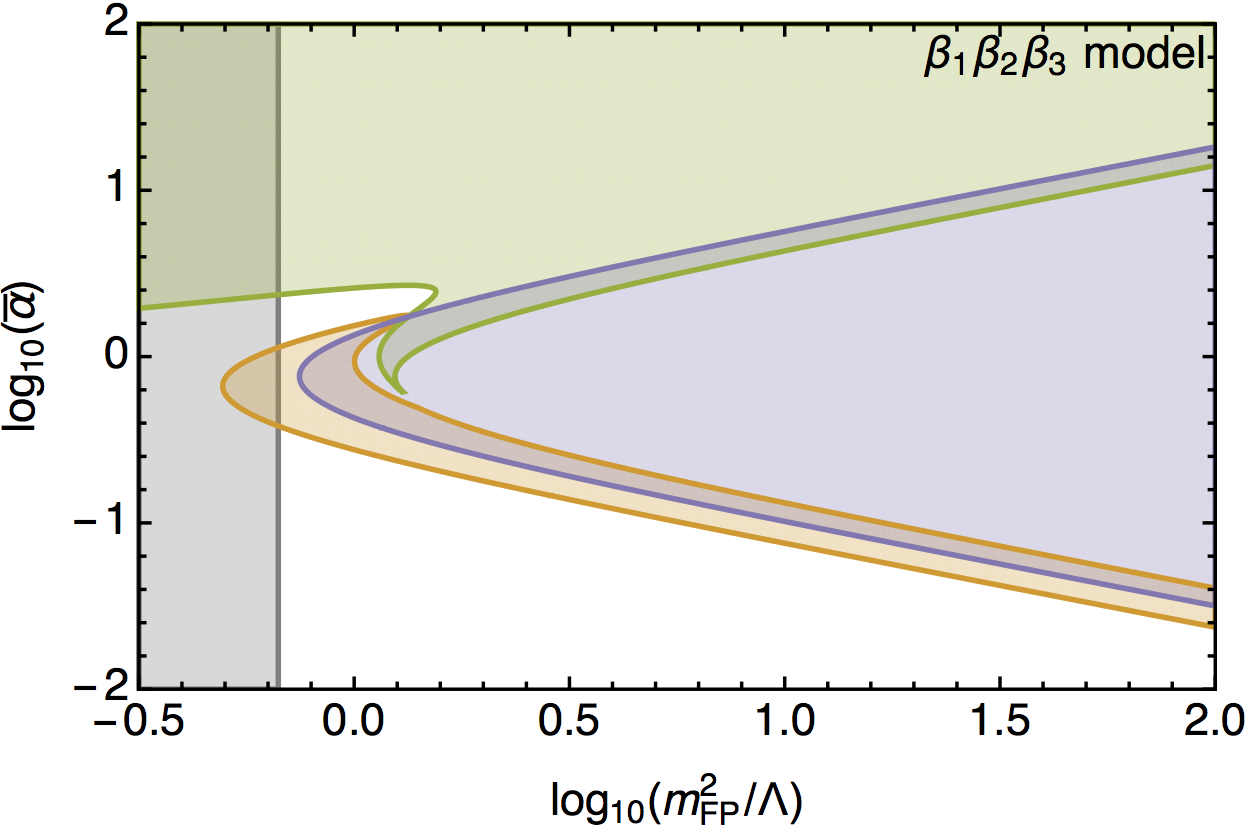}
\caption{Exclusion plots for a working Vainshtein mechanism for the
$\beta_1\beta_2\beta_3$-model.
\textit{Left panel:} In the blue-shaded region $\beta < d_1/d_2$ is implied while in
the yellow-shaded region $d_2<0$ holds.
In the overlap, there is no everywhere real solution for the St\"uckelberg field $\mu$ that incorporates
the Vainshtein mechanism at small radii while giving rise to the correct asymptotics.
\textit{Right panel:} In the green-shaded region $\mu\ll1$ is no solution for large radii while in the purple-shaded region $\gamma<1$.
The yellow-shaded region shows the overlap of the bounds presented in the left panel.
In the gray shaded region the Higuchi bound is violated.
In summary, only outside the shaded regions, the bimetric model incorporates a working Vainshtein mechanism.}
\label{fig:Vainshtein-cond}
\end{figure}

This point was studied in detail in~\cite{Enander:2015kda}.
Firstly, the nonlinear equation~(\ref{eq:mu-polynomial}) determining $\mu$ must give rise to a solution with $\mu\ll 1$ for $r\gg r_{\rm V}$, which matches the linearised solution.
This is the case only if
\begin{equation}\label{eq:consistent-linear-sol}
	\beta  < \sqrt{\gamma}
\end{equation}
is satisfied.
This bound is stated in~\cref{eq:consistent-asymptotics} and ensures a consistent asymptotic behavior of the nonlinear solutions.
This is the condition ensuring consistent asymptotics as stated in~\cref{eq:consistent-asymptotics}.
Secondly, the linearised solution can be smoothly connected only to the nonlinear solution with $\mu=-1/\sqrt{\gamma}$, which is required to be larger than the solution with $\mu=-1$.
Therefore, a consistent screening regime requires the parameters to satisfy
\begin{equation}\label{eq:pos-gamma}
	\gamma >1
\end{equation}
as stated in~\cref{eq:pos-b3}.
Thirdly, there must be a solution that is real-valued for every $r$ and interpolates between the two aforementioned asymptotic limits without branch cuts.
This is the case if the parameters satisfy the following bound:
\begin{equation}\label{eq:Vainshtein-Yukawa-cond}
	\beta  > \frac{d_1}{d_2} \quad {\rm if} \quad d_2<0\,,
\end{equation}
where
\begin{gather}
\begin{split}
	d_1 & = -1+6(1+\bar\alpha^2)\sqrt{\gamma}(1+\gamma)-(13+12\bar\alpha^2)\gamma\,,\\
	d_2 & = 1+3\bar\alpha^2-6(1+\bar\alpha^2)\sqrt{\gamma}+3(1+\bar\alpha^2)\gamma\,.
\end{split}
\end{gather}

The bound in~\cref{eq:pos-gamma} implies $\beta_3>0$.
Combining~\cref{eq:consistent-linear-sol,eq:pos-gamma} then implies $\beta_2<0$.
A theoretically consistent expansion history requires $\beta_1>0$.
Therefore, the $\beta_1\beta_2\beta_3$-model is the only bimetric model with up to three free parameters that allows for Vainshtein screening and has a consistent background cosmology.
In~\cref{fig:Vainshtein-cond} the bounds in~\cref{eq:consistent-linear-sol,eq:pos-gamma,eq:Vainshtein-Yukawa-cond} are presented in terms of the physical parameters $\ba$ and $\mFP^2/\Lambda$ for that model.
To express the interaction in terms of the physical parameters, we use the parameter relations reported in~\cref{sec:details-physical-parametrisation}.
In the left panel the violation of the bound~(\ref{eq:Vainshtein-Yukawa-cond}) is shown.
In the overlap of the blue- and yellow-shaded regions, the Vainshtein-Yukawa solution does not exist.
In the right panel, we combine this bound with the other two.
The shaded regions represent, where the bounds are violated and hence where the Vainshtein screening mechanism does not work.
Only the region left white gives rise to screening.
In~\cref{fig:b123-param-space}, the Vainshtein bounds are combined with the bounds ensuring a consistent background cosmology.

\bibliographystyle{jhep}
\bibliography{CosmoTests}

\end{document}